\documentclass{aa}
\usepackage{graphicx, subcaption, color, amssymb, ulem, url, bm, mathtools, amsmath, natbib}
\usepackage[varg]{txfonts}
\newcommand{\D}{{\rm d}}

\begin{document}

\title{Relativistic theory for time and frequency transfer through flowing media with an application to the atmosphere of Earth}
\author{Jan Ger{\v s}l}
\institute{Czech Metrology Institute, Okru{\v z}n{\' i} 31, 63800 Brno, Czech Republic\\
\email{jgersl@cmi.cz}}

\abstract
{Several space missions that will use atomic clocks on board of an Earth-orbiting satellite are planned for the near future, such as the Atomic Clock Ensemble in Space (ACES) or the Space Optical Clock on the International Space Station (I-SOC). The increasing accuracies of the developed clocks and of the links connecting them with ground stations impose corresponding accuracy requirements for theoretical models of electromagnetic signal propagation through the atmosphere of Earth and for the related time and frequency transfer corrections. For example, the fractional frequency accuracy of the optical lattice clock for the I-SOC project is about $10^{-17}$.}
{We develop a relativistic model of one- and two-way time and frequency transfer. In addition to the gravitational effects, it also includes the effects of atmospheric refractivity and atmospheric flows within the relativistic framework.}
{The model is based on an analytical solution of the equation of motion of a light ray in spacetime filled with a medium: the null geodesic equation of Gordon's optical metric.} 
{Explicit formulas for one- and two-way time and frequency transfer corrections are given using realistic fields of the gravitational potential, the refractive index, and the wind speed, taking nonstationarity and deviations from spherical symmetry into account. Numerical examples are provided that focus on two-way ground-to-satellite transfer, with satellite parameters similar to those of the International Space Station. The effect of the atmospheric refractive index increases as the satellite position moves from zenith to horizon, and it is shown that the effect ranges from 0~ps to 5~ps for two-way time transfer and from $10^{-17}$ to $10^{-13}$ for two-way frequency transfer, with a steep increase as the satellite approaches the horizon. The effect of the wind contribution is well below 1~ps for the two-way time transfer for normal atmospheric conditions, but for the two-way frequency transfer, the effect can be significant:  A contribution of $10^{-17}$ is possible for a horizontal wind field with a velocity magnitude of about 11~m/s.}
{The atmospheric effects including the effect of wind should be considered in the forthcoming clock-on-satellite experiments such as ACES or I-SOC.}

\keywords{relativistic processes -- atmospheric effects -- time -- reference systems -- methods: analytical} 
\titlerunning{Relativistic theory for time and frequency transfer through flowing media}

\maketitle

\section{Introduction}

Electromagnetic signals propagating through the atmosphere of Earth are used for time and frequency transfer between distant atomic clocks located on the surface of the Earth or at satellites with applications in fundamental science \citep{Delva_review, Safronova, Delva2, Beloy}, geodesy \citep{Muller, Mehlstaubler, Delva3}, and metrology \citep{Fujieda, Hachisu, Riedel, Pizzocaro}. Future perspectives in this field are described, for example, by \citet{Alonso} and \citet{Derevianko}.

Recent developments in optical atomic clocks that reach fractional systematic uncertainty in frequency on the order of $10^{-18}$ in case of stationary clocks \citep{Ludlow, McGrew, Lisdat} or $10^{-17}$ in case of transportable clocks \citep{Koller, Cao, Origlia}, together with developments in free-space optical links (see \citet{Bodine} and references therein and also \citet{Djerroud}, \citet{Gozzard}, \citet{Kang}, \citet{Dix-Matthews} and \citet{Shen}) enable increasing accuracies of the experiments. Therefore, the atmospheric effects originating from spatial and temporal variations of the refractive index of air or from winds start to play a more important role.

In case of the ground-to-satellite time and frequency transfer, two projects are going to be realized in the near future that will place an atomic clock on the International Space Station (ISS), namely the Atomic Clock Ensemble in Space (ACES), with a Cs clock and H-maser \citep{Lilley, Cacciapuoti}, and the Space Optical Clock on the ISS (I-SOC), with an optical lattice clock \citep{Schiller, Origlia}. In the ACES mission, the absolute frequency accuracy of the on-board clock is about $10^{-16}$ in fractional frequency, whereas the same parameter for the I-SOC mission is about $10^{-17}$. This gives an uncertainty limit to one of the main objectives of the experiments, that is, the test of the gravitational redshift. To compare time and frequency between various clocks, ACES will use a microwave link (MWL) and an optical link of the European Laser Timing (ELT) experiment \citep{Schreiber}. When used for time comparisons between ground clocks, the MWL is expected to provide absolute synchronization of the ground clock timescales with an uncertainty of 100 ps. For the ELT, the overall planned accuracy of the time transfer is 50 ps \citep{Turyshev}. The I-SOC proposal uses an improved version of the MWL as well as the optical link of ELT, aiming at an accuracy of a few ps in the clock synchronization \citep{Schiller}. Therefore, based on the planned accuracies of these experiments, a relativistic model of one- and two-way time and frequency transfer is needed that includes the atmospheric effects and has an accuracy of approximately 1 ps for the time transfer and a relative accuracy of the lower multiples of $10^{-18}$ for the frequency transfer.

\citet{Blanchet} developed the relativistic theory of one- and two-way time and frequency transfer in vacuum, including terms up to the order of $c^{-3}$ to cover contributions relevant for experiments with an uncertainty of $5\times 10^{-17}$ in frequency transfer and 5 ps in time transfer. This theory is based on a solution of the null geodesic equation for a light ray in vacuum, static, spherically symmetric spacetime.

Relativistic theory of the propagation of electromagnetic signals in the broader context of astrometric measurements in the Solar System is discussed, for example, in \citet{Kopeikin_book} and references therein. The propagation time and frequency shift of a signal in the gravitational field of multiple moving bodies were studied especially in \citet{Kopeikin1} and \citet{Kopeikin2}, where the Li\'{e}nard-Wiechert representation of the metric tensor up to the first post-Minkowskian order was used.

\citet{Linet} applied the Synge world function formalism to the gravitational field of an axisymmetric rotating body, and they showed that certain $c^{-4}$ terms in frequency shift approach the $10^{-18}$ value for frequency transfer from a ground station to the ISS. They also discussed the influence of the quadrupole moment of the Earth $J_2$ at the $c^{-3}$ order. The theory of \citet{Linet} was further developed by \citet{LePoncin} and \citet{Teyssandier}, where the formalism of time transfer functions was introduced. \citet{Hees1} and \citet{Hees2} employed this formalism to compute various observables relevant for actual space missions in the Solar System, including the coordinate propagation time and the frequency shift of light. 

The relativistic theoretical works mentioned above considered the vacuum case alone and did not take the atmospheric effects into account. The theory of \citet{Teyssandier} was extended to include the atmospheric effects by \citet{Bourgoin2}. Here, a general formalism of time transfer functions was developed using the Gordon optical metric \citep{Gordon}, and it was applied to a case of stationary optical spacetime. The effects of refractivity and motion of a neutral atmosphere were added to the gravitational effects, and the light-dragging effect due to the steady rotation of the atmosphere of Earth was discussed. The models of \citet{Linet} and \citet{Bourgoin2} focused on one-way time and frequency transfer only. The results of \citet{Bourgoin2} were later used to model the atmospheric occultation experiments \citep{Bourgoin3}. In this work, explicit formulas for time and frequency transfer were derived for a steadily rotating spherically symmetric atmosphere up to the first post-Minkowskian order, with the remark that the method described in the paper easily allows extending the results to higher orders and beyond the spherical symmetry. \citet{Feng} used the formalism of Gordon's optical metric in the context of global navigation satellite systems to include the atmospheric effects in the relativistic framework. The null geodesics of the Gordon metric were solved numerically. The effect of motion of the atmosphere of Earth on the speed of light (the Fresnel-Fizeau effect) and the related propagation time delay were discussed also by \citet{Kopeikin3} in the context of geodetic very long baseline interferometry.

Several effects originating from the atmospheric refractivity have been addressed in the literature in a nonrelativistic framework. The influence of the refractive index fluctuations given by atmospheric turbulence has been discussed by \citet{Sinclair}, \citet{Robert}, \citet{Sinclair2}, \citet{Belmonte}, \citet{Swann}, and \citet{Taylor}. Nonrelativistic analytical solutions of light rays in planetary atmospheres were discussed by \citet{Bourgoin1}. Propagation-time nonreciprocity in two-way ground-to-satellite time transfer due to the distribution of the atmospheric refractive index was discussed by \citet{Stuhl}. Earlier findings on tropospheric and ionospheric corrections were summarized, for example, in \citet{IERS}. A two-way time transfer model including the effect of ionospheric dispersion was discussed by \citet{Duchayne}, who also provided uncertainty requirements on the orbit determination of space clocks.

The aim of this paper is to develop a relativistic model of one- and two-way time and frequency transfer that includes gravitational and atmospheric effects for realistic fields of the gravitational potential, the refractive index, and the wind speed, assuming neither stationarity nor spherical or axial symmetry. The intended accuracy of the model is given by the experimental needs described above: approximately 1~ps for time transfer, and lower multiples of $10^{-18}$ for frequency transfer. The model is based on the solution of the equation of motion for a light ray, that is, the null geodesic equation in Gordon's optical metric, which is given by the fields of the gravitational potential, the atmospheric refractive index, and the wind speed in the vicinity of the Earth. Dispersion of the medium and free electric charges or currents are not included in the model.

The paper is organized as follows. After defining the notation and conventions in Sect.~\ref{Notation}, we introduce the coordinates, the spacetime metric, and the fields describing the medium in Sect.~\ref{SceneActors}. Then, Sect.~\ref{Summary} summarizes the main results of the paper for easy reference. In the Sect.~\ref{Summary}, the level of approximation of the model is explained and the solution for the light rays in the spherically symmetric, static case is derived because this solution is needed to express the results for the general case without the symmetries, the formulas for one- and two-way time and frequency transfer are given, and several examples are shown that quantify the atmospheric effects for some typical situations whose parameters are similar to the experiments at the ISS. A complete derivation of the results is then given in Sect.~\ref{sec_Theory}. In the appendix, some basic facts about the refractive index of air and its distribution in the atmosphere of Earth are summarized.

\section{Notation and conventions} \label{Notation}

In this paper, $c$ is the speed of light in a vacuum, and $G$ is the Newtonian gravitational constant. The spacetime metric signature $(- + + +)$ is used. Small Latin indices run from 1 to 3, and small Greek indices run from 0 to 3. Partial derivatives with respect to spacetime coordinates are denoted as $\partial_\alpha=\partial/\partial x^\alpha$ and also $\partial_t=\partial/\partial t$. Einstein's summation convention is used. The Euclidean norm of a triplet of components $a^i$ is denoted as $|a^i|\equiv\sqrt{\delta_{ij}a^ia^j}$, with $\delta_{ij}$ being the Kronecker delta. Quantities $\omega^i, v_R^i, v^i, V^i, {{A}}^i, h^i, \chi^i, \varsigma^i,\text{ and } \gamma^i$ in this paper are treated as Euclidean three-vectors when their index is lowered or raised, for example, $V_i=\delta_{ij}V^j$. The three-dimensional antisymmetric Levi-Civita symbol is denoted as $\epsilon_{ijk}$, eventually $\epsilon^i_{\ jk}=\delta^{im}\epsilon_{mjk}$.

\section{Coordinates, fields, and observers} \label{SceneActors}

\subsection{Background metric and coordinate system} \label{sec_background}

The Geocentric Celestial Reference System (GCRS) is a coordinate system in the surroundings of the Earth that is centered in the center of mass of the Earth and is nonrotating with respect to distant stars. We denote the GCRS coordinates as $\tilde{x}^\alpha=(\tilde{x}^0, \tilde{x}^i)$, where $\tilde{x}^0/c=t$ is the Geocentric Coordinate Time (TCG), and $\tilde{x}^i$ are the spatial coordinates of the system. 
The components of the spacetime metric in the GCRS coordinates  up to the order that is sufficient for the analysis further are \citep{Soffel}
\begin{subequations}
\begin{eqnarray}
\tilde{g}_{00}&=&-1+\frac{2W}{c^2}+O(c^{-4})\\
\tilde{g}_{0i}&=&O(c^{-3})\\
\tilde{g}_{ij}&=&\delta_{ij}\left(1+\frac{2W}{c^2}\right)+O(c^{-4}),
\end{eqnarray}
\label{gGCRS}
\end{subequations}
with $W$ being the scalar gravitational potential defined by \citet{Soffel} with the convention $W\geq~\!\!0$. 

To evaluate the time and frequency transfer, however, we use coordinates that rotate with respect to the GCRS system, with a constant angular velocity vector approximating the real angular velocity vector of the rotation of the Earth around its axis, such that the surface of the Earth is nearly at rest in the new coordinates, and we can use the advantage of nearly vanishing velocities of stationary observers on the ground. The use of this corotating system leads to more complicated formulas for one-way time and frequency transfer, but, on the other hand, the fact that the position of the stationary observer on the ground is almost constant makes the calculation of two-way corrections more straightforward.  

The rotating coordinate system $x^\alpha=(x^0, x^i)$ is related to the GCRS system as $x^0=\tilde{x}^0,\ x^i=R^i_{\ j}\tilde{x}^j$, where $R^i_{\ j}(t)\in{\rm SO(3)}$, or inversely, $\tilde{x}^i=R^{\ i}_{\!j}x^j$, where $R^{\ i}_{\!j}=\delta_{jk}\delta^{il}R^k_{\ l}$. The time-dependent rotation matrix $R^i_{\ j}(t)$ corresponds to the rigid uniform rotation with a constant angular velocity vector that approximates the real angular velocity vector of the Earth. Therefore, we have
\begin{equation}
\frac{\D R^{\ i}_{\!n}}{\D t}=\epsilon_{ijk}\tilde{\omega}^jR^{\ k}_{\!n},
\end{equation}
where $\tilde{\omega}^j$ are constant components of the angular velocity vector of the coordinate rotation expressed in the $\tilde{x}^i$ frame. Since $\tilde{\omega}^j$ is defined to be constant and the coordinate rotation is rigid, the irregularities in the rotation of the Earth and the deformations of the Earth given mainly by the solid Earth tides are observed as tiny movements of the surface of the Earth in the coordinates $x^i$. We denote $\tilde{v}_{R}^i=\D \tilde{x}^i/\D t=\epsilon_{ijk}\tilde{\omega}^j \tilde{x}^k$ the velocity of a point that is fixed in the rotating frame with respect to the GCRS frame. In terms of components in the rotating frame, we have $v_{R}^i=R^i_{\ j}\tilde{v}_R^j=\epsilon_{ijk}\omega^j x^k$  with  $\omega^i=R^i_{\ j}\tilde{\omega}^j$.

The components of the metric (\ref{gGCRS}) in the rotating coordinates are
\begin{subequations}
\begin{eqnarray}
g_{00}&=&-1+\frac{2W}{c^2}+\frac{v_R^2}{c^2}+O(c^{-4})\\
g_{0i}&=&\frac{1}{c}v_{R}^i+O(c^{-3})\\
g_{ij}&=&\delta_{ij}\left(1+\frac{2W}{c^2}\right)+O(c^{-4}),
\end{eqnarray}
\label{grot}
\end{subequations}
where $v_R^2=\delta_{ij}v_R^iv_R^j$. The monopole part of the scalar gravitational potential is denoted as $\hat{W}(r)$, and the remaining part containing the higher multipole moments and the time-dependent components as $\Delta W(x^\alpha)$ such that
\begin{eqnarray}
W(x^\alpha)&=&\hat{W}(r)+\Delta W(x^\alpha)\ \ {\rm with}\label{def_dW}\\
\hat{W}(r)&=&\frac{GM}{r},\label{monopol}
\end{eqnarray}
where $r=|x^i|$ is the radial coordinate, and $M$ is the mass of the gravitating object. For details of the gravitational potential $W(x^\alpha)$ in the vicinity of the Earth, see, for example, \citet{Wolf} and \citet{Voigt}.

\subsection{Medium}

In our model, the atmosphere of Earth is represented by a nondispersive medium, and we assume that its optical properties are given by a refractive index field $n(x^\alpha)$. This simplification can be applied, for example, in the case of a monochromatic signal by setting the frequency (or vacuum wavelength) variable in the refractive index of real air to a fixed value, assuming that the slight frequency variations along the signal trajectory, as observed from the local rest frames of the medium, have a negligible dispersion effect. We also assume that the medium is electrically neutral and has a vanishing density of the electric current.

The refractive index of air as a function of temperature, pressure, composition, and vacuum wavelength is discussed, for example, by \citet{Owens} and \citet{Ciddor} (see also part~\ref{appendix1} of the appendix). When the fields of the atmospheric temperature, the pressure, and the composition are known, the refractive index field $n(x^\alpha)$ for a given vacuum wavelength can be determined.

The atmospheric temperature and pressure as functions of altitude show a systematic behavior at large scales that is discussed, for example, in \citet{ICAO} and \citet{USatm} (see also part \ref{appendix2} of the appendix). Spatial and temporal local fluctuations are present as well. Therefore, it is reasonable to decompose the atmospheric refractive index field as 
\begin{equation}
n(x^\alpha)=\hat{n}(r)(1+\alpha(x^\alpha)),\label{def_alpha}
\end{equation}
where $\hat{n}(r)$ is the part of the refractive index field that depends on the radial coordinate alone (i.e., it is static and spherically symmetric), and $\alpha(x^\alpha)$ is its relative correction, which can be time dependent and in general has no spatial symmetry. 

In our model, we consider general fields $\hat{n}(r), \alpha(x^\alpha)$. When the model is to be applied to a specific case, these fields must be specified. A detailed discussion of the possible definitions of these fields is beyond scope of this paper, but some guidelines for determining $\hat{n}(r)$ for altitudes below 80~km are given in the appendix. The correction $\alpha(x^\alpha)$ then consists of various more or less predictable components, such as the deviation from spherical symmetry caused by the centrifugal potential, the effects of the changing position of Sun as a heat source, the effects of large-scale meteorological structures, or components coming from variations in local temperature, pressure, and composition, and from atmospheric turbulence.

The total refractivity $N(x^\alpha)$ and the static, spherically symmetric refractivity $\hat{N}(r)$ are then defined as deviations of the corresponding refractive index from its vacuum value,
\begin{eqnarray}
N&=&n-1,\\
\hat{N}&=&\hat{n}-1.
\end{eqnarray}
Denoting $x^i_M(t,x_0^\alpha)$ the trajectory of the medium element in the rotating coordinates that passes through a spacetime point $x_0^\alpha=(ct_0,x^i_0)$, the winds in the atmosphere are described by their velocity field $V^i(x^\alpha)$, which is defined by
\begin{equation}
V^i(x_0^\alpha)=\frac{\partial x^i_M}{\partial t}(t,x_0^\alpha)|_{t=t_0} \label{Vdef}
\end{equation} 
for any $x_0^\alpha$ where the medium is present. We also define the following quantity that is useful when the optical effects of the winds are evaluated:
\begin{equation}
{{A}}^i = (1-n^2)V^i.\label{Ai_def}
\end{equation}

\subsection{Observers}

For a given trajectory $x^i(t)$ of an observer in the rotating coordinates, we denote $v^i=\D x^i/\D t$ the observer's velocity in these coordinates.

\section{Time and frequency transfer through the atmosphere of Earth. Summary of the results} \label{Summary}

This section summarizes the main results of the paper: the corrections for one- and two-way time and frequency transfer through the atmosphere of Earth. It also provides numerical examples that give an idea about the magnitude of the atmospheric effects. The details of the derivation of the results are given in Sect.~\ref{sec_Theory}.

\subsection{Approximation level of the model}\label{sec_accuracy}

As mentioned in the introduction, a model accuracy of approximately 1~ps for the time transfer and relative accuracy of lower multiples of $10^{-18}$ for the frequency transfer is needed in order to meet the requirements of the forthcoming clock-on-satellite experiments. The corresponding approximation level of the model should be defined. 

The influencing quantities entering the equation of motion and the evaluation of the time and frequency transfer in the rotating frame originate from gravitation given by the Newtonian potential $W/c^2$ (higher-order gravitational effects such as the Lense-Thirring effect are negligible considering our intended accuracy), from inertial effects given by the rotation velocity $v^i_R/c$, from atmospheric effects given by the air refractivity $\hat{N}$ together with the correction $\alpha$ and by the wind speed $V^i/c$ and from the observers' velocities $v^i/c$ at the emission and reception side.

We define the "order" of a term in an expansion according to the following procedure: a) we express all formulas in terms of the refractivity $\hat{N}(r)$ and its correction $\alpha(x^\alpha)$, b)~we denote $\hat{N}_M$ the maximum value of $\hat{N}$ and $\alpha_M$ the maximum value of $|\alpha|$, c) we define normalized functions $\hat{N}_*=\hat{N}/\hat{N}_M$ and $\alpha_*=\alpha/\alpha_M$ and in all formulas we express $\hat{N}=\hat{N}_M\hat{N}_*$ and  $\alpha=\alpha_M \alpha_*$, d) the order of a term in an expansion is then given by its factor $(c^{-1})^{p_{c^{-1}}}(\hat{N}_M)^{p_{\hat{N}}}(\alpha_M)^{p_\alpha}$, where the powers $p_{c^{-1}}, p_{\hat{N}},\text{ and } p_\alpha$ are nonnegative integers, and finally, e) the derivative of the fields $\partial_0=c^{-1}\partial_t$ increases the power of $c^{-1}$ by one. The approximation level of an expansion is then given by a selection of terms defined by a subset in the grid of possible triplets of the exponents $(p_{c^{-1}}, p_{\hat{N}}, p_{\alpha})$. The expansion of $n$ in terms of $\hat{N}_M, \hat{N}_*, \alpha_M$, and $\alpha_*$ serves for the purpose of defining the order, but is not shown explicitly in formulas where it is not practical.

First, considering terms on the order of $(c^{-1})^{p_{c^{-1}}}(\hat{N}_M)^{p_{\hat{N}}}$ (i.e., $p_\alpha =0$) in the time and frequency transfer formulas, we take $p_{c^{-1}}\leq 3$ corresponding to the vacuum model of \citet{Blanchet}, and we take an arbitrary value of $p_{\hat{N}}$ for $p_{c^{-1}}\leq 2$. Terms on the order of $c^{-3}(\hat{N}_M)^{p_{\hat{N}}}$ with $p_{\hat{N}}\geq 1$ are neglected. Some of the terms on the order $c^{-3}\hat{N}_M$ can slightly exceed the $10^{-18}$ level in case of ground-to-ISS frequency transfer with a near horizon position of the satellite. The $c^{-4}$ terms, which we neglect as well, can approach the $10^{-18}$ level for ground-to-ISS frequency transfer, as was shown by \citet{Linet}. Then, taking the field $\alpha$ into account, its effect is considered to be minor compared to the spherical, static part of the refractivity $\hat{N}$. We therefore only include the field $\alpha$ to a lower order. Namely, for $p_\alpha\geq 1$, we take $p_{c^{-1}}+p_{\hat{N}}+p_\alpha \leq 3$. Possible magnitudes of the neglected terms with $p_\alpha\geq 1$ have not been evaluated, however.

Next, we denote the total $p_{c^{-1}}+p_{\hat{N}}+p_\alpha = p_T$, and we can summarize the approximation level of the model as follows: the expressions for the signal propagation time, the relative frequency shift, and the related two-way time and frequency transfer corrections in this paper include all terms with the order from a set $\mathcal{P}(3)$, which is given as
\begin{equation}
\mathcal{P}(3)=\{(p_{c^{-1}}, p_{\hat{N}}, p_{\alpha})\!: p_T\leq 3 \vee (p_{c^{-1}}\leq 2, p_\alpha=0)\}. \nonumber
\end{equation}
The only exception in which we express the results up to a lower order only are the terms in the two-way corrections that originate from the motion of a stationary observer on the ground in the corotating frame during the back and forth propagation of the signal. This motion is given by the deviations of the surface motion of Earth from rigid uniform rotation and therefore is very small. The  approximation level of these terms is discussed together with the particular formulas.

Next we introduce the notation $O(q)$ for expressions containing terms with $p_T\geq q$ only. Expressions containing terms with $p_{\hat{N}}=p_\alpha=0$ and $p_{c^{-1}}\geq k$ only are denoted $O(c^{-k})$ as usual. It is also useful to define the following subset of the exponent triplets:
\begin{equation}
\mathcal{P}(2)=\{(p_{c^{-1}}, p_{\hat{N}}, p_{\alpha})\!: p_T\leq 2 \vee (p_{c^{-1}}\leq 1, p_\alpha=0)\}. \nonumber
\end{equation}

\subsection{Strategy of solving the problem}

The approach used in this paper is based on the formulation and on an approximate solution of the equation of motion for a light ray in a flowing medium and a gravitational field. This equation is given as the null geodesic equation of the corresponding Gordon optical metric \citep{Gordon}. Details of the solution are given in Sect.~\ref{sec_Theory}.

In summary, our strategy of solving the equation of motion is the following. We formulate the equation in the Newtonian form in the corotating coordinate system with an Euclidean length of the light ray (in the same system) as a parameter, and we split the equation into its static, spherically symmetric part plus a correction. First, an exact solution of the static, spherically symmetric part is found with the given spatial positions of the emission and reception events. We denote this solution as $\hat{x}^i(\hat{l})$, with $\hat{l}$ being a parameter given by the Euclidean length of this path from its initial point. Then, an approximate linearized equation is solved for the path correction $\Delta x^i(\hat{l})$ caused by the deviations of the refractive index and the gravitational potential from sphericity and staticity, by winds and by the Coriolis and centrifugal forces. Finally, a separate first-order differential equation for the coordinate time $t(\hat{l})$ is solved.

The main results of this paper are expressed in terms of the integrals along the path $\hat{x}^i(\hat{l}).$ We therefore discuss the solution of the static, spherically symmetric part of the equation of motion in detail in this summary.

\subsection{Light rays in a static, spherically symmetric atmosphere} \label{sec_solution_hatx}

The equation of motion for a light ray in a static, spherically symmetric distribution of the refractive index $\hat{n}(r)$ and gravitational potential $\hat{W}(r)$ is given as (see Eq. (\ref{eqmX0}))
\begin{equation}
\frac{\D^2 \hat{x}^i}{\D \hat{l}^2}=\left(\frac{1}{\hat{n}}\partial_j\hat{n}+\frac{2}{c^2}\partial_j\hat{W}\right)\left(\delta^{ij}-\frac{\D \hat{x}^i}{\D \hat{l}}\frac{\D \hat{x}^j}{\D \hat{l}}\right),
\label{eqmX0a}
\end{equation}
where the fields $\hat{n}$, $\partial_j\hat{n,}$ and $\partial_j\hat{W}$ are evaluated in $\hat{x}^k(\hat{l})$, and the solution has a unit tangent satisfying
\begin{equation}
\delta_{ij}\frac{\D \hat{x}^i}{\D \hat{l}}\frac{\D \hat{x}^j}{\D \hat{l}}=1,
\label{Enorm0a}
\end{equation}
corresponding to the fact that $\hat{l}$ is its Euclidean length parameter. We assume that the initial and final points of the path are known as input parameters, and we later relate them to spatial coordinates of the emission and reception events of the one- or two-way time and frequency transfer setups. We denote the total path length between the initial and final points as $\hat{L,}$ such that the parameter has the range $\hat{l}\in[0,\hat{L}]$.

Defining a field $w$ as
\begin{equation}
w=\hat{n}\exp\left({2\hat{W}}/{c^2}\right),
\label{def_smw}
\end{equation}
which depends on the spatial coordinates through the radius $r=|x^i|$ only and introducing a new parameter $\zeta$ with a function dependence $\zeta(\hat{l})$ satisfying 
\begin{equation}
\frac{\D \zeta}{\D \hat{l}}=\frac{1}{w(\hat{x}^i(\hat{l}))},
\label{defs}
\end{equation}
Eq. (\ref{eqmX0a}) can be written as
\begin{equation}
\frac{\D^2\hat{x}^i(\zeta)}{\D \zeta^2}=\frac{1}{2}\partial_i {w}^2(\hat{x}^j(\zeta)),
\label{eqmX0s}
\end{equation}
where we briefly denote $\hat{x}^i(\zeta)\equiv\hat{x}^i(\hat{l}(\zeta))$. Thus, we obtain an equation of motion of the signal in the central potential field. The method of solving an analogous equation in mechanics is known from the literature (see, e.g., \citet{Landau}). It is based on two conservation laws. One law corresponds to the total energy and can be obtained from Eqs. (\ref{Enorm0a}) and (\ref{defs}) in our case, and it is given as
\begin{equation}
\delta_{ij}\frac{\D\hat{x}^i(\zeta)}{\D \zeta}\frac{\D\hat{x}^j(\zeta)}{\D \zeta}={w}^2(\hat{x}^k(\zeta)).
\label{zze}
\end{equation}
The second law corresponds to the angular momentum and can be obtained by taking a cross product of Eq. (\ref{eqmX0s}) with the signal position vector $\hat{x}^i(\zeta)$ and using the assumption of the central field, which leads to
\begin{equation}
\epsilon^i_{\ jk}\hat{x}^j(\zeta)\frac{\D \hat{x}^k(\zeta)}{\D \zeta}=h^i,
\label{zzmh}
\end{equation}
with $h^i$ being a constant vector. Equation (\ref{zzmh}) shows that for $|h^i|\neq 0$, the signal propagates in a plane that intersects the origin $r=0$ and has a normal aligned with $h^i$ (for $|h^i|=0$, the signal path is just a straight radial line). Denoting $\psi(\hat{l})$ the angle between the tangent ${\D \hat{x}^i(\hat{l})}/{\D \hat{l}}$ and the position vector $\hat{x}^i(\hat{l})$ and taking the magnitude of Eq. (\ref{zzmh}) with use of Eq. (\ref{zze}), we obtain
\begin{equation}
r(\hat{l})\ w(\hat{x}^i(\hat{l}))\sin\psi(\hat{l})=h={\rm const.},
\label{Snell}
\end{equation}
where $r(\hat{l})=|\hat{x}^i(\hat{l})|$ and $h=|h^i|$. This is the analog of the Snell law in static, spherically symmetric media with a gravitational field of the same symmetry. For the nongravitational case $(\hat{W}=0)$,  Eq. (\ref{Snell}) gives the Bouguer formula of classical geometrical optics (see Eq.~(7) in Sect.~3.2 of \citet{BornWolf}). Understanding $r$ as the trajectory parameter, with $\hat{l}(r)$ being the inverse function to $r(\hat{l})$ (assuming $\D r/\D\hat{l}\neq 0$ all along the trajectory), Eq. (\ref{Snell}) can be written as
\begin{equation}
r w(r)\sin\psi(\hat{l}(r))=h.
\label{Snell2}
\end{equation}

Selecting a right-handed Cartesian coordinate system $y^i$ with the origin at $r=0$, the $y^1$-axis in $\hat{x}^i(0)$ direction, and the $y^3$-axis in the $h^i$ direction (for $h=0$, we can take any direction perpendicular to $\hat{x}^i(0)$), defining polar coordinates by $y^1=r\cos\varphi$ and $y^2=r\sin\varphi$ and expressing the trajectory $\hat{x}^i(\hat{l})$ in the polar coordinates as $\varphi(r)$, with the radius as the parameter, it follows from the definition of $\psi$ that
\begin{equation}
r\frac{\D\varphi}{\D r}=\tan\psi (\hat{l}(r)). \label{tanPsi}
\end{equation}
Integrating this equation with $\tan\psi$ expressed from the Snell law (\ref{Snell2}), we obtain
\begin{equation}
\varphi(r)=\pm \int_{{r}_I}^r\!\!\D r^\prime \frac{h}{r^\prime}\left(r^{\prime 2}w(r^\prime)^2-h^2\right)^{\!\!-\frac{1}{2}},
\label{phir}
\end{equation}
with $\pm=+$ for $\D r/\D \hat{l}>0$ (ascending trajectory) and $\pm=-$ for $\D r/\D \hat{l}<0$ (descending trajectory).

Thus, up to the final integration, we obtain the solution of Eq. (\ref{eqmX0a}) in polar coordinates parameterized by $r$. The remaining step is to express the constant $h$ in terms of the input parameters, which, in this case, are the boundary values of the trajectory $\hat{x}^i(\hat{l})$. 

\begin{figure}[t]
  \centering
 \includegraphics[width=0.8\linewidth]{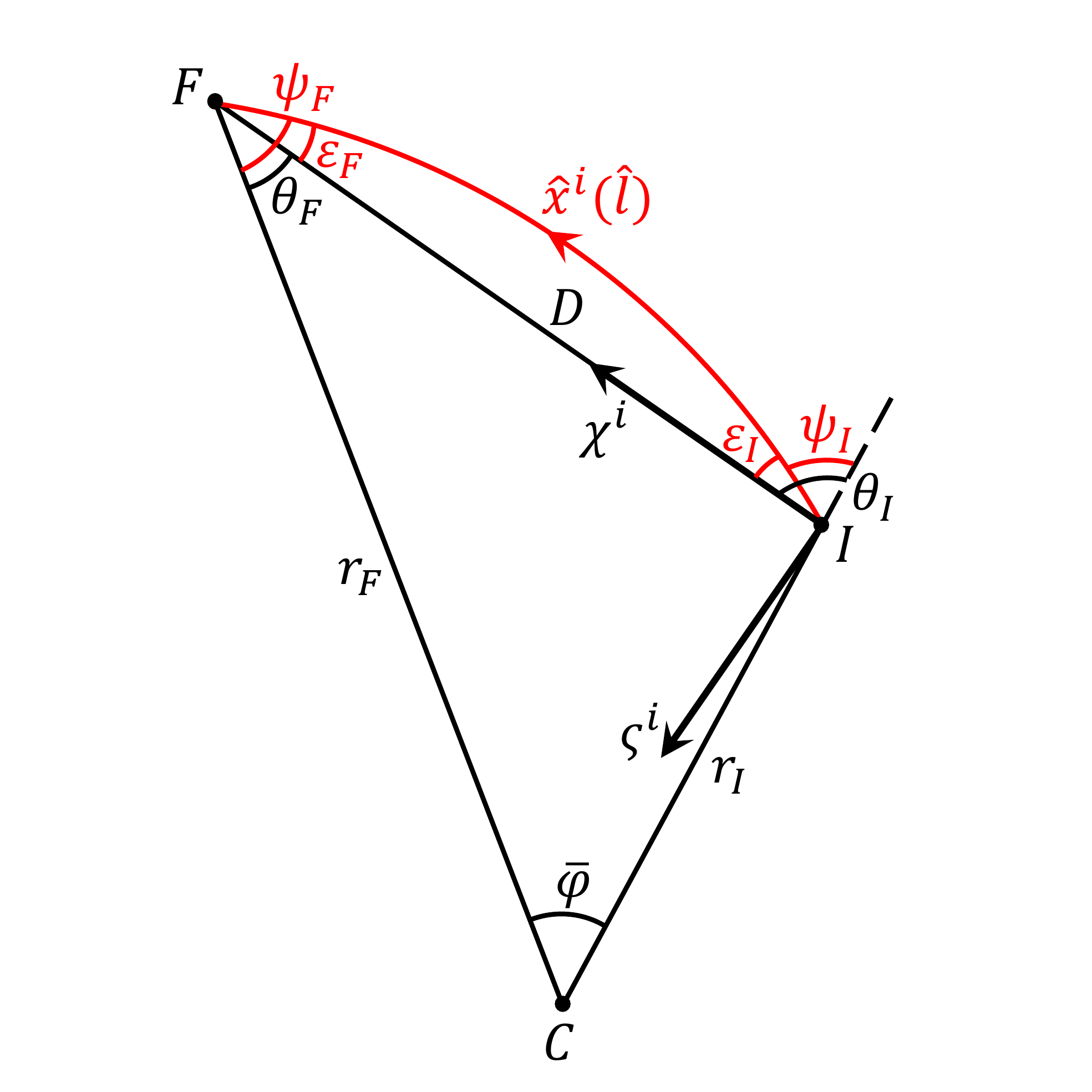}
   \caption{Definition of angles, basis vectors, and other quantities.}  
  \label{figAng0}
\end{figure}

In the following, we denote quantities related to the initial point $\hat{x}^i(0)$ by the index $I$ and to the final point $\hat{x}^i(\hat{L})$ by the index $F$, for example, $\hat{x}^i_I=\hat{x}^i(0)$, $r_I=|\hat{x}^i_I|$, $w_I=w(\hat{x}^i_I)$, $\psi_I=\psi(0)$, and $\hat{x}^i_F=\hat{x}^i(\hat{L})$, $r_F=|\hat{x}^i_F|$, $w_F=w(\hat{x}^i_F)$, and  $\psi_F=\psi(\hat{L})$. We also define the angle $\bar{\varphi}$ between $\hat{x}^i_F$ and $\hat{x}^i_I$, the angle $\theta_I$ between $\hat{x}^i_F-\hat{x}^i_I$ and $\hat{x}^i_I$, and the angle $\theta_F$ between $\hat{x}^i_F-\hat{x}^i_I$ and $\hat{x}^i_F$, and we denote $D=|\hat{x}^i_F-\hat{x}^i_I|$ (see Fig.\ref{figAng0}). The following relations hold:
\begin{eqnarray}
r_F\sin\theta_F&=&r_I\sin\theta_I,\\
r_F\cos\theta_F&=&r_I\cos\theta_I + D.
\end{eqnarray}

The boundary value problem is more complicated than the initial value problem, in which case $h$ is given simply as $h=r_Iw_I\sin\psi_I$. In the boundary value problem, we need to express $h$ as a function of $r_I, r_F$ and one of the angles $\bar{\varphi}$, $\theta_I$, or $\theta_F$. One possible approach is presented below. We denote as $\varepsilon_I$ and $\varepsilon_F$ the angular deviations of the path tangents $\D\hat{x}^i/\D\hat{l}|_{0}$ and $\D\hat{x}^i/\D\hat{l}|_{\hat{L}}$, respectively, from the vector $\hat{x}^i_F-\hat{x}^i_I$ (see Fig.\ref{figAng0}), so that we have
\begin{eqnarray}
\varepsilon_I&=&\psi_I-\theta_I,\label{def_eI}\\
\varepsilon_F&=&\psi_F-\theta_F. \label{def_eF}
\end{eqnarray}
Evaluating the Snell law (\ref{Snell2}) in the boundary points and using Eqs. (\ref{def_eI}) and (\ref{def_eF}), we obtain
\begin{equation}
h= r_Iw_I\sin(\theta_I+\varepsilon_I)=r_Fw_F\sin(\theta_F+\varepsilon_F).\label{Snell_bp}
\end{equation}
Now we need to express the angle $\varepsilon_I$ or $\varepsilon_F$. For this purpose, we introduce a total bending angle of the path $\varepsilon_T$ that is the angle between $\D\hat{x}^i/\D\hat{l}|_{0}$ and $\D\hat{x}^i/\D\hat{l}|_{\hat{L}}$, which satisfies
\begin{equation}
\varepsilon_F-\varepsilon_I=\varepsilon_T. \label{etot_2}
\end{equation} 

Solving Eq. (\ref{Snell_bp}) together with Eq. (\ref{etot_2}) for the angles $\varepsilon_I$ and $\varepsilon_F$, we obtain
\begin{eqnarray}
\tan\varepsilon_I&=&\frac{\frac{w_F}{w_I}\sin(\theta_F+\varepsilon_T)-\sin\theta_F}{-{D}/{r_F}-\frac{w_F}{w_I}\cos(\theta_F+\varepsilon_T)+\cos\theta_F}, \label{tan_eI}\\
\tan\varepsilon_F&=&\frac{\frac{w_I}{w_F}\sin(\theta_I-\varepsilon_T)-\sin\theta_I}{{D}/{r_I}-\frac{w_I}{w_F}\cos(\theta_I-\varepsilon_T)+\cos\theta_I}. \label{tan_eF}
\end{eqnarray}
The total bending angle also satisfies
\begin{equation}
\varepsilon_T = \bar{\varphi}+\psi_F-\psi_I, \label{etot_def}
\end{equation}
where the angle $\bar{\varphi}$ can be expressed as $\bar{\varphi}=\varphi(r_F)$ using Eq. (\ref{phir}), and we can also write
\begin{equation}
\psi_F-\psi_I=\int_{r_I}^{r_F}\!\!\D r \frac{\D\psi (\hat{l}(r))}{\D r},
\end{equation}
where on the right-hand side (RHS), we use Eq. (\ref{Snell2}). Thus, we obtain\begin{equation}
\varepsilon_T(h) = \mp \int_{r_I}^{r_F}\!\!\D r \frac{h}{w(r)}\frac{\D w(r)}{\D r}\left(r^2w(r)^2-h^2\right)^{\!\!-\frac{1}{2}} \label{etot}
\end{equation}
with $\mp = -$ for the ascending path and $\mp = +$ for the descending path.

Then, the constant $h$ can be determined by the following iterative procedure: (a) we estimate $h=r_Fw_F\sin\theta_F$, corresponding to $\varepsilon_F=0$ in Eq. (\ref{Snell_bp}), (b) we calculate $\varepsilon_T(h)$ according to Eq.~(\ref{etot}), (c) we calculate $\varepsilon_F$ according to Eq. (\ref{tan_eF}), (d) we calculate $h$ according to Eq. (\ref{Snell_bp}) and start another loop from (b). As a result, not only the constant $h$ is obtained, but also the angles $\varepsilon_F$, $\varepsilon_T$, and $\varepsilon_I=\varepsilon_F-\varepsilon_T$. Alternatively, the procedure can be performed using the angle $\varepsilon_I$ given by Eq. (\ref{tan_eI}) instead of $\varepsilon_F$. We also note that for the ground-to-satellite or satellite-to-ground transfer, the $\varepsilon_{F/I}$ angle at the satellite position is at least one order of magnitude smaller than the corresponding angle at the ground position. Therefore, neglecting the $\varepsilon_{F/I}$ angle at the satellite position gives a better initial estimate of $h$ according to Eq. (\ref{Snell_bp}) than neglecting the corresponding angle at the ground position.

To retrieve the solution in the coordinates $x^i$, we express the components of the first two basis vectors of $y^i$ in the coordinate system $x^i$. Assuming $\bar{\varphi}\neq 0$, we have 
\begin{equation}
e_1^i=\frac{\hat{x}^i_I}{{r}_I},\ \ \ \ e_2^i=\frac{1}{\sin\bar{\varphi}}\left(\frac{\hat{x}^i_F}{{r}_F}-\cos\bar{\varphi}\frac{\hat{x}^i_I}{{r}_I}\right).
\end{equation}
The solution parameterized by $r$ is then given as
\begin{equation}
\hat{x}^i(\hat{l}(r))=r\cos\varphi(r)e_1^i+r\sin\varphi(r)e_2^i.
\end{equation}
For $\bar{\varphi}=0$, it is simply $\hat{x}^i(\hat{l}(r))=r e_1^i$.

The reparameterization function $\hat{l}(r)$ can be expressed with the use of
\begin{equation}
\frac{\D \hat{l}}{\D r}=\pm\sqrt{1+r^2\left(\frac{\D\varphi}{\D r}\right)^{\!\!2}}=\frac{1}{\cos\psi(\hat{l}(r))},
\label{dldr}
\end{equation}
where the $\pm$ sign has the same meaning as in Eq. (\ref{phir}). Integrating this formula using the Snell law (\ref{Snell2}), we obtain\begin{equation}
\hat{l}(r)=\pm\int_{r_I}^r \!\!\D r^\prime\left(1-\frac{h^2}{r^{\prime 2} w(r^\prime)^2}\right)^{\!\!-\frac{1}{2}}\!\!. \label{hatlr}
\end{equation}
For $\hat{L}$ we then get $\hat{L}=\hat{l}(r_F)$.

It is also useful to define an orthonormal right-handed basis ($\chi^i, \varsigma^i, \gamma^i$) with $\chi^i$ pointing in direction from $\hat{x}^i_I$ to $\hat{x}^i_F$, $\gamma^i$ pointing in direction of the normal to the propagation plane $h^i$ (assuming $h\neq 0$), and $\varsigma^i$ completing the triplet. We have (see also Fig.\ref{figAng0})
\begin{eqnarray}
\chi^i&=&\frac{1}{D}(\hat{x}^i_F-\hat{x}^i_I),\label{def_chi}\\
\varsigma^i&=&\frac{1}{D}(\hat{x}^i_F\cot\theta_I - \hat{x}^i_I\cot\theta_F),\\
\gamma^i&=&\frac{\epsilon^i_{\ jk}\hat{x}^j_I\hat{x}^k_F}{r_Ir_F\sin\bar{\varphi}}.
\end{eqnarray}

Defining the angle $\varepsilon (\hat{l})$ between the path tangent $\D \hat{x}^i/\D\hat{l}$ and the vector $\chi^i$ satisfying $\varepsilon(0)=\varepsilon_I$ (negative value) and $\varepsilon(\hat{L})=\varepsilon_F$ (positive value), we can write
\begin{equation}
\frac{\D\hat{x}^i}{\D\hat{l}}=\chi^i\cos\varepsilon (\hat{l}) + \varsigma^i\sin\varepsilon (\hat{l}). \label{tanL_eF}
\end{equation}
In some formulas, the lowest-order approximation of the tangent $\D\hat{x}^i/\D\hat{l}$ and of the length $\hat{L}$ is sufficient. Expanding Eq. (\ref{tanL_eF}) in $\varepsilon$ and using $|\varepsilon|\leq\varepsilon_T=O(1)$, which follows from Eq. (\ref{etot}), we obtain
\begin{equation}
\frac{\D\hat{x}^i}{\D\hat{l}}=\chi^i+O(1). \label{tan_ap}
\end{equation}
Next we obtain
\begin{equation}
D=\int_0^{\hat{L}}\!\!\D\hat{l}\ \chi_i \frac{\D\hat{x}^i}{\D\hat{l}} = \int_0^{\hat{L}}\!\!\D\hat{l}\ \cos\varepsilon(\hat{l})=\hat{L}+O(2), \label{L_ap}
\end{equation}
where the expansion of $\cos\varepsilon$ was used in the last step. 

In the following text, we refer to the solution $\hat{x}^i(\hat{l})$ of Eq. (\ref{eqmX0a}) with the given boundary conditions as the {\it \textup{base path}}.

\subsection{Coordinate time transfer}

\subsubsection{One-way time transfer}

In one-way time transfer, we wish to express a coordinate time $t_F-t_I$ of the propagation of the signal emitted from a position $x^i_I$ at a time $t_I$ and received in a position $x^i_F$ at a time $t_F$. We consider a base path $\hat{x}^i(\hat{l})$ with boundary points $\hat{x}^i(0)=x^i_I$ and $\hat{x}^i(\hat{L})=x^i_F$. 

Based on the analysis given in Sect.~\ref{sec_Theory}, we obtain the following formula for the propagation time, including all terms on the order of $\mathcal{P}(3)$:
\begin{eqnarray}
t_F\!-\!t_I\!\!\!\!&=&\!\!\!\!\frac{1}{c}\int_0^{\hat{L}}\!\!\!\D\hat{l}\ n +\frac{1}{c^3}\int_0^{\hat{L}}\!\!\!\D\hat{l}\ 2W \nonumber\\
&+&\!\!\!\!\frac{1}{c^2}\int_0^{\hat{L}}\!\!\D\hat{l}\left(v_{Ri}+{{A}}_i\right)\frac{\D\hat{x}^i}{\D\hat{l}}\nonumber\\
&+&\!\!\!\!\frac{1}{c^3}\frac{D}{2}v_{RI}^i v_{RF}^j (\delta_{ij}+\chi_i\chi_j) \nonumber\\
&+&\!\!\!\!\frac{1}{c^2}\epsilon^i_{\ jk}\chi^j\omega^k \int_0^{\hat{L}}\!\!\D\hat{l}\ \hat{l}(\hat{l}-\hat{L})\partial_i\alpha\nonumber\\
&+&\!\!\!\!\frac{1}{c^2}\int_0^{\hat{L}}\!\D\hat{l}\ \hat{l} \partial_t \alpha\nonumber\\
&+&\!\!\!\!\frac{1}{2c}(\delta^{ij}\!-\!\chi^i\chi^j)\!\!
\int_0^{\hat{L}}\!\!\!\!\D\hat{l}\!\int_0^{\hat{L}}\!\!\!\!\D\hat{l}^{\prime}D(\hat{l},\hat{l}^{\prime})\partial_i\alpha(t_I,\hat{x}^k(\hat{l}))\partial_j\alpha(t_I,\hat{x}^k(\hat{l}^\prime)),\nonumber\\
&&\label{dt2_su}
\end{eqnarray}
where the fields $n, \alpha, W, v_{Ri}, {{A}}_i$ and their partial derivatives with respect to the space-time coordinates are evaluated in $t=t_I$ and $x^i=\hat{x}^i(\hat{l})$ if not stated explicitly otherwise, the velocities $v_{RI}^i$ and $v_{RF}^i$ are defined as $v_R^i(\hat{x}^k_I)$ and $v_R^i(\hat{x}^k_F)$, respectively, and the function $D(\hat{l},\hat{l}^\prime)$ is given by Eq. (\ref{defDij}). 
Equation (\ref{dt2_su}) is expressed in corotating coordinates, which means that all components of the fields are expressed as functions of the corotating coordinates, the wind speed $V^i$ in ${{A}}^i$ is the speed in the corotating frame, and the base path is defined by the boundary conditions that are given by the spatial coordinates of the emission and reception events in the corotating frame as well. 

Equation (\ref{dt2_su}) corresponds to Eq. (5) (or (A.39)) of \citet{Blanchet}, transformed to the rotating frame and generalized by including the atmospheric effects. Its first term is the leading Newtonian term in the refractive medium, the second term is the Shapiro delay, the first term in the second line is the Sagnac correction, which appears due to the use of the rotating frame as well as the term in the third line, and the second term in the second line is the effect of the wind, corresponding to the Fresnel-Fizeau effect of dragging of light by a medium. The remaining terms containing $\alpha$ are the effects of nonsphericity and nonstationarity of the refractive index field.
Numerical integration of the terms in Eq. (\ref{dt2_su}) can be performed, for example, by transforming the integration variable $\hat{l}$ into the radial coordinate $r$ using Eq. (\ref{hatlr}) and by expressing the integrated functions in spherical coordinates with the base path given by Eq. (\ref{phir}).

\subsubsection{Two-way time transfer}

\begin{figure}[t]
\centering
\includegraphics[width=0.8\linewidth]{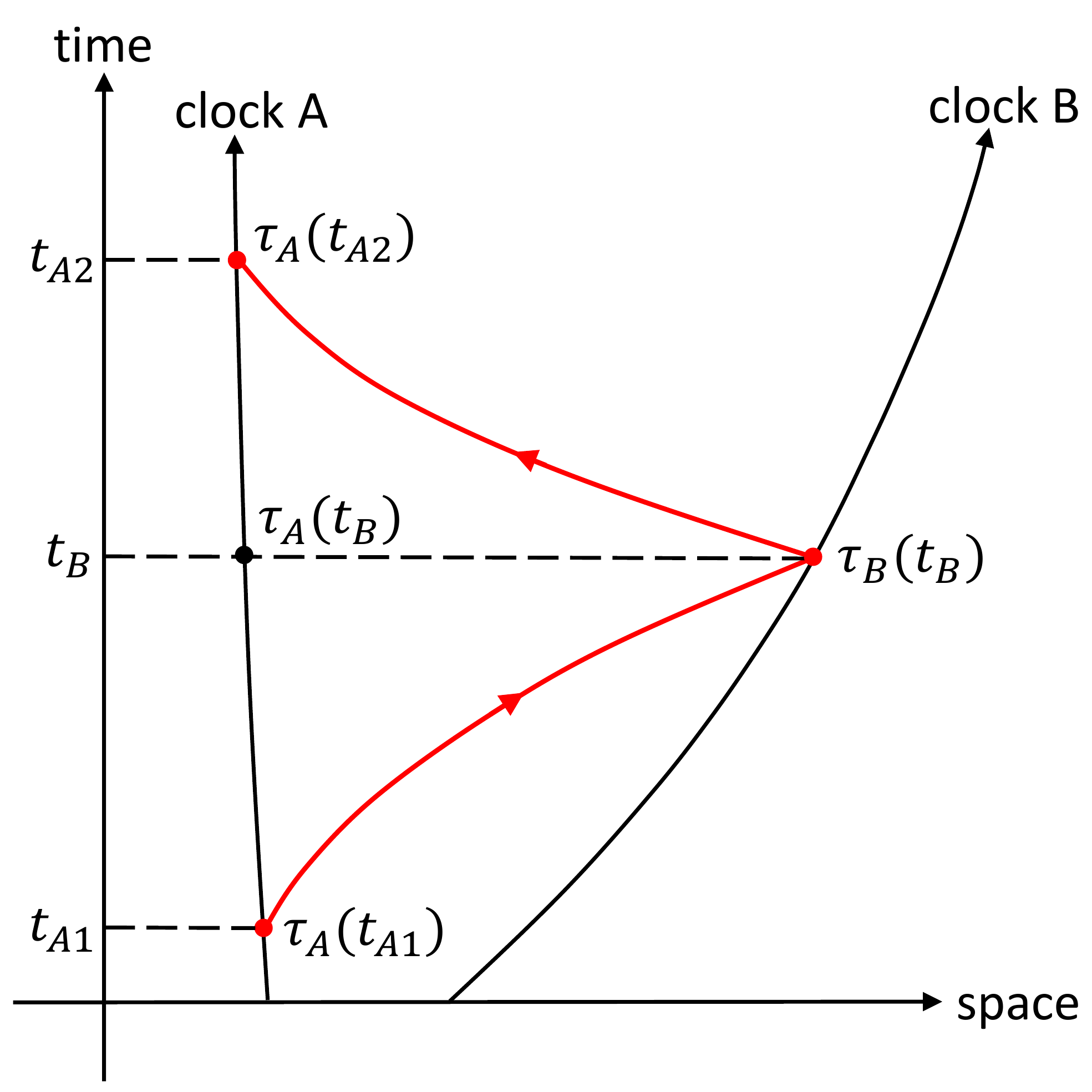}
\caption{Scheme of the two-way time transfer. The signal leaves observer $A$ at a coordinate time $t_{A1}$, corresponding to the proper time $\tau_{A}(t_{A1})$ of clock $A$, is reflected from observer $B$ at a coordinate time $t_B$, corresponding to the proper time $\tau_{B}(t_B)$ of clock $B$ and $\tau_{A}(t_B)$ of clock $A$, and is finally
received by observer $A$ at the coordinate time $t_{A2}$, corresponding to the proper time $\tau_{A}(t_{A2})$ of clock $A$.
\label{fig:2way}}
\end{figure}

In two-way transfer, we consider the signal emitted from a stationary observer on the ground $A$ at the coordinate time $t_{A1}$, which
corresponds to the proper time $\tau_{A}(t_{A1})$ of clock $A$. The spatial coordinates of this event are denoted $x^i_{A1}$. The signal is then received by observer
$B$ (ground or satellite) at the coordinate time $t_B$, which corresponds to the proper time $\tau_{B}(t_B)$ of clock $B$ (see Fig.~\ref{fig:2way}). The spatial coordinates of this event are denoted $x^i_{B}$. The pseudo-time-of-flight $\tau_{B}(t_B) - \tau_{A}(t_{A1})$ is obtained from the measurements. Then a signal is sent from observer $B$
at the coordinate time $t_B$ and is received by observer $A$ at the coordinate time $t_{A2}$, which corresponds to the proper time $\tau_{A}(t_{A2})$ of clock $A$. The spatial coordinates of this event are denoted $x^i_{A2}$. This signal can either be the same signal reflected or another signal that is synchronously sent when the one-way signal is received by observer $B$. We call this set-up the $\Lambda$-configuration. The pseudo-time-of-flight $\tau_{A}(t_{A2}) - \tau_{B}(t_B)$ is also obtained from measurements.
%

Using the coordinate-time synchronization convention, the desynchronization of the clocks is defined as the difference $\tau_{B}(t_B)-\tau_{A}(t_B)$, where $\tau_A (t_B)$ is the proper time measured by observer $A$ at the coordinate time $t_B$. In case of the two-way time transfer, it can be expressed as
\begin{eqnarray}
\tau_{B}(t_B)-\tau_{A}(t_B)&=&\frac{1}{2}(\tau_{B}(t_B)-\tau_{A}(t_{A1}))\nonumber\\
&-&\frac{1}{2}(\tau_{A}(t_{A2})-\tau_{B}(t_B))\nonumber\\
&+&\frac{1}{2}(\Delta\tau_- -\Delta\tau_+),\label{m:desyn}
\end{eqnarray}
where $\Delta\tau_+ =\tau_{A}(t_B)-\tau_{A}(t_{A1})$ and $\Delta\tau_- =\tau_{A}(t_{A2})-\tau_{A}(t_B)$. 
The first two lines of Eq. (\ref{m:desyn}) are the measured pseudo-times-of-flight, and the difference $\Delta\tau_- -\Delta\tau_+$ needs to be computed. 

Denoting $\Delta t_+ = t_B-t_{A1}$ the coordinate time of the signal propagation from $A$ to $B$ and $\Delta t_- =  t_{A2}-t_B$ the coordinate time of the signal propagation from $B$ to $A$, the computed term can be approximated as 
\begin{equation}
\Delta\tau_- -\Delta\tau_+\approx \Delta t_- -\Delta t_+\ . \label{dtauapp}
\end{equation}
The error of this approximation is several orders below the 1~ps level in terrestrial conditions.

We also note that in the corotating coordinates we use, the position of the stationary observer on the ground $A$ is nearly fixed and changes only due to the deformations of Earth, which are given, for example, by the solid Earth tides or by the irregularity of the rotation of Earth. Thus, the difference between $x^i_{A1}$ and $x^i_{A2}$ is very small, and we include its effect up to the largest order only using the velocity $v^i_{A1}$ of observer $A$ in the corotating system at the emission event.

The propagation times $\Delta t_+$ and $\Delta t_-$ can be calculated using Eq. (\ref{dt2_su}) or its slightly extended version, Eq. (\ref{dt2}). In their difference, the leading term in the first line of Eq. (\ref{dt2_su}) and several other terms are compensated for. Denoting $\hat{x}^i_+(\hat{l})$ the base path with the boundary points $\hat{x}^i_+(0)=x^i_{A1}$ and $\hat{x}^i_+(\hat{L})=x^i_B$, we arrive at the following result for the two-way time transfer correction, including all terms on the order of $\mathcal{P}(3)$ except the $p_T=3$ order terms originating from $x^i_{A2}-x^i_{A1,}$
\begin{eqnarray}
\Delta t_{-}-\Delta t_{+}&=&-\frac{2}{c^2}\int_0^{\hat{L}}\!\!\D\hat{l}\ (v_{Ri+}+{{A}}_{i+})\frac{\D \hat{x}^i_+}{\D\hat{l}}\nonumber\\
&+&\frac{2}{c^2}\epsilon^i_{\ jk}\chi^j_+\omega^k\int_0^{\hat{L}}\!\!\D\hat{l}\ \hat{l}(\hat{L}-\hat{l})\partial_i\alpha_{+}\nonumber\\
&+&\frac{2}{c^2}\int_0^{\hat{L}}\!\!\D\hat{l}(\hat{L}-\hat{l})\partial_t \alpha_{+}\nonumber\\
&-&\frac{2}{c^2}\hat{L}\chi_{i+}v^i_{A1},\label{TWTT}
\end{eqnarray}
where the fields $v_{Ri}, {{A}}_i, \partial_i\alpha$, and $\partial_t\alpha$ with the index + are evaluated in $t=t_B$ and $x^i=\hat{x}^i_+(\hat{l})$, and $\chi^i_+$ is the vector $\chi^i$ corresponding to the path $\hat{x}^i_+(\hat{l})$ (i.e., it is the unit vector in the direction from $x^i_{A1}$ to $x^i_B$).

The main contribution to the two-way time transfer correction in Eq. (\ref{TWTT}) is given by the first term in the first line, which is the well-known Sagnac effect. We denote
\begin{equation}
(\Delta t_{-}-\Delta t_{+})_{\rm Sag}=-\frac{2}{c^2}\int_0^{\hat{L}}\!\!\D\hat{l}\ v_{Ri+}\frac{\D \hat{x}^i_+}{\D\hat{l}}. \label{Sagnac_def}
\end{equation}

\begin{figure}[t]
 \centering
 \includegraphics[width=0.8\linewidth]{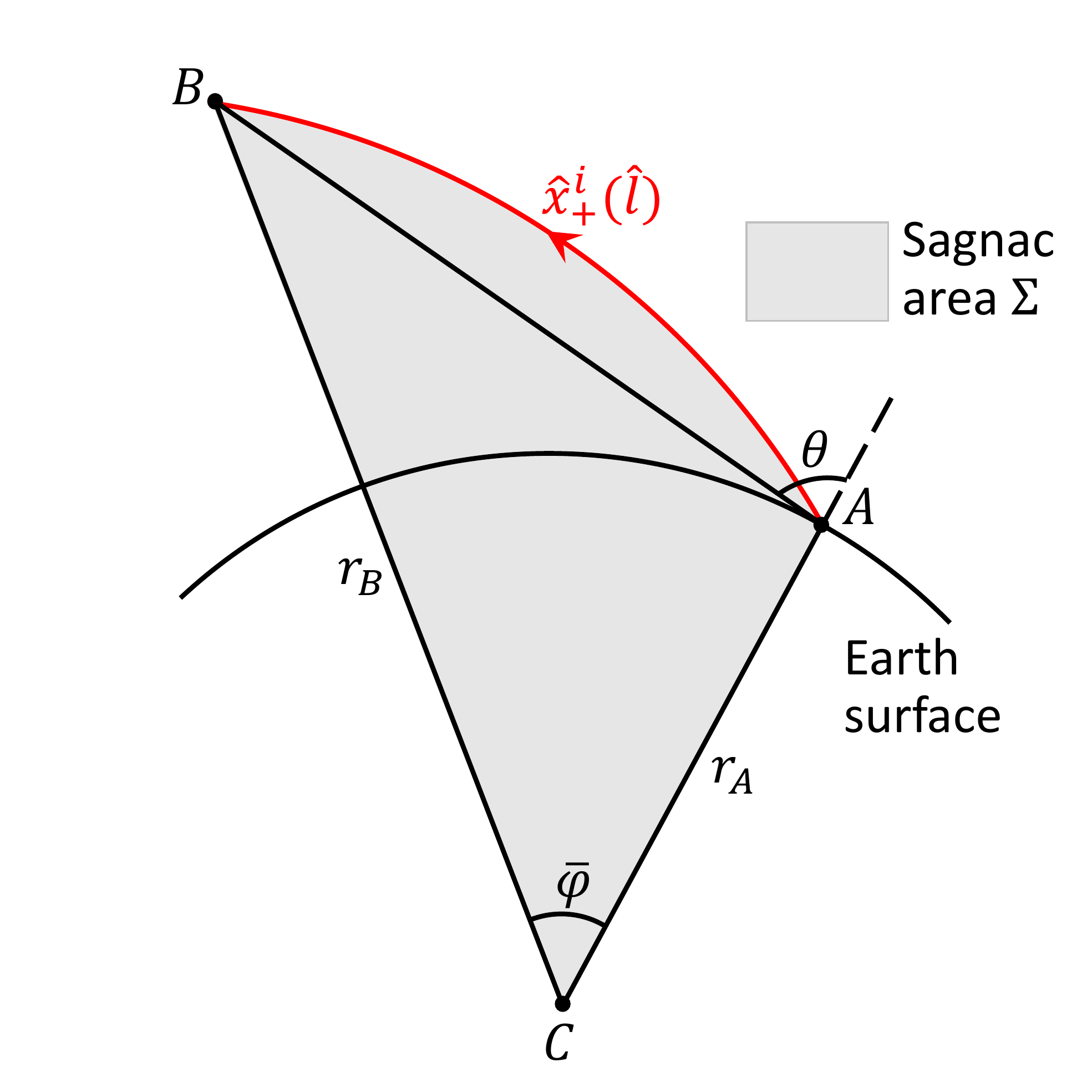}
   \caption{Visualization of the Sagnac area $\Sigma$. The part of the Sagnac effect that is caused by the atmosphere is given by the change in the area $\Sigma$ when the path $\hat{x}^i_+(\hat{l})$ changes from the case including atmospheric refraction and gravitation to a vacuum case that only includes gravitation ($\hat{n}=1$ in Eq. (\ref{def_smw})).}
  \label{figAng}
\end{figure}

Then, taking into account that $v_{Ri+}=\epsilon_{ijk}\omega^j \hat{x}^k_+(\hat{l})$, the correction given by Eq.~(\ref{Sagnac_def}) can be expressed as
\begin{equation}
(\Delta t_{-}-\Delta t_{+})_{\rm Sag}=-\frac{4}{c^2}\omega^j\Sigma_j ,
\label{Sagnac}
\end{equation}
where $\Sigma_j$ is defined as
\begin{equation}
\Sigma_j=\frac{1}{2}\int_0^{\hat{L}}\!\!\D\hat{l}\ \epsilon_{jki} \hat{x}^k_{+}(\hat{l}) \frac{\D \hat{x}^i_+(\hat{l})}{\D\hat{l}}.
\label{def_Sigma_Sag}
\end{equation}
The magnitude $\Sigma=|\Sigma_j|$ equals the area of the plane section surrounded by three lines: 1) the straight line from the origin of the coordinate system (the Earth center) to $x^i_{A1}$, 2) the path $\hat{x}^i_{+}(\hat{l})$ from $x^i_{A1}$ to $x^i_B$, and 3) the straight line from $x^i_B$ to the origin of the coordinate system (see Fig.~\ref{figAng}). The direction of $\Sigma_j$ is normal to this plane.
The difference between the Sagnac effect in the Earth atmosphere and in vacuum is then given by the change in the area $\Sigma$ when the path $\hat{x}^i_+(\hat{l})$ changes from the case including atmospheric refraction and gravitation to a case that only includes gravitation.

\begin{figure*}
\begin{subfigure}[b]{0.488\linewidth}
\includegraphics[width=\linewidth]{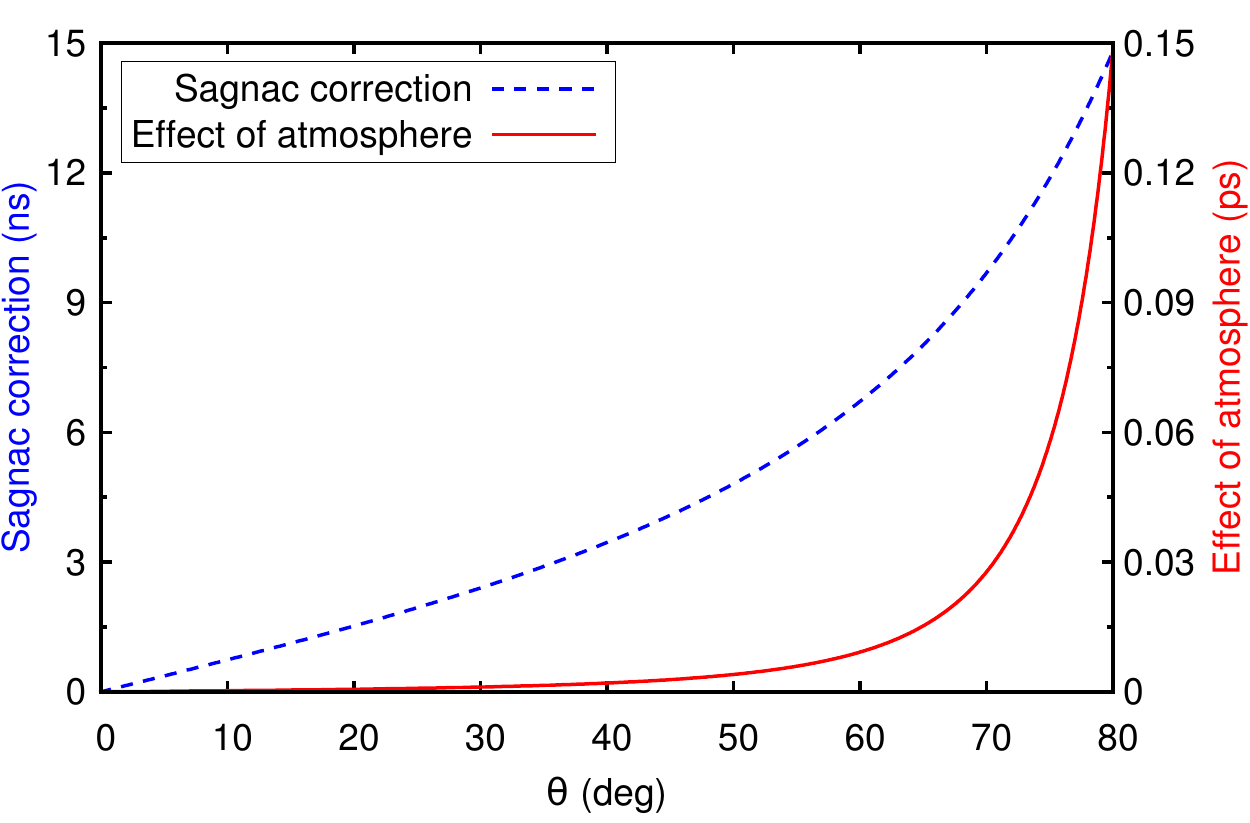}
\end{subfigure}
\hspace{1.2mm}
\begin{subfigure}[b]{0.488\linewidth}
\includegraphics[width=\linewidth]{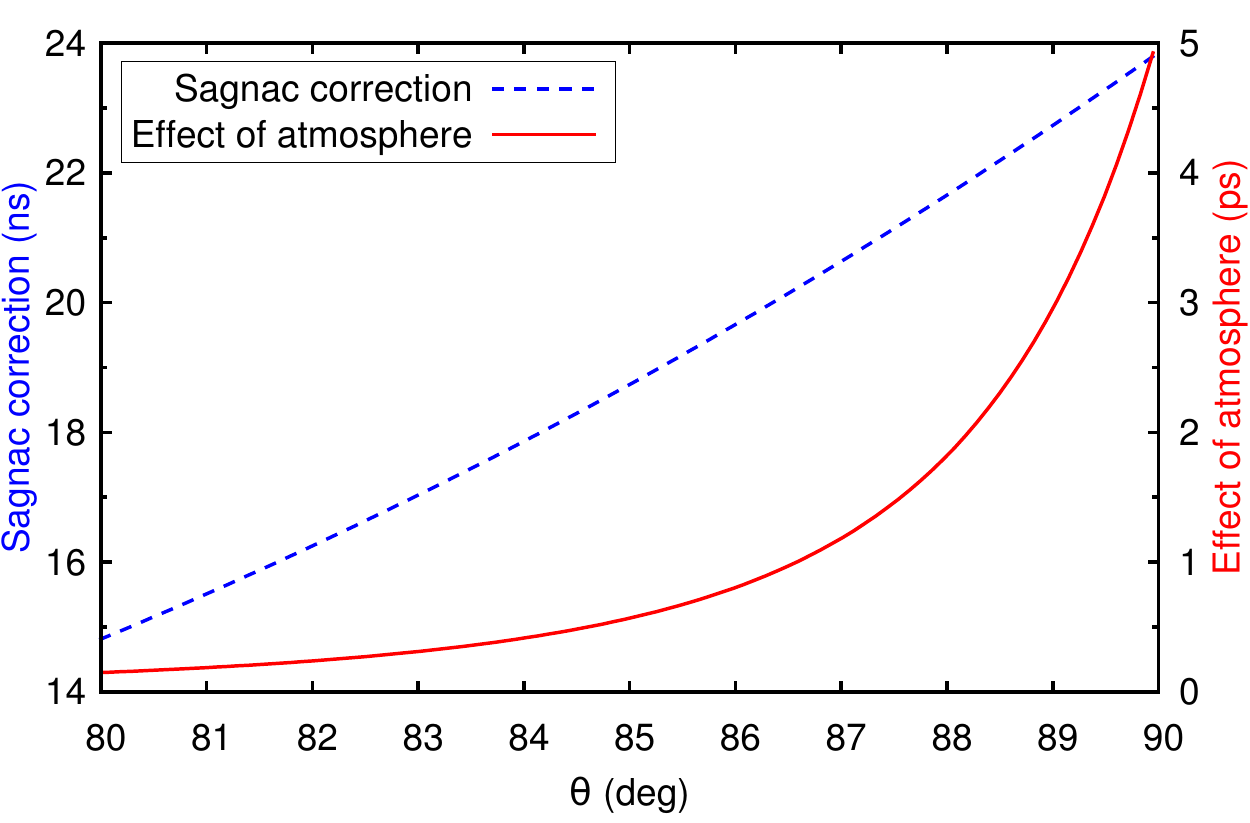}
\end{subfigure}
\caption{Sagnac correction, Eq. (\ref{Sagnac}), for an isothermal atmosphere and its deviation from a vacuum value denoted as effect of the atmosphere as functions of the satellite zenith angle $\theta$. Separate plots for $\theta\in[0^\circ,80^\circ]$ ({\it left}) and $\theta\in[80^\circ,90^\circ]$ ({\it right}) are presented to show details of the effect of the atmosphere curve, which grows steeply for $\theta$ approaching $90^\circ$. The example was evaluated for ground-to-satellite transfer in the equatorial plane, with $r_A=$ 6371~km, $r_B-r_A=$ 408~km, and $\chi^i_+$ inclined against the rotation of the Earth.}
\label{figSag}
\end{figure*}

A convenient formula for $\Sigma_j$ can be obtained when the integrand of Eq. (\ref{def_Sigma_Sag}) is expressed using Eq. (\ref{zzmh}) together with Eq. (\ref{defs}) and using the definition $\gamma_i=h_i/h$. Then, transforming the integration variable to $r$ using Eq. (\ref{hatlr}), we obtain
\begin{equation}
\Sigma_j=\frac{1}{2}\gamma_{j+}h\int_{r_A}^{r_B}\!\!\D r\left(w(r)^2-\frac{h^2}{r^2}\right)^{\!\!-\frac{1}{2}},
\label{Sigma_Sag}
\end{equation} 
where $r_A=|x^i_{A1}|$, $r_B=|x^i_B|$, $\gamma_{j+}$ is the vector $\gamma_j$ corresponding to the path $\hat{x}^i_+(\hat{l})$, and $w(r)$ is given by Eq. (\ref{def_smw}). This integral can be computed numerically for any refractive index profile $\hat{n}(r)$ of interest.

The second term in the first line of Eq. (\ref{TWTT}) is the effect of the wind, and the following two lines give the effects of nonsphericity and nonstationarity of the refractive index field. The last line of Eq. (\ref{TWTT}) then describes the effect of motion of the stationary observer on the ground in the corotating frame. 

Possible magnitudes of the atmospheric contribution to the Sagnac effect and of the effect of wind are discussed in the next section. A numerical evaluation of the other effects is left for future research.

\subsection{Time transfer. Examples and magnitudes of the effects} \label{TT_examples}

\subsubsection{Sagnac correction} \label{example_Sagnac}

In this section, we evaluate the Sagnac correction, Eq. (\ref{Sagnac}), for the example of ground-to-satellite transfer through an isothermal atmosphere at a temperature $T_0$ with an unchanging composition corresponding to a molar mass of air $M_a$. We assume that the atmosphere rotates rigidly with the Earth and is in hydrostatic equilibrium with its gravitational and centrifugal forces. In this case, the refractivity $\hat{N}(r)=\hat{n}(r)-1$ is given by Eq. (\ref{hatN_ex}), which gives the following spherical refractive index field:
\begin{equation}
\hat{n}(r)=1+N_{\!A}\exp\left(\frac{M_a}{RT_0}(\hat{W}-\hat{W}_{\!A})\right), \label{hatn_isoT}
\end{equation}
where $\hat{W}=GM/r$, $\hat{W}_{\!A}=GM/r_A$, $R$ is the universal gas constant, and $N_{\!A}$ is the air refractivity in the reference position $x^i_{A1}$, which also equals $\hat{N}(r_A)$ (for details, see the appendix). For the sake of simplicity, we use Eq. (\ref{hatn_isoT}) in the full range of the radius between the ground station and the satellite. The function $w(r)$ in Eq. (\ref{Sigma_Sag}) is then calculated according to Eq. (\ref{def_smw}).

We consider $r_A=6371$~km, corresponding to the surface of Earth, $r_B=r_A+408$~km, approximately corresponding to the altitude of the ISS, and $T_0=288.15$~K ($15~^\circ$C) and $M_a=28.964\times 10^{-3}$~kg/mol, corresponding to dry air with a carbon dioxide molar fraction $x_c=450$~ppm. For an atmospheric pressure of 101~325~Pa in the position $x^i_{A1}$ and a light signal with a vacuum wavelength of $1\ \mu{\rm m,}$ we obtain (see, e.g., \citet{Ciddor})
\begin{equation}
N_{\!A}\approx 2.742\times 10^{-4}.\label{N0val}
\end{equation}

In two-way transfer, we denote the $\theta_I$ angle of the $\hat{x}^i_+(\hat{l})$ path simply as $\theta$. It represents the zenith angle of the satellite at the reception/re-emission event (see Fig.~\ref{figAng}). 

Based on Eqs. (\ref{Sagnac}) and (\ref{Sigma_Sag}), the Sagnac correction can be calculated as a function of $\theta$. In Fig.~\ref{figSag} this function is shown (the dashed blue curve) together with the effect of the atmosphere (the red full curve), which is defined as the Sagnac correction in the atmosphere ($N_{\!A}$ given by Eq. (\ref{N0val})) minus the Sagnac correction in vacuum ($N_{\!A}=0$) for the same value of $\theta$. The values in the plots in Fig.~\ref{figSag} are obtained for the case $\omega^i \gamma_{i+}=-\omega$, which corresponds to the paths $\hat{x}^i_+(\hat{l})$ propagating in the equatorial plane against the rotation of the Earth. This is the case with the highest positive values of the Sagnac correction.

Fig.~\ref{figSag} shows that the Sagnac correction increases from 0 to approximately 24~ns, with $\theta$ increasing from 0 to $90^\circ$. The effect of the atmosphere is below 0.1~ps for $\theta < 78^\circ$, it reaches 1~ps for $\theta\approx 86.6^\circ$, and it increases to approximately 5~ps as $\theta$ approaches $90^\circ$. Therefore, for the time transfer at an accuracy level of 1~ps, this effect is only significant for large angles $\theta$ .

Since the gravitational effect to $\hat{x}^i_+(\hat{l})$ is much smaller than the effect of atmospheric refraction, the path $\hat{x}^i_+(\hat{l})$ can be approximated by a straight line for angles $\theta$ where the effect of the atmosphere is negligible. Thus, for example, for $\theta\in[0^\circ,78^\circ]$, the following formula for the Sagnac correction has a bias below 0.1~ps for the example studied in this section,
\begin{equation}
(\Delta t_{-}-\Delta t_{+})_{\rm Sag}=-\frac{2}{c^2}\omega^i\gamma_{i+} r_Ar_B\sin\bar{\varphi}.
\end{equation}

\subsubsection{Effect of the wind}

We evaluate the effect of the wind speed in the two-way time transfer correction (\ref{TWTT}), which is given by
\begin{equation}
(\Delta t_{-}-\Delta t_{+})_{\rm wind}=-\frac{2}{c^2}\int_0^{\hat{L}}\!\!\D\hat{l}\ {{A}}_{i+}\frac{\D \hat{x}^i_+}{\D\hat{l}}.
\end{equation}
For a rough estimate of the magnitude of the effect, we assume that the signal propagates in the part of the atmosphere in which the refractive index $n$ is approximately constant along the path $\hat{x}^i_+(\hat{l})$, as well as the component of the air velocity $V^i$ (in the corotating frame) in the direction of the path tangent $\D\hat{x}^i_+/\D\hat{l,}$ which we denote $V$. The approximately constant $n$ corresponds, for example, to the horizontal signal path from the ground to the ground or to the initial part of the signal path from the ground to the satellite when the satellite is in a position near the horizon. In this case, we obtain
\begin{equation}
(\Delta t_{-}-\Delta t_{+})_{\rm wind}\approx\frac{2}{c^2}(n^2-1)V\hat{L}.
\end{equation}      
For a 1~ps correction in the air with the refractivity value of Eq. (\ref{N0val}), which is typical on the surface of the Earth, we need $V\hat{L}\approx 8.2\times 10^7\ {\rm m^2/s,}$ which, for example, for the path length of $\hat{L}=100$~km leads to $V\approx 820$~m/s. Thus, the effect is negligible for time transfer at an accuracy level of 1~ps under normal atmospheric conditions. However, it can be detected by interferometric methods, as was shown, for example, by \citet{Tselikov}, who used the effect in a different context for flow metering applications in laboratory conditions.

\subsection{Frequency transfer}

\subsubsection{One-way frequency transfer}

In one-way frequency transfer, the goal is to express the ratio of the proper frequency $\nu_I$ of a signal emitted by an observer with velocity $v^i_I$ from a position $x^i_I$ at a time $t_I$ and a proper frequency $\nu_F$ of the same signal as is received by the observer with a velocity $v^i_F$ in the position $x^i_F$ at the time $t_F$. Again, we consider a base path $\hat{x}^i(\hat{l})$ with the boundary points $\hat{x}^i(0)=x^i_I$ and $\hat{x}^i(\hat{L})=x^i_F$, and we denote $W_I\!=\!W(t_I,x^i_I)$, $W_F\!=\!W(t_F,x^i_F)$, $v_{RI}^i\!=\!v_R^i(x^j_I)$, and $v_{RF}^i\!=\!v_R^i(x^j_F)$.

Based on the analysis included in Sect.~\ref{sec_Theory}, we obtain the following formula for the frequency ratio: 
\begin{equation}
\frac{\nu_I}{\nu_F}=\frac{1+\frac{W_I}{c^2}+\frac{1}{2c^2}|{v}^i_{RI}+{v^i_I}|^2}{1+\frac{W_F}{c^2}\!+\frac{1}{2c^2}|{v}^i_{RF}\!+\!{v^i_F}|^2}\ \frac{1-\frac{1}{c}v^i_I(\bar{l}_{i})_I}{1\!-\!\frac{1}{c} v^i_F(\bar{l}_{i})_F}\ \exp I ,
\label{fratioX_su}
\end{equation}
where $(\bar{l}_{i})_I$ and $(\bar{l}_{i})_F$ are quantities related to tangents of the light-ray trajectory at the emission and reception events, and they are given by Eqs. (\ref{li_split}) to (\ref{li_dx}) with $\eta = 0$ for $(\bar{l}_{i})_I$ and $\eta = 1$ for $(\bar{l}_{i})_F$. In these formulas, we set $\Delta x^i_F=0$ and $\hat{l}_0=0$ corresponding to $\hat{x}^\alpha(\hat{l})=(ct_I, \hat{x}^i(\hat{l}))$. The details of the expressions for $(\bar{l}_{i})_I$ and $(\bar{l}_{i})_F$ are given in the Theory section, but the corresponding contribution to two-way transfer is discussed further in this summary. The exponent $I$ in the last term of Eq. (\ref{fratioX_su}) is given as
\begin{eqnarray}
I\!\!\!&=&\!\!\!\int_0^{\hat{L}}\!\!\D\hat{l}\left(\frac{1}{c}\partial_t n+\frac{1}{c^2}(\partial_t {{A}}_i)\frac{\D \hat{x}^i}{\D \hat{l}}+\frac{2}{c^3}\partial_t W\right) \nonumber\\
&+&\!\!\!\frac{1}{c^2}\epsilon^i_{\ jk}\chi^j\omega^k\int_0^{\hat{L}}\!\!\D\hat{l}\ \hat{l}(\hat{l}-\hat{L})\partial_i\partial_t \alpha\nonumber\\
&+&\!\!\!\frac{1}{c^2}\int_0^{\hat{L}}\!\!\D\hat{l}\ \hat{l}\partial_t\partial_t\alpha\nonumber\\
&+&\!\!\!\frac{1}{c}(\delta^{ij} \!-\! \chi^i\chi^j)\!\!
\int_0^{\hat{L}}\!\!\!\!\D\hat{l}\!\int_0^{\hat{L}}\!\!\!\!\D\hat{l}^\prime D(\hat{l},\hat{l}^\prime)\ \partial_i\partial_t\alpha(t_I,\hat{x}^k(\hat{l}))\ \partial_j\alpha(t_I,\hat{x}^k(\hat{l}^\prime))\nonumber\\
&&\label{I2_su}
\end{eqnarray}
where the fields $n, \alpha, {{A}}_i, W$, and their partial derivatives with respect to the space-time coordinates are evaluated in $t=t_I$ and $x^i=\hat{x}^i(\hat{l})$ if not stated explicitly otherwise.

Equation (\ref{fratioX_su}) together with Eqs. (\ref{li_split}) to (\ref{li_dx}) and Eq. (\ref{I2_su}) gives the frequency ratio including all terms on the order of $\mathcal{P}(3)$. The first fraction in Eq. (\ref{fratioX_su}) contains the gravitational redshift effect and the second-order Doppler effect, the second fraction contains the first-order Doppler effect, and the exponential term is the effect of nonstationarity.

\subsubsection{Two-way frequency transfer}

In two-way frequency transfer, a stationary observer on the ground $A$ emits a signal with the proper frequency $\nu_{A1}$ from the position $x^i_{A1}$ at the time $t_{A1}$. The signal is received by observer $B$ (ground or satellite) with the proper frequency $\nu_B$ in the position $x^i_B$ at the time $t_B$ and is immediately transponded back with the same frequency to observer $A$, where it is received with the proper frequency $\nu_{A2}$ in the position $x^i_{A2}$ at the time $t_{A2}$. The velocities of observers $A$ and $B$ at the particular emission and reception events are denoted as $v^i_{A1}$, $v^i_B$, and $v^i_{A2}$. For the stationary (fixed at the ground) observer $A$, the velocities $v^i_{A1}$ and $v^i_{A2}$ in the corotating coordinates and the corresponding spatial shift $x^i_{A2}-x^i_{A1}$ are given just by the deformations in the body of the Earth, for example, the solid tides, or by irregularities in the rotation of Earth. Thus, these quantities are very small, and the main contribution to the Doppler shift comes from the satellite part and not from the ground.

The frequency ratio $\nu_{A2}/\nu_{B}$, which is needed for the frequency comparison, is then expressed in terms of the ratio $\nu_{A2}/\nu_{A1}$, which is {\it \textup{measured}} by the reference clock at ground station $A$, and the correction, which must be \textup{{\it \textup{computed}}.} \citet{Ashby} and \citet{Blanchet} defined the correction $\Delta$ by the following relation:
\begin{equation}
\frac{\nu_{A2}}{\nu_{B}}=\frac{1}{2}\frac{\nu_{A2}}{\nu_{A1}}+\Delta +\frac{1}{2}.
\label{defDelta}
\end{equation}
The correction $\Delta$, including all terms on the order of $\mathcal{P}(3)$ except the $p_T\geq 3$ order terms originating from the motion of observer $A$ in the corotating frame and except the terms quadratic in $v^i_{A1}/c$, is given by Eqs. (\ref{Delta2}) to (\ref{Ic2}) in the Theory section. In this summary, we present a simpler result for $\Delta$ that is obtained by neglecting terms of higher than the third order ($p_T>3$) and terms proportional to $c^{-2}N_B$, which are negligible for satellite applications because the atmospheric refractivity $N_B$ in the satellite position is very low for satellite altitudes of several hundred kilometers. 

The fields evaluated in the spacetime points $x^\alpha_{A1}$, $x^\alpha_{B}$, or $x^\alpha_{A2}$ of the emission and reception events are denoted by the corresponding index, for example, $W_{A1}=W(t_{A1},x^j_{A1})$ and $v^i_{R|A1}=v^i_R(x^j_{A1})$. The base path $\hat{x}^i_+(\hat{l})$ is again defined by the boundary points $\hat{x}^i_+(0)=x^i_{A1}$ and $\hat{x}^i_+(\hat{L})=x^i_{B}$, and the fields denoted with the index~+ are evaluated in $t=t_B$ and $x^i=\hat{x}^i_+(\hat{l})$, for example, $\partial_b\alpha_+=(\partial_b\alpha)(t_B,\hat{x}^i_+(\hat{l}))$. The basis $\chi^i_+$, $\varsigma^i_+$, $\gamma^i_+$ is associated with the path $\hat{x}^i_+(\hat{l})$, and $\varepsilon_B$ is the angle $\varepsilon_F$ of the path $\hat{x}^i_+(\hat{l})$. The acceleration of observer $A$ at the time $t_{A1}$ in the rotating frame is denoted $a^i_{A1}$ (i.e., $a^i_{A1}=\D^2 x^i_{A}/\D t^2|_{t_{A1}}$). Using this notation, we obtain
\begin{eqnarray}
\Delta&=&\frac{1}{2c^2}\!\left[W_{A1}\!+\!W_{A2}\!-\!2W_B\!+\!|v_{R|A1}^i|^2\!+\!2v_{Bi}v^i_{R|A1}-|v_{R|B}^i+v^i_{B}|^2\right]\nonumber\\
&\times&\left(1-\frac{1}{c}v_B^j\chi_{j+}\right)\nonumber\\
&+&\frac{1}{c^2}D\chi_{i+}(a_{A1}^i+\epsilon^i_{\ jk}\omega^j v_{A1}^k)+\frac{1}{c^2}v_{Bi}v_{A1}^i\nonumber\\
&+&\frac{1}{c}v_B^i\beta_{i(2)}+I_{+(c^{-2})}-\frac{1}{c^2}v_B^i\chi_{i+}\int_0^{\hat{L}}\!\!\!\D\hat{l}\ \partial_t n_+ ,\label{Delta2_su}
\end{eqnarray}
where the atmospheric effects are included in the last line, in which
\begin{eqnarray}
\beta_{i(2)}&=&\frac{1}{c}\chi^i_+ D\omega_j\gamma^j_+\varepsilon_B \nonumber\\
&+&\frac{1}{c}(\delta^{ij}\!\!-\!\!\chi^i_+\chi^j_+)\epsilon_{\ jk}^{b}\omega^k\!\frac{1}{\hat{L}}\int_0^{\hat{L}}\!\!\!\!\D\hat{l}\ \hat{l}(\hat{l}-\hat{L})\partial_b\hat{n}_+\nonumber\\
&+&\frac{1}{c}(\delta^{ia}\!\!-\!\!\chi^i_+\chi^a_+)\epsilon_{\ jk}^{b}\chi_+^j\omega^k\!\frac{1}{\hat{L}}\int_0^{\hat{L}}\!\!\!\!\D\hat{l}\ \hat{l}^2(\hat{l}\!-\!\hat{L})\partial_a\partial_b\hat{n}_+\nonumber\\
&+&\frac{1}{c}\epsilon_{ijk}\chi_+^j\frac{1}{\hat{L}}\int_0^{\hat{L}}\!\!\!\!\D\hat{l}\ \hat{l}\ {\rm curl}{{A}}_+^k\nonumber\\
&+&\frac{1}{c}(\delta^{ij}\!+\! 2\chi^i_+\chi^j_+)\epsilon_{\ jk}^{b}\omega^k\!\frac{1}{\hat{L}}\int_0^{\hat{L}}\!\!\!\!\D\hat{l}\ \hat{l}(\hat{l}-\hat{L})\partial_b \alpha_+\nonumber\\
&+&\frac{1}{c}\epsilon_{\ jk}^{b}\chi_+^j\omega^k\!\frac{1}{\hat{L}}\int_0^{\hat{L}}\!\!\!\!\D\hat{l}\ \hat{l}^2(\hat{l}-\hat{L})\partial_i\partial_b\alpha_+\nonumber\\
&+&\frac{1}{c}(\delta^{ij}\!-\! \chi^i_+\chi^j_+)\frac{1}{\hat{L}}\int_0^{\hat{L}}\!\!\!\!\D\hat{l}\ \hat{l}(\hat{l}-\hat{L})\partial_t\partial_j \alpha_+\label{beta(2)}
\end{eqnarray}
and
\begin{eqnarray}
I_{+(c^{-2})}&=&\frac{1}{c^2}\int_0^{\hat{L}}\!\!\!\D\hat{l}\ (\partial_t {{A}}_{i+})\frac{\D \hat{x}^i_+}{\D \hat{l}} \nonumber\\
&+&\frac{1}{c^2}\epsilon^i_{\ jk}\chi^j_+\omega^k\int_0^{\hat{L}}\!\!\D\hat{l}\ \hat{l}(\hat{l}-\hat{L})\partial_i\partial_t \alpha_+\nonumber\\
&+&\frac{1}{c^2}\int_0^{\hat{L}}\!\!\D\hat{l}\ (\hat{l}-\hat{L})\partial_t\partial_t\alpha_+ .\label{delta_I2_su}
\end{eqnarray}
When the atmospheric effects vanish, the correction (\ref{Delta2_su}) reduces to the vacuum correction of \citet{Blanchet}, as can be verified by transforming this correction into the rotating coordinates.

The examples in the next section focus on the numerical evaluation of the atmospheric effects originating from the spherically symmetric, static part of the refractive index $\hat{n}(r)$ (the first three lines of Eq.~(\ref{beta(2)})) and from the wind (the fourth line of Eq.~(\ref{beta(2)})). The estimation of the remaining atmospheric terms describing the effects of nonsphericity and nonstationarity of the refractive index and the nonstationarity of the wind field is left for future research.

\subsection{Frequency transfer. Examples and magnitudes of the effects} \label{FT_examples}

\subsubsection{Effect of the atmospheric refractivity. The spherical part}

First, we evaluate the contribution of the spherical part of the atmospheric refractive index $\hat{n}(r)$ to the two-way frequency transfer correction (\ref{Delta2_su}), which is given by the first three lines of Eq.~(\ref{beta(2)}). We denote
\begin{subequations}
\begin{eqnarray}
\Delta_{\rm At}^{(1)}\!\!\!&=&\!\!\!\frac{1}{c^2}v_{Bi}\chi^i_+ D\omega_j\gamma^j_+ \varepsilon_B,\\
\Delta_{\rm At}^{(2)}\!\!\!&=&\!\!\!\frac{1}{c^2}v_{Bi}(\delta^{ij}\!\!-\!\!\chi^i_+\chi^j_+)\epsilon_{\ jk}^{b}\omega^k\!\frac{1}{\hat{L}}\int_0^{\hat{L}}\!\!\!\!\D\hat{l}\ \hat{l}(\hat{l}-\hat{L})\partial_b\hat{n}_+,\\
\Delta_{\rm At}^{(3)}\!\!\!&=&\!\!\!\frac{1}{c^2}v_{Bi}(\delta^{ia}\!\!-\!\!\chi^i_+\chi^a_+)\epsilon_{\ jk}^{b}\chi_+^j\omega^k\!\frac{1}{\hat{L}}\int_0^{\hat{L}}\!\!\!\!\D\hat{l}\ \hat{l}^2(\hat{l}\!-\!\hat{L})\partial_a\partial_b\hat{n}_+,\\
\Delta_{\rm At}^{(S)}\!\!\!&=&\!\!\!\Delta_{\rm At}^{(1)}+\Delta_{\rm At}^{(2)}+\Delta_{\rm At}^{(3)}.
\end{eqnarray}
\end{subequations}

As an example, we consider an  observer on the ground, located at the equator with $r_A=$ 6371~km and a satellite moving on a circular orbit in the equatorial plane at a radius of $r_B=r_A$+408~km in the direction of the rotation of the Earth. The corresponding satellite velocity in the co-rotating system is $v_B=$ 7.17~km/s. We consider the spherical refractive index field given by Eq.~(\ref{hatn_isoT}), which corresponds to the isothermal atmosphere in hydrostatic equilibrium with a constant air composition, and we use the same parameters as given in the initial part of Sect.~\ref{example_Sagnac} up to Eq.~(\ref{N0val}).

\begin{figure}[t]
 \centering
 \includegraphics[width=1\linewidth]{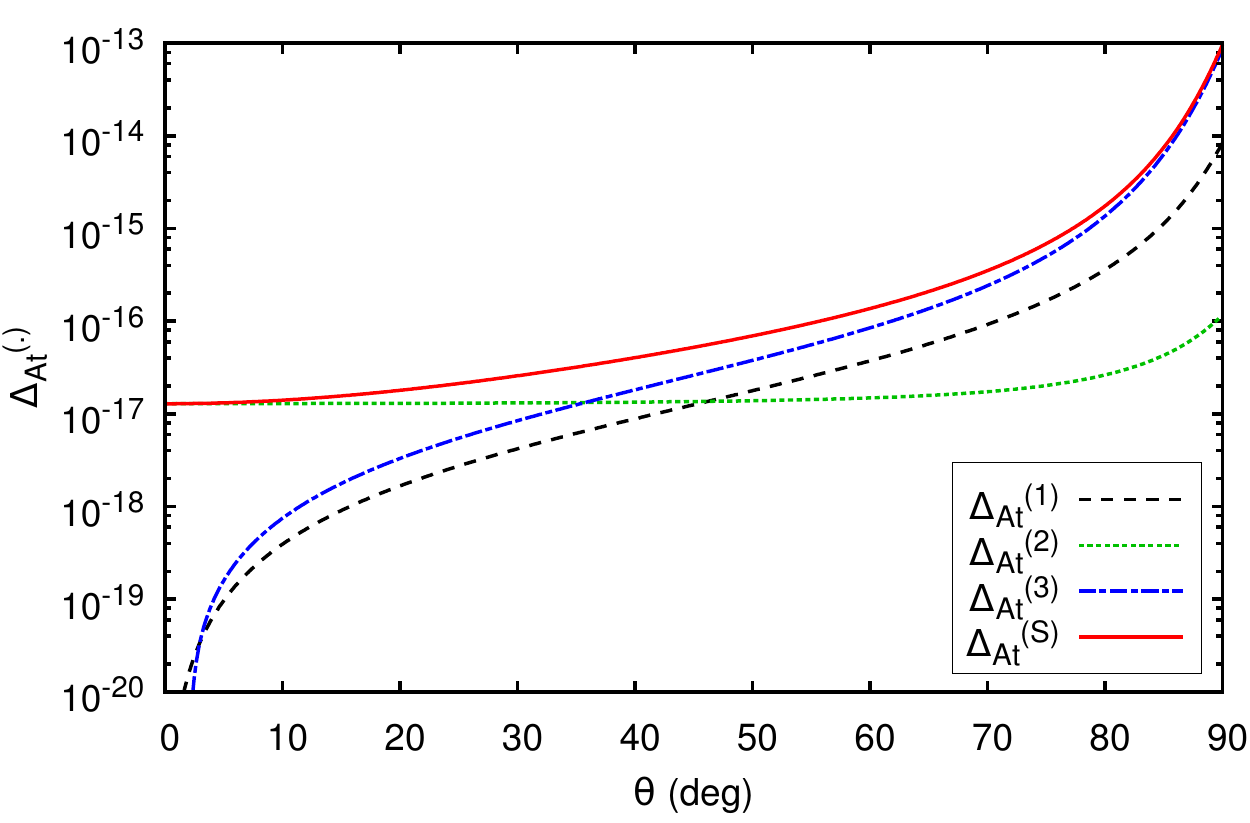}
   \caption{Three contributions of the spherically symmetric, static part of the atmospheric refractivity $\Delta_{\rm At}^{(1)}$, $\Delta_{\rm At}^{(2)}$, and $\Delta_{\rm At}^{(3)}$ to the two-way frequency transfer correction (\ref{Delta2_su}) and their sum $\Delta_{\rm At}^{(S)}$ as functions of the satellite zenith angle $\theta$. The example is evaluated for ground-to-satellite transfer through an isothermal atmosphere in hydrostatic equilibrium with an observer on the ground located at the equator with $r_A=$ 6371~km and a satellite moving on a circular orbit in the equatorial plane at a radius of $r_A$+408~km in the direction of the rotation of the Earth.}
  \label{fig_DAt_123}
\end{figure}

The resulting corrections $\Delta_{\rm At}^{(1)}$ to $\Delta_{\rm At}^{(3)}$ and their sum $\Delta_{\rm At}^{(S)}$ as functions of the satellite zenith angle $\theta$ are shown in Fig.~\ref{fig_DAt_123}. One value of $\theta$ corresponds to two possible positions of the satellite at the reception event $B$: one before the satellite zenith, and one after. The value of all the three corrections is the same for both satellite positions. The corrections $\Delta_{\rm At}^{(2)}$ and $\Delta_{\rm At}^{(3)}$ were calculated by transforming the integration variable into $r$ using Eq.~(\ref{hatlr}).

Fig.~\ref{fig_DAt_123} shows that the sum of the corrections starts at $10^{-17}$ for $\theta=0$, reaches $10^{-16}$ for $\theta\approx 56^\circ$, and ends at $10^{-13}$ for $\theta$ approaching $90^\circ$. Therefore, the corrections should be taken into account in the forthcoming clock-on-satellite experiments.

\subsubsection{Effect of the wind}

In this section, we evaluate the effect of the wind, which enters the correction (\ref{Delta2_su}) through the fourth line of Eq.~(\ref{beta(2)}). In case of a nonstationary ${{A}}_i$, another effect is given by the first line of Eq.~(\ref{delta_I2_su}), which is not discussed here. We denote
\begin{equation}
\Delta_{\rm At}^{\rm wind}=\frac{1}{c^2}v_B^i\chi_+^j\epsilon_{ijk}\frac{1}{\hat{L}}\int_0^{\hat{L}}\!\!\!\!\D\hat{l}\ \hat{l}\ {\rm curl}{{A}}_+^k.
\label{DBwind3}
\end{equation}

To show how large the effect can be, we evaluate Eq.~(\ref{DBwind3}) for the simple situation with the satellite position $x^i_B$ radially above the emission point on the ground $x^i_{A1}$, in which case, the path $\hat{x}^i_+(\hat{l})$ is just a straight vertical line connecting $x^i_{A1}$ with $x^i_{B}$ (i.e., $\hat{x}^i_+(\hat{l})=x^i_{A1}+\chi_+^i\hat{l}$ and $\hat{L}=r_B-r_A$). Next, we assume that the wind field $V_j$ at $t=t_B$ is perpendicular to the path tangent $\chi^i_+$ in the neighborhood of $\hat{x}^i_+(\hat{l})$ (horizontal flow). In this case, using Eq.~(\ref{Ai_def}) and writing
\begin{equation}
\chi_+^j\epsilon_{ijk} {\rm curl}{{A}}^k_+=\chi_+^j\partial_i {{A}}_{j+}-\chi_+^j\partial_j {{A}}_{i+},\label{dA}
\end{equation}
the first term on the RHS of Eq.~(\ref{dA}) vanishes, and the second term, which equals $-\D {{A}}_{i+}/\D\hat{l}$, can be integrated by parts to obtain
\begin{equation}
\Delta_{\rm At}^{\rm wind}=\frac{1}{c^2}v^i_{B}\frac{1}{\hat{L}}\int_{0}^{\hat{L}}\!\!\!\D \hat{l}\ {{A}}_{i+}, \label{DBwind4}
\end{equation}
where the term containing ${{A}}_{i+}|_{\hat{L}}={{A}}^i_B$, which is on the order of $c^{-2}N_B$, was neglected in the integration by parts.
Expansion of Eq.~(\ref{Ai_def}) in the refractivity $N$ gives
\begin{equation}
{{A}}_i = -2NV_i+O(2).\label{Aj_expanded}
\end{equation}

For the numerical evaluation of Eq.~(\ref{DBwind4}), we further assume that in the neighborhood of the base path $\hat{x}^i_+(\hat{l})$, the wind speed $V^i$ is constant and has a direction opposite to the satellite velocity $v_B^i$, which is itself perpendicular to $\chi_+^i$ (i.e., $-v_B^iV_i=v_BV$ with $v_B=|v_B^i|$ and $V=|V^i|$). Next, we consider the refractive index field approximated by Eq.~(\ref{hatn_isoT}). For the position of the observer on the ground, $r_A=6371$~km, and the satellite position $r_B=r_A+408$~km, we obtain
\begin{equation}
\int_{0}^{\hat{L}}\!\!\!\D \hat{l}\ N_+\approx\int_{r_A}^{r_B}\!\!\!\D r \hat{N}(r)\approx 2.32\ {\rm m}.
\end{equation}
This value can also be used to estimate the effect for higher satellite positions because the refractivity of air is negligible for altitudes above 400~km. 

Finally, considering the satellite velocity magnitude in the corotating frame $v_B= 7.36$~km/s, which approximately corresponds to the ISS, Eq.~(\ref{DBwind4}) gives
\begin{equation}
\Delta_{\rm At}^{\rm wind}=0.93 \times 10^{-18} ({\rm m/s})^{-1} \times V.
\end{equation}
This means that a horizontal wind speed $V > 11$~m/s can cause an effect that exceeds the $10^{-17}$ level.

We also note that in case of negligible air composition variations, Eq. (\ref{n_simple}) holds, and using Eq.~(\ref{Aj_expanded}), we obtain 
\begin{equation}
{{A}}_i =-2\frac{N_0}{\rho_0}\rho V_i+O(2), \label{Aj_rho}
\end{equation}
with $\rho$ being the air density field and $N_0$, $\rho_0$ being air refractivity and density in a certain spacetime point.
At the same time, the mass of air that passes through a 2D spatial surface $\Omega$ per unit of time (mass flow rate) is given as
\begin{equation}
\dot{m}=\int_\Omega \rho V_i  n^i \D\Omega , \label{Qdef}
\end{equation}
with $n^i$ being a unit normal field of the surface, and $\D\Omega$ its volume element (both in the Euclidean sense). Therefore, the correction (\ref{DBwind4}) is related to the mass flow rate of air through a narrow band $\Omega$ defined by dragging the vertical path $\hat{x}^i_+(\hat{l})$ in the direction perpendicular to both $v_B^i$ and $\chi^i_+$ by the small Euclidean distance $d$. In this case, taking $n^i$ in the direction of $v_B^i$, Eq.~(\ref{DBwind4}) leads to
\begin{equation}
\Delta_{\rm At}^{\rm wind}= -\frac{2}{c^2}\frac{N_0}{\rho_0}v_B \lim_{d \to 0}\frac{\dot{m}}{\hat{L}d}.
\end{equation}
However, this relation between the correction $\Delta_{\rm At}^{\rm wind}$ and the mass flow rate $\dot{m}$ is not valid for general wind fields and base paths, and it requires special conditions, such as those assumed to derive Eq.~(\ref{DBwind4}).

\section{Theory}\label{sec_Theory}

In this section, the underlying theory is described, and a complete derivation of the results presented in Sect.~\ref{Summary} is given. This section may be skipped when derivation details are not of interest.

\subsection{Light rays in flowing media}

We assume that an electromagnetic signal propagates in spacetime with the metric $g_{\alpha\beta}$, filled with a medium that moves with a four-velocity field $U^\alpha$ satisfying $U^\alpha U_\alpha=-1$. Next, we assume that the electromagnetic properties of the medium are linear, isotropic, transparent, and nondispersive, such that they are described by two scalar functions,  the permittivity $\epsilon(x^\alpha)$, and the permeability $\mu(x^\alpha)$. The refractive index of the medium is then defined as  $n(x^\alpha)=c\sqrt{\epsilon(x^\alpha)\mu(x^\alpha)}$. 

According to \citet{Gordon}, the Maxwell equations in the medium can be reformulated using the optical metric, which is defined as
\begin{equation}
\bar{g}_{\alpha\beta}\equiv g_{\alpha\beta}+\left(1-\frac{1}{n^2}\right)U_\alpha U_\beta .
\label{optMet}
\end{equation}
A summary of this reformulation and of the geometrical optics approximation is given, for example, by \citet{Chen1}, and we follow this reference in this paragraph. A detailed treatment of the ray optics in relativistic media is given by \citet{Synge} and \citet{Perlick}.

We denote by ${\nabla}_\alpha$ the covariant derivative associated with the metric $g_{\alpha\beta}$ and $\bar{\nabla}_\alpha$ the covariant derivative associated with the optical metric  $\bar{g}_{\alpha\beta}$. We introduce a notation with the bar for tensors, obtained by lowering or raising indices with the optical metric $\bar{g}_{\alpha\beta}$ or its inverse $\bar{g}^{\alpha\beta}$. For example, for the electromagnetic tensor, we have $\bar{g}^{\alpha\mu}\bar{g}^{\beta\nu}F_{\mu\nu}=\bar{F}^{\alpha\beta}$. The Maxwell equations in the medium with vanishing free four-current then read
\begin{equation}
\partial_{[\alpha}F_{\beta\gamma]}=0, \hspace{0.8cm} \bar{\nabla}_\beta\left(\sqrt{\epsilon/\mu}\ \bar{F}^{\beta\alpha}\right)=0.
\label{Maxwell}
\end{equation}

We use the ansatz of geometrical optics,
\begin{equation}
F_{\alpha\beta}=\mathfrak{R}\left\{ e^{iS/\lambdabar} \left(A_{\alpha\beta}+O(\lambdabar)\right) \right\},
\label{GO}
\end{equation}
where $\lambdabar$ is a wavelength related parameter, $S(x^\alpha)$ is the so-called eikonal function, and $\mathfrak{R}$ denotes the real part. We define 
\begin{equation}
k_\alpha=\partial_\alpha S.
\label{defk}
\end{equation}
The Maxwell equations (\ref{Maxwell}) then give to the order $\lambdabar^{-1}$
\begin{equation}
A_{[\alpha\beta}k_{\gamma]}=0, \hspace{1cm}\bar{A}^{\alpha\beta} k_\beta=0.
\end{equation}
Contracting the first equation with $\bar{g}^{\gamma\nu}k_\nu$ and using the second equation, we obtain
\begin{equation}
\bar{g}^{\alpha\beta}k_\alpha k_\beta = 0,
\label{knull}
\end{equation}
therefore, $\bar{k}^\alpha$ is a null vector of the optical metric.
At the same time, from Eq.~(\ref{defk}), it follows that
\begin{equation}
\partial_{[\alpha}k_{\beta]}=\bar{\nabla}_{[\alpha}k_{\beta]}=0.
\label{dk}
\end{equation}
Contracting Eq.~(\ref{dk}) with $\bar{k}^\alpha$ and using Eq.~(\ref{knull}), we obtain
\begin{equation}
\bar{k}^\alpha \bar{\nabla}_\alpha \bar{k}^\beta=0,
\label{geod}
\end{equation}
which means that $\bar{k}^\alpha$ is tangent to the null geodesic of the optical metric. The electromagnetic signal trajectory or the light ray is defined as the integral line of the field $\bar{k}^\alpha$. Therefore, it is the null geodesic of the optical metric.   

To derive the equation of motion for a light ray, we might use the geodesic equation (\ref{geod}), but a more convenient way is to directly start with Eqs.~(\ref{knull}) and (\ref{dk}). Taking a partial derivative of Eq.~(\ref{knull}) with respect to the spacetime coordinates, using Eq.~(\ref{dk}), and assuming the field $\bar{k}^\beta$ to be tangent to the trajectory $x^\beta(a)$ we search for, that is,
\begin{equation}
\frac{\D x^\beta}{\D a} = \bar{k}^\beta=\bar{g}^{\alpha\beta}k_\alpha,\label{Ham1_af}
\end{equation} 
we obtain the equation of motion with an affine parameterization
\begin{equation}
\frac{\D k_\gamma}{\D a}+\frac{1}{2}\left(\partial_\gamma\bar{g}^{\alpha\beta}\right)k_\alpha k_\beta=0,
\label{Ham_af}
\end{equation}
where the inverse optical metric is given as
\begin{equation}
\bar{g}^{\alpha\beta}=g^{\alpha\beta}+(1-n^2)U^\alpha U^\beta .\label{gopt_inv}
\end{equation}

\subsection{Frequency shift in nonstationary flowing media}

The period of the electromagnetic wave given by Eq. (\ref{GO}) as observed by an observer moving along the trajectory $x^\alpha(\tau)$ parameterized by its proper time $\tau$ and having a four-velocity $u^\alpha=\D x^\alpha/\D c\tau$ is given as the proper time $\Delta\tau$ during which the phase $S(x^\alpha(\tau))/\lambdabar$ changes by $2\pi$ in a linear order (see, e.g., \citet{Chen1} and \citet{Synge}), so that we have
\begin{equation}
2\pi=-\frac{1}{\lambdabar}\frac{\D S(x^\alpha(\tau))}{\D\tau}\Delta\tau =-\frac{c\Delta\tau}{\lambdabar}u^\alpha k_\alpha ,
\end{equation}
where the $\text{minus}$ sign is there to obtain a positive $\Delta\tau$ for future oriented $\bar{k}^\alpha$. Therefore, the proper frequency $\nu=1/\Delta\tau$ is given as
\begin{equation}
\nu=-\frac{c}{2\pi\lambdabar}u^\alpha k_\alpha .
\end{equation}

We consider a signal emitted with the proper frequency $\nu_I$ by an observer with the four-velocity $u^\alpha_I$ from the spacetime point $x^\alpha_I$ that is received with the proper frequency $\nu_F$ by an observer with the four-velocity $u^\alpha_F$ in the spacetime point $x^\alpha_F$ lying on the signal trajectory given by the solution $x^\alpha(a)$ of the Eqs.~(\ref{Ham1_af}) and (\ref{Ham_af}), satisfying $x^\alpha(a_I)=x^\alpha_I$ and $x^\alpha(a_F)=x^\alpha_F$. The ratio of the frequencies then is
\begin{equation}
\frac{\nu_{I}}{\nu_{F}}=\frac{(u^\alpha k_{\alpha})_I}{(u^\alpha k_{\alpha})_F} ,
\label{fratio}
\end{equation}
where $(k_{\alpha})_I$ and $(k_{\alpha})_F$ are the values of $k_\alpha$ in the emission and reception events, which can be calculated from Eq.~(\ref{Ham1_af}) when the solution $x^\alpha(a)$ is known.

As the first step, we replace the affine parameter $a$ that is given by solution of the system of Eqs. (\ref{Ham1_af}) and (\ref{Ham_af}) by another parameter $\lambda$ that is defined by an additional equation. Namely, we consider the vector field $\xi^\alpha$ with $k_\alpha{\xi}^\alpha\neq0$ (specified below), and we define the parameter $\lambda$ such that the corresponding geodesic tangent $l^\alpha=\D x^\alpha(a(\lambda))/\D\lambda$ satisfies  
\begin{equation}
\bar{g}_{\alpha\beta}l^\alpha\xi^\beta=-1
\label{norm}
\end{equation} 
all along the light ray. We denote $\kappa(\lambda)=\D\lambda/\D a$ the factor converting $l^\alpha$ to the affine tangent $\bar{k}^\alpha$, that is,
\begin{equation}
\bar{k}^\alpha = \kappa(\lambda) l^\alpha .
\label{ldef}
\end{equation}
The frequency ratio then is
\begin{equation}
\frac{\nu_I}{\nu_F}=\frac{(u^\alpha \bar{l}_{\alpha})_I}{(u^\alpha \bar{l}_{\alpha})_F}\frac{\kappa_I}{\kappa_F}.
\label{fratio2}
\end{equation} 
Now we express the ratio $\kappa_I/\kappa_F$ in terms of the fields along the signal trajectory. The geodesic equation (\ref{geod}) gives
\begin{equation}
\bar{k}^\alpha \bar{\nabla}_\alpha \bar{k}^\beta=(\kappa l^\alpha\bar{\nabla}_\alpha\kappa)l^\beta+\kappa^2 l^\alpha\bar{\nabla}_\alpha l^\beta=0
\end{equation}
or
\begin{equation}
\frac{\D\kappa}{\D\lambda}l^\beta+\kappa l^\alpha\bar{\nabla}_\alpha l^\beta=0.
\label{sigmaEq1}
\end{equation}

Contracting Eq.~(\ref{sigmaEq1}) with $\bar{\xi}_\beta$, we obtain
\begin{eqnarray}
\frac{1}{\kappa}\frac{\D\kappa}{\D\lambda}&=&l^\alpha\bar{\xi}_\beta\bar{\nabla}_\alpha l^\beta \nonumber\\
&=&-l^\alpha l^\beta\bar{\nabla}_\alpha \bar{\xi}_\beta ,\label{sigmaEq2}
\end{eqnarray}
where in the second line we used $l^\alpha\bar{\nabla}_\alpha(l^\beta\bar{\xi}_\beta)=0$, which follows from Eq.~(\ref{norm}). Taking into account that the Lie derivative of $\bar{g}_{\alpha\beta}$ with respect to the field $\xi^\alpha$ is given as
\begin{equation}
L_\xi\bar{g}_{\alpha\beta}=\bar{\nabla}_\alpha \bar{\xi}_\beta+\bar{\nabla}_\beta \bar{\xi}_\alpha ,
\end{equation}
we obtain
\begin{equation}
\frac{1}{\kappa}\frac{\D\kappa}{\D\lambda}=-\frac{1}{2}l^\alpha l^\beta L_{\xi}\bar{g}_{\alpha\beta}.\label{kappa_Lie}
\end{equation}

Integrating Eq.~(\ref{kappa_Lie}) along the signal trajectory from $\lambda_I$ to $\lambda_F$, we obtain
\begin{equation}
\frac{\kappa_I}{\kappa_F}=\exp\int\limits^{\lambda_F}_{\lambda_I}\frac{1}{2}l^\alpha l^\beta L_{\xi}\bar{g}_{\alpha\beta}\ \D\lambda .
\end{equation}
Equation~(\ref{fratio2}) then leads to the following result for the frequency ratio:
\begin{equation}
\frac{\nu_I}{\nu_F}=\frac{(u^\alpha \bar{l}_{\alpha})_I}{(u^\alpha \bar{l}_{\alpha})_F}\exp\int\limits^{\lambda_F}_{\lambda_I}\frac{1}{2}l^\alpha l^\beta L_{\xi}\bar{g}_{\alpha\beta}\ \D\lambda .
\label{fratio3}
\end{equation}

Next, we select $\xi^\alpha$ to be the coordinate basis vector field corresponding to the time coordinate (i.e., considering the coordinate system $(x^0, x^i)$ with $x^0=ct$ on spacetime, we take $\xi^\alpha=(\partial_{0})^\alpha$). In this case, the condition (\ref{norm}) leads to
\begin{equation}
\bar{l}_0=-1.\label{barl0}
\end{equation}
Denoting with $v^i=\D x^i/\D t$ the coordinate velocity of an observer, the four-velocity of the observer is expressed as
\begin{equation}
u^\alpha = u^0(1, v^i/c),\label{udef}
\end{equation}
and we obtain
\begin{equation}
u^\alpha\bar{l}_\alpha=u^0(-1+\tfrac{1}{c}v^i\bar{l}_i).
\end{equation}
Moreover, expressing all tensor components in the coordinate basis of the selected coordinate system, the Lie derivative in the integral reduces to the partial derivative with respect to $x^0$. Equation~(\ref{fratio3}) then gives
\begin{equation}
\frac{\nu_I}{\nu_F}=\frac{u^0_I}{u^0_F}\ \frac{1-\frac{1}{c}(v^i\bar{l}_{i})_I}{1-\frac{1}{c}(v^i\bar{l}_{i})_F}\ \exp\!\int\limits^{\lambda_F}_{\lambda_I}\frac{1}{2}l^\alpha l^\beta\partial_0 \bar{g}_{\alpha\beta}\ \D \lambda.
\label{fratioX}
\end{equation}
This is an analogy of the formula (A.46) of \citet{Blanchet}, generalized to the nonstationary spacetime filled with a nonstationary medium. We also note that for $\xi^\alpha=(\partial_{0})^\alpha$, we have $\kappa=-k_0$ and $\bar{l}_i=-k_i/k_0$.

The first fraction in Eq.~(\ref{fratioX}) contains the gravitational redshift and the second-order Doppler shift. The second fraction is the first-order Doppler shift in the flowing medium. Here, effects of the medium such as bending of the light ray due to the refractive index gradient or dragging of the light by the motion of the medium enter most significantly. The integral in the exponent in Eq.~(\ref{fratioX}) gives the correction for nonstationarity of the optical metric, or in other words, for the time dependence of the refractive index, the flow velocity of the medium and the gravitational field along the signal path.

In order to proceed with Eq.~(\ref{fratioX}) that is used for the frequency transfer and also with the calculation of the signal propagation time used for the time transfer, we now need to find the solution of the system of Eqs.~(\ref{Ham1_af}) and (\ref{Ham_af}), reparameterized to the parameter $\lambda$ or another related parameter. Using Eqs.~(\ref{ldef}) and (\ref{kappa_Lie}), we transform the system of Eqs.~(\ref{Ham1_af}) and (\ref{Ham_af}) into the following set of equations of motion in Hamiltonian form:
\begin{eqnarray}
\frac{\D x^\alpha}{\D \lambda}&=&\bar{g}^{\alpha\beta}\bar{l}_\beta ,  \label{Hamx}\\
\frac{\D \bar{l}_\gamma}{\D \lambda}&=&-\frac{1}{2}\left(\partial_\gamma\bar{g}^{\alpha\beta}+\bar{l}_\gamma L_\xi\bar{g}^{\alpha\beta}\right)\bar{l}_\alpha\bar{l}_\beta .\label{Haml}
\end{eqnarray}
With the choice $\xi^\alpha=(\partial_{0})^\alpha$, which leads to Eq.~(\ref{barl0}), the time component of Eq.~(\ref{Haml}) is satisfied identically.

In the next sections, we solve this set of equations for the conditions in the Earth atmosphere, and we express the time and frequency transfer corrections based on this solution.

\subsection{Optical metric in the vicinity of the Earth}

The approximation level of the model as described in Sect.~\ref{sec_accuracy} requires the Doppler terms $\bar{l}_iv^i/c$ in Eq.~(\ref{fratioX}) to be evaluated including all terms on the order of $\mathcal{P}(3)$, and the same accuracy is required for the propagation time $\Delta t=\Delta x^0/c$. For this approximation level, it is sufficient to expand the metric components $\bar{g}^{\alpha\beta}$ on the RHS of Eqs.~(\ref{Hamx}) and (\ref{Haml}) to include all terms with the power of $c^{-1}$ lower by one, which are the terms on the order of $\mathcal{P}(2)$. This can be deduced retrospectively from the procedure leading to the final results.

We solve the problem in the rotating coordinates, where the components of the spacetime metric are given by Eq.~(\ref{grot}) and the components of the four-velocity of the medium $U^\alpha$ are
\begin{equation}
U^\alpha=U^0(1,{V}^i/c) , \label{defU}
\end{equation}
where $V^i(x^\alpha)$ is the velocity field of the medium in the rotating coordinates given by Eq.~(\ref{Vdef}), and
\begin{equation}
U^0=1+\frac{W}{c^2}+\frac{1}{2c^2}|{v}^i_R+{V}^i|^2+O(c^{-4}).\label{U0}
\end{equation}
Using the definition (\ref{Ai_def}), the contravariant components of the optical metric (\ref{gopt_inv}) in the rotating coordinates to the approximation level described above are
\begin{subequations}
\begin{eqnarray}
\bar{g}^{00}&=&-n^2-\frac{2W}{c^2},\\
\bar{g}^{0i}&=&\frac{1}{c}(v_R^i + {{A}}^i),\\
\bar{g}^{ij}&=&\delta^{ij}\left(1-\frac{2W}{c^2}\right)-\frac{1}{c^2}v_R^i v_R^j.
\end{eqnarray}
\label{goptrot_up}
\end{subequations}
The covariant components of the optical metric (\ref{optMet}) in the rotating coordinates to the same approximation level are 
\begin{subequations}
\begin{eqnarray}
\bar{g}_{00}&=&-\frac{1}{n^2}+\frac{2W}{c^2}+\frac{v_R^2}{c^2},\\
\bar{g}_{0i}&=&\frac{1}{cn^2}(v_{Ri}+{{A}}_i),\\
\bar{g}_{ij}&=&\delta_{ij}\left(1+\frac{2W}{c^2}\right).
\end{eqnarray}
\label{goptrot}
\end{subequations}

\subsection{Equation of motion for a light ray in the atmosphere of Earth} \label{EoM}

We express Eqs. (\ref{Hamx}) and (\ref{Haml}) using the optical metric (\ref{goptrot_up}) and Eq.~(\ref{barl0}). The Lie derivative in Eq.~(\ref{Haml}) is given as $L_\xi\bar{g}^{\alpha\beta}=\partial_0 \bar{g}^{\alpha\beta}$.
Expanding the RHSs of Eqs.~(\ref{Hamx}) and (\ref{Haml}) such that they include all terms on the order of $\mathcal{P}(2)$ (we also keep the $\partial_0 {{A}}^i$ term to determine how it affects the equation of motion even if it is on the order that we otherwise neglect), we obtain
\begin{eqnarray}
\frac{\D \bar{l}_i}{\D \lambda}&=&\frac{1}{2}\partial_i (n^2)+\frac{2}{c^2}\partial_i W+\frac{1}{c}\partial_i (v^j_{R})\bar{l}_j\left(1+\frac{1}{c}v_R^k\bar{l}_k\right)\nonumber\\
&+&\frac{1}{c}\bar{l}_j\partial_i {{A}}^j+\frac{1}{2}\partial_0(n^2)\bar{l}_i+\frac{1}{c}\bar{l}_i\bar{l}_j\partial_0 {{A}}^j\label{Haml_rot}\\
\frac{\D x^i}{\D \lambda}&=&\bar{l}_i\left(1\!-\!\frac{2W}{c^2}\right)\!-\!\frac{1}{c}(v_R^i \!+\! {{A}}^i) -\frac{1}{c^2}v_R^i v_R^j \bar{l}_j\label{Hamx_rot}\\
\frac{\D x^0}{\D \lambda}&=&n^2+\frac{2W}{c^2}+\frac{1}{c}(v_R^i+{{A}}^i)\bar{l}_i,\label{Hamx0_rot}
\end{eqnarray}
where we used the fact that
\begin{equation}
\delta^{ij}\bar{l}_i\bar{l}_j=1+O(1),\label{barl_norm}
\end{equation}
which can be obtained when we express $\bar{l}_k=(\delta^{ij}\bar{l}_i\bar{l}_j)^{\!1\!/\!2}a_k$ with $\delta^{ij}a_ia_j=1$, and we solve the null condition $\bar{g}^{\alpha\beta}\bar{l}_\alpha\bar{l}_\beta =0$ for $(\delta^{ij}\bar{l}_i\bar{l}_j)^{\!1\!/\!2}$.

Equation (\ref{Hamx_rot}) can be inverted to give $\bar{l}_i$ as a function of $\D x^i/\D\lambda$. Thus, we obtain the following formula including all terms on the order of $\mathcal{P}(2)$:
\begin{equation}
\bar{l}_i=\frac{\D x^i}{\D\lambda}\left(1+\frac{2W}{c^2}\right)+\frac{1}{c}(v_R^i+{{A}}^i) +\frac{1}{c^2}v_R^i v_{Rj} \frac{\D x^j}{\D\lambda}.\label{Hamx_inv}
\end{equation}

Differentiating Eq.~(\ref{Hamx_rot}) with respect to $\lambda$ and using Eqs.~(\ref{Haml_rot}), (\ref{Hamx0_rot}), and (\ref{Hamx_inv}) we arrive at the second-order Newton-type equation of motion for a light ray, with the RHS including all terms on the order of $\mathcal{P}(2)$,
\begin{eqnarray}
\frac{\D^2 x^i}{\D\lambda^2}&=&\frac{1}{2}\partial_i(n^2)+\frac{2}{c^2}\partial_j W\left(\delta^{ij}-\frac{\D x^i}{\D\lambda}\frac{\D x^j}{\D\lambda}\right)+\frac{1}{2c^2}\partial_i v_R^2\nonumber\\
&-&\frac{1}{c}\partial_0 {{A}}_j\left(\delta^{ij}-\frac{\D x^i}{\D\lambda}\frac{\D x^j}{\D\lambda}\right)+\frac{1}{2}\partial_0(n^2)\frac{\D x^i}{\D\lambda}\nonumber\\
&+&\frac{2}{c}\left(\partial_{[i} v_{Rj]}+\partial_{[i}{{A}}_{j]}\right)\frac{\D x^j}{\D\lambda}\nonumber\\
&+&\frac{1}{c^2}\partial_i (v_{Rj} v_{Rk})\frac{\D x^j}{\D\lambda}\frac{\D x^k}{\D\lambda},\label{eqm_ray_lambda}
\end{eqnarray}
where the antisymmetrization is defined as $\partial_{[i}{{A}}_{j]}=\frac{1}{2}(\partial_i {{A}}_j-\partial_j {{A}}_i)$. 
Inserting Eq.~(\ref{Hamx_inv}) into Eq.~(\ref{Hamx0_rot}), we obtain the time component of the equation of motion, with the RHS including all terms on the order of $\mathcal{P}(2)$,
\begin{equation}
\frac{\D x^0}{\D\lambda}=n^2+\frac{2W}{c^2}+\frac{v_R^2}{c^2}+\frac{1}{c}\left(v_{Ri}+{{A}}_i \right)\frac{\D x^i}{\D\lambda}. \label{eqm_ray_t2}
\end{equation} 

For the purpose of analyzing the two-way transfer, it is convenient to change the parameter $\lambda$ in the equation of motion into the Euclidean length of the light ray $l_E$, which is defined by 
\begin{equation}
\frac{\D l_E}{\D \lambda}=\sqrt{\delta_{ij}\frac{\D x^i}{\D \lambda}\frac{\D x^j}{\D \lambda}}=|l^i|.
\label{dledla}
\end{equation}
In this case, the spatial part of the light ray $x^i(l_E)$ has a unit tangent:
\begin{equation}
\delta_{ij}\frac{\D x^i}{\D l_E}\frac{\D x^j}{\D l_E}=1.
\label{Enorm}
\end{equation}

Using $l^i=|{l^i}|\D x^i/\D l_E$ and solving the null-vector condition $\bar{g}_{\alpha\beta}l^\alpha l^\beta=0$ together with Eq.~$(\ref{norm})$ for $l^0$ and $|{l^i}|$, we obtain the following formulas including all terms on the order of $\mathcal{P}(2)$:
\begin{eqnarray}
l^0&=&n^2+\frac{2W}{c^2}+\frac{v_R^2}{c^2}+\frac{n}{c}\left(v_{Ri}+{{A}}_i\right)\frac{\D x^i}{\D l_E},\label{l^0}\\
|{l^i}|&=&n+\frac{1}{2c^2}v_{Ri}v_{Rj}\left(\delta^{ij}-\frac{\D x^i}{\D l_E}\frac{\D x^j}{\D l_E}\right).\label{mag_l}
\end{eqnarray}
The transformed Eqs.~(\ref{eqm_ray_lambda}) and (\ref{eqm_ray_t2}) with RHSs including all terms on the order of $\mathcal{P}(2)$ then are
\begin{eqnarray}
\frac{\D^2 x^i}{\D l_E^2}\!\!\!&=&\!\!\!\left(\frac{1}{n}\partial_j n+\frac{2}{c^2}\partial_j W+\frac{1}{2c^2}\partial_j v_R^2-\frac{1}{c}\partial_0 {{A}}_j\right)\left(\delta^{ij}-\frac{\D x^i}{\D l_E}\frac{\D x^j}{\D l_E}\right)\nonumber\\
&+&\!\!\!\frac{2}{nc}\left(\partial_{[i} v_{Rj]}+\partial_{[i}{{A}}_{j]}\right)\frac{\D x^j}{\D l_E}\nonumber\\
&+&\!\!\!\frac{1}{c^2}\partial_i (v_{Rj}v_{Rk})\frac{\D x^j}{\D l_E}\frac{\D x^k}{\D l_E}\label{eqm_ray_lE}
\end{eqnarray}
and
\begin{equation}
\frac{\D x^0}{\D l_E}=n+\frac{2W}{c^2}+\frac{v_R^2}{2c^2}+\frac{1}{c}\left(v_{Ri}+{{A}}_i \right)\frac{\D x^i}{\D l_E}
+\frac{1}{2c^2}v_{Ri}v_{Rj}\frac{\D x^i}{\D l_E}\frac{\D x^j}{\D l_E}.\label{eqm_ray_t2_lE}
\end{equation}
 
The terms in the first line of Eq.~(\ref{eqm_ray_lE}) correspond to optical acceleration due to the refractive index gradient, gravitational acceleration, centrifugal acceleration, and acceleration due to the time variation in the field ${{A}}_j$. These accelerations are projected onto a plane perpendicular to the light-ray direction by the projection tensor at the end of the line. This corresponds to the selected parameterization by the Euclidean length, which does not allow changes in "velocity" magnitude (see Eq.~(\ref{Enorm})). The second line is an analog of a magnetic force, with $v_{Rj}$ being the magnetic potential, leading to Coriolis acceleration, and ${{A}}_j$ being an additional magnetic potential originating from the wind speed and the refractive index (see Eq.~(\ref{Ai_def})). It is remarkable that the effect of the velocity field of the medium to light rays is similar to the effect of the magnetic potential to charged particles. This has been pointed out in the literature (see, e.g., \citet{Leonhardt}).

We expressed the equation of motion in the coordinates rotating together with the Earth, but a simpler form of the equation in the GCRS coordinates can be obtained by selecting ${v}^i_R=0$ if needed.

\subsection{Solving the equation of motion for a light ray}

In this section, we solve the system of differential equations (\ref{eqm_ray_lE}) and (\ref{eqm_ray_t2_lE}) assuming that we know the spatial coordinates of an emitter $x^i_I$ at the event of the emission, the spatial coordinates $x^i_F$ of the receiver at the event of reception, and either the coordinate time of the emission $x^0_I=ct_I$ or the coordinate time of the reception $x^0_F=ct_F$. Denoting by $L$ the Euclidean length of the light path between the emission and reception events, the range of the parameter $l_E$ is $l_E\in[0,L]$, and we assume the following boundary conditions: 
\begin{eqnarray}
x^i(0)=x^i_I &\ \ {\rm and}\ \ & x^i(L)=x^i_F,\label{bcX}\\
x^0(0)=ct_I &\ \ {\rm or}\ \ & x^0(L)=ct_F.\label{bcT}
\end{eqnarray}
For the sake of brevity, we write Eqs.~(\ref{eqm_ray_lE}) and (\ref{eqm_ray_t2_lE}) in the form 
\begin{eqnarray}
\frac{\D^2 x^i}{\D l_E^2}\!\!\!\!&=&\!\!\!\!E_j(x^\alpha)\left(\delta^{ij}-\frac{\D x^i}{\D l_E}\frac{\D x^j}{\D l_E}\right)+E_{ij}(x^\alpha)\frac{\D x^j}{\D l_E}+E_{ijk}(x^i)\frac{\D x^j}{\D l_E}\frac{\D x^k}{\D l_E}, \nonumber\\
&& \label{eqmX}\\
\frac{\D x^0}{\D l_E}\!\!\!\!&=&\!\!\!\!\Phi(x^\alpha)+\Phi_i(x^\alpha)\frac{\D x^i}{\D l_E}+\Phi_{ij}(x^\alpha)\frac{\D x^i}{\D l_E}\frac{\D x^j}{\D l_E},
\label{eqmT}
\end{eqnarray} 
where we introduced the coefficients
\begin{eqnarray}
E_i(x^\alpha)&=&\frac{1}{n}\partial_i n+\frac{2}{c^2}\partial_i W+\frac{1}{2c^2}\partial_i v_R^2\!-\!\frac{1}{c}\partial_0 {{A}}_i,\label{Ei}\\
E_{ij}(x^\alpha)&=&\frac{2}{nc}\left(\partial_{[i} v_{Rj]}+\partial_{[i}{{A}}_{j]}\right)\nonumber\\
&=&\frac{1}{nc}\epsilon_{ijk}\left(2\omega^k+{\rm curl}{{A}}^k\right),\label{Eij}\\
E_{ijk}(x^i)&=&\frac{1}{c^2}\partial_i (v_{Rj}v_{Rk}),\label{Eijk}
\end{eqnarray}
\begin{eqnarray}
\Phi(x^\alpha)&=&n+\frac{2W}{c^2}+\frac{v_R^2}{2c^2},\\
\Phi_i(x^\alpha)&=&\frac{1}{c}(v_{Ri}+{{A}}_i),\\
\Phi_{ij}(x^\alpha)&=&\frac{1}{2c^2}v_{Ri}v_{Rj}\ .
\end{eqnarray}

\subsubsection{Split of the spatial equation into a spherically symmetric static part and a correction}

The strategy for the approximate solution of the spatial part of the system given by Eq.~(\ref{eqmX}) with the boundary conditions (\ref{bcX}) is to split the RHS of Eq.~(\ref{eqmX}) into a spherically symmetric static part plus a correction to solve the spherically symmetric static problem and to formulate and solve the equation for a correction of this solution up to a linear order in this correction. Using Eqs.~(\ref{def_dW}) and (\ref{def_alpha}), we decompose Eq.~(\ref{Ei}) as
\begin{equation}
E_i(x^\alpha)=\hat{E}_i(x^j)+\Delta E_i(x^\alpha),
\label{defDE}
\end{equation} 
with
\begin{eqnarray}
\hat{E}_i&=&\frac{1}{\hat{n}}\partial_i\hat{n}+\frac{2}{c^2}\partial_i\hat{W},\label{def_hatE}\\
\Delta E_i&=&\frac{\partial_i\alpha}{1+\alpha}+\frac{2}{c^2}\partial_i\Delta W+\frac{1}{2c^2}\partial_i v_R^2\!-\!\frac{1}{c}\partial_0 {{A}}_i.
\end{eqnarray}

We denote by $\hat{x}^i(\hat{l})$ the solution of the equation
\begin{equation}
\frac{\D^2 \hat{x}^i}{\D \hat{l}^2}=\hat{E}_j(\hat{x}^k(\hat{l}))\left(\delta^{ij}-\frac{\D \hat{x}^i}{\D \hat{l}}\frac{\D \hat{x}^j}{\D \hat{l}}\right),
\label{eqmX0}
\end{equation}
with $\hat{l}\in[0,\hat{L}]$ satisfying the boundary conditions
\begin{equation}
\hat{x}^i(0)=x^i_I-\Delta x^i_I,\ \  \hat{x}^i(\hat{L})=x^i_F-\Delta x^i_F ,\label{bcX0}
\end{equation}
where we admit certain deviations $\Delta x^i_I, \Delta x^i_F$ of the boundary points from the boundary points of the complete solution given by Eq.~(\ref{bcX}), which will be useful for two-way time and frequency transfer. Practically, we use $\Delta x^i_I=0$ and either $\Delta x^i_F=0$ or $\Delta x^i_F=x^i_{A2}-x^i_{A1}$, which is a spatial shift of the observer on the ground in corotating frame during the two-way back and forth propagation of the signal.

Since Eq.~(\ref{eqmX0}) is a special case of Eq.~(\ref{eqm_ray_lE}) with vanishing fields $\alpha, \Delta W, v_R^i$, and ${{A}}^i$, the condition (\ref{Enorm}) also applies to the solution of Eq.~(\ref{eqmX0}), so that we have
\begin{equation}
\delta_{ij}\frac{\D \hat{x}^i}{\D \hat{l}}\frac{\D \hat{x}^j}{\D \hat{l}}=1,
\label{Enorm0}
\end{equation}
which means that $\hat{x}^i(\hat{l})$ is the path of a light ray between the points $x^i_I-\Delta x^i_I$ and $x^i_F-\Delta x^i_F$ for vanishing $\alpha, \Delta W, v_R^i$, and ${{A}}^i$, parameterized by its Euclidean length, and $\hat{L}$ is the total Euclidean length of the path, which can differ from $L$. Reparameterizing the solution $x^i(l_E)$ of Eq.~(\ref{eqmX}) to the range $[0,\hat{L}]$ by introducing a parameter $\hat{l}=(\hat{L}/L)l_E$ on $x^i(l_E)$, the solution of the complete Eq.~(\ref{eqmX}) can be expressed as
\begin{equation}
x^i(\hat{l}{L}/{\hat{L}})=\hat{x}^i(\hat{l})+\Delta x^i(\hat{l}),
\label{defDx}
\end{equation}
which defines the correction $\Delta x^i(\hat{l})$. Similarly, if $x^0(l_E)$ is the solution of Eq.~(\ref{eqmT}) for the time coordinate, the reparameterized solution can be expressed as
\begin{equation}
x^0(\hat{l}{L}/{\hat{L}})=\hat{x}^0+\Delta x^0(\hat{l})=c\hat{t}+c\Delta t(\hat{l}),
\label{defDt}
\end{equation}
where $\hat{x}^0=c\hat{t}$ is a constant with $\hat{t}\in[t_I,t_F]$ and $\Delta t(\hat{l})$ is the function to be found. 

The next step is to express the differential equation for the correction $\Delta x^i(\hat{l})$. To do this, we need to express the ratio $L/\hat{L}$ in order to proceed with the derivatives $\D/\D l_E=(\hat{L}/L)\D/\D\hat{l}$. Taking a derivative of Eq.~(\ref{defDx}) with respect to $\hat{l}$ and using Eqs.~(\ref{Enorm}) and (\ref{Enorm0}), we obtain
\begin{equation}
\frac{L}{\hat{L}}=\sqrt{1+2\delta_{ij}\frac{\D \hat{x}^i}{\D \hat{l}}\frac{\D \Delta{x}^j}{\D \hat{l}}+\delta_{ij}\frac{\D \Delta{x}^i}{\D \hat{l}}\frac{\D \Delta{x}^j}{\D \hat{l}}}.\label{LtoLhat}
\end{equation}
Inserting Eqs.~(\ref{defDx}) and (\ref{defDt}) into Eq.~(\ref{eqmX}), expanding the equation up to the linear order in $\Delta x^\alpha$ and $\D\Delta x^i/\D\hat{l}$ and using Eqs.~(\ref{defDE}) and (\ref{eqmX0}) and the linear expansion of Eq.~(\ref{LtoLhat}) in $\D \Delta x^i/\D\hat{l}$, we arrive at the equation of the form
\begin{equation}
\frac{\D^2 \Delta{x}^i}{\D \hat{l}^2}=S^i(\hat{l})+B^i_j(\hat{l})\Delta{x}^j+C^i_j(\hat{l})\frac{\D \Delta x^j}{\D\hat{l}},\label{EoMdx_SBC}
\end{equation} 
with the boundary conditions 
\begin{equation}
\Delta x^i(0)=\Delta x^i_I,\ \ \Delta x^i(\hat{L})=\Delta x^i_F,
\label{bcdx}
\end{equation}
which follow from Eqs.~(\ref{bcX}), (\ref{bcX0}), and (\ref{defDx}), and with the coefficients given by
\begin{eqnarray}
S^i(\hat{l})&=&S_0^i(\hat{l})+S_t^i(\hat{l}) \Delta t(\hat{l}),\label{defS}\\
S_0^i(\hat{l})&=& \Delta E_j(\hat{x}^\alpha(\hat{l}))\left(\delta^{ij}-\frac{\D \hat{x}^i}{\D\hat{l}}\frac{\D \hat{x}^j}{\D\hat{l}}\right) \nonumber\\
&+&E_{ij}(\hat{x}^\alpha(\hat{l}))\frac{\D \hat{x}^j}{\D\hat{l}}\nonumber\\
&+&E_{ijk}(\hat{x}^i(\hat{l}))\frac{\D \hat{x}^j}{\D\hat{l}}\frac{\D \hat{x}^k}{\D\hat{l}},\label{def_S0}\\
S_t^i(\hat{l})&=& \partial_t \Delta E_j(\hat{x}^\alpha(\hat{l}))\left(\delta^{ij}-\frac{\D \hat{x}^i}{\D\hat{l}}\frac{\D \hat{x}^j}{\D\hat{l}}\right) \nonumber\\
&+&\partial_tE_{ij}(\hat{x}^\alpha(\hat{l}))\frac{\D \hat{x}^j}{\D\hat{l}},\label{def_St}
\end{eqnarray}
\begin{eqnarray}
B^i_j(\hat{l})&=& \partial_j{E}_k(\hat{x}^\alpha(\hat{l}))\left(\delta^{ik}-\frac{\D \hat{x}^i}{\D\hat{l}}\frac{\D \hat{x}^k}{\D\hat{l}}\right)\nonumber\\
&+&\partial_j E_{ik}(\hat{x}^\alpha(\hat{l}))\frac{\D \hat{x}^k}{\D\hat{l}}\nonumber\\
&+&\partial_j E_{ikl}(\hat{x}^i(\hat{l}))\frac{\D \hat{x}^k}{\D\hat{l}}\frac{\D \hat{x}^l}{\D\hat{l}},\label{def_B}
\end{eqnarray}
\begin{eqnarray}
C^i_j(\hat{l})&=&E_k(\hat{x}^\alpha(\hat{l}))\left(2\delta^{ik}\frac{\D \hat{x}^j}{\D\hat{l}}-\delta^k_j \frac{\D \hat{x}^i}{\D\hat{l}}-\delta^i_j \frac{\D \hat{x}^k}{\D\hat{l}}\right)\nonumber\\
&+&E_{ik}(\hat{x}^\alpha(\hat{l}))\left(\delta^k_j+\frac{\D \hat{x}^k}{\D\hat{l}}\frac{\D \hat{x}^j}{\D\hat{l}}\right)\nonumber\\
&+&2E_{ijk}(\hat{x}^i(\hat{l}))\frac{\D \hat{x}^k}{\D\hat{l}}.\label{def_C}
\end{eqnarray}

We proceed with the solution of the particular equations. The solution $\hat{x}^i(\hat{l})$ of the zeroth-order Eq.~(\ref{eqmX0}) with $\hat{E}_j$ given by Eq.~(\ref{def_hatE}) has been discussed in Sect.~\ref{sec_solution_hatx}. We therefore continue with the solution of Eq.~(\ref{EoMdx_SBC}) for the spatial path correction $\Delta x^i(\hat{l})$ with the boundary conditions (\ref{bcdx}).

\subsubsection{Solution for $\Delta x^i(\hat{l})$}

Integrating Eq.~(\ref{EoMdx_SBC}) twice, we obtain
\begin{equation}
\Delta{x}^i(\hat{l})=c_1^i+c_2^i\hat{l}
\!+\int_0^{\hat{l}}\!\!\!\D \hat{l}^\prime \!\! \int_0^{\hat{l}^\prime}\!\!\!\D \hat{l}^{\prime\prime}\left[S^i\!+\!B^i_j\Delta{x}^j\!+\!C^i_j\frac{\D \Delta x^j}{\D\hat{l}}\right](\hat{l}^{\prime\prime}),
\end{equation}
with $c_1^i, c_2^i$ being integration constants, and $[\ \ ](\hat{l}^{\prime\prime})$ denoting that the function in the square brackets is evaluated in $\hat{l}^{\prime\prime}$. Determining $c_1^i, c_2^i$ from the boundary conditions (\ref{bcdx}), integrating by parts using 
\begin{equation}
\int_0^{\hat{l}^\prime}\D \hat{l}^{\prime\prime}{f}^i(\hat{l}^{\prime\prime})=\frac{\D}{\D \hat{l}^\prime}\left(\hat{l}^\prime \int_0^{\hat{l}^\prime}\D \hat{l}^{\prime\prime}{f}^i(\hat{l}^{\prime\prime})\right)-\hat{l}^\prime {f}^i(\hat{l}^\prime)
\label{perpartes}
\end{equation}
and expressing the integrals over the range $\hat{l}^\prime\in[0, \hat{l}]$ or $\hat{l}^\prime\in[\hat{l}, \hat{L}]$ as integrals over the full range $\hat{l}^\prime\in[0,\hat{L}]$ using the Heaviside step function $H(\hat{l}^\prime-\hat{l})$, which is defined as $H(\hat{l}^\prime-\hat{l})=0$ for $\hat{l}^\prime<\hat{l}$, $H(\hat{l}^\prime-\hat{l})=1$ for $\hat{l}^\prime>\hat{l}$ and $H(\hat{l}^\prime-\hat{l})=1/2$ for $\hat{l}^\prime=\hat{l}$, we obtain
\begin{eqnarray}
\Delta{x}^i(\hat{l})&=&\Delta x^i_I+(\Delta x^i_F-\Delta x^i_I)\frac{\hat{l}}{\hat{L}} \nonumber\\
&+&\int^{\hat{L}}_0\!\!\D \hat{l}^\prime\left\{\left(({\hat{l}}/{\hat{L}}-1)\hat{l}^\prime+(\hat{l}^\prime-\hat{l}) H(\hat{l}^\prime-\hat{l}) \right)\right. \nonumber\\
&&\left.\times \left[S^i+B^i_j\Delta{x}^j+C^i_j\frac{\D \Delta x^j}{\D\hat{l}}\right](\hat{l}^{\prime})\right\}.\label{dx_int1}
\end{eqnarray}

Integrating by parts again, we can express the term containing $\D \Delta x^j/\D\hat{l}$ in terms of $\Delta x^j$. Thus, we obtain the equation of the form
\begin{eqnarray}
\Delta{x}^i(\hat{l})&=&\Delta x^i_I+(\Delta x^i_F-\Delta x^i_I)\frac{\hat{l}}{\hat{L}}+\int^{\hat{L}}_0\!\D \hat{l}^\prime D(\hat{l},\hat{l}^\prime) S^i(\hat{l}^\prime) \nonumber\\
&+&\int^{\hat{L}}_0\!\D \hat{l}^\prime K^i_j(\hat{l},\hat{l}^\prime)\Delta x^j(\hat{l}^\prime) \label{EoMdx_int}
\end{eqnarray}
with
\begin{eqnarray}
D(\hat{l},\hat{l}^\prime)\!\!\!&=&\!\!\!({\hat{l}}/{\hat{L}}-1)\hat{l}^\prime+(\hat{l}^\prime-\hat{l}) H(\hat{l}^\prime-\hat{l}),\label{defDij}\\
K^i_j(\hat{l},\hat{l}^\prime)\!\!\!&=&\!\!\!D(\hat{l},\hat{l}^\prime)\left[B^i_j-\frac{\D}{\D\hat{l}}C^i_j\right](\hat{l}^\prime)\nonumber\\
&+&\!\!\!\left(-{\hat{l}}/{\hat{L}}+ H(\hat{l}-\hat{l}^\prime) \right)C^i_j(\hat{l}^\prime).\label{defKij}
\end{eqnarray}
The functions have the following properties, which will be useful later:
\begin{eqnarray}
&&D(0,\hat{l}^\prime)=D(\hat{L},\hat{l}^\prime)=0,\label{Dbc}\\
&&K^i_j(0,\hat{l}^\prime)=K^i_j(\hat{L},\hat{l}^\prime)=0,\label{Kbc}\\
&&D(\hat{l},\hat{l}^\prime)=D(\hat{l}^\prime,\hat{l}),\label{Dsymm}\\
&&D(\hat{L}-\hat{l},\hat{L}-\hat{l}^\prime)=D(\hat{l},\hat{l}^\prime).\label{Dback}
\end{eqnarray}

The solution of Eq.~(\ref{EoMdx_int}) can be found in terms of infinite series corresponding to an infinite number of iterations of Eq.~(\ref{EoMdx_int}). To be able to evaluate convergence of this series, we need to introduce the notion of a norm on the space of the possible functions $\Delta x^i(\hat{l})$, and then we can use some results of the linear operator theory in Banach spaces (see, e.g., \citet{Naylor}). 

We define the vector space of all continuous functions $f^i\!: [0, \hat{L}]\to\mathbb{R}^3$, and we define the norm on this vector space as
\begin{equation}
\|f^i\|=\max\limits_{\hat{l}\in[0,\hat{L}]}\left(\max\left\{|f^1(\hat{l})|, |f^2(\hat{l})|, |f^3(\hat{l})|\right\}\right),
\label{norm_B}
\end{equation}
where $f^1, f^2, f^3$ are the corresponding component functions $[0, \hat{L}]\to\mathbb{R}$, which are all continuous.
This vector space together with the norm (\ref{norm_B}) forms a Banach space that we denote $\mathsf{B}^3$ (see, e.g., \citet{Naylor} for the one-dimensional case). The subspace of $\mathsf{B}^3$ that is formed by functions satisfying $f^i(0)=f^i(\hat{L})=0_{\mathbb{R}^3}$ is denoted $\mathsf{B}^3_0$.

The norm on the Banach space $\mathsf{B}^3$ induces an operator norm on the linear operators from $\mathsf{B}^3$ to itself. We denote by $\mathcal{A}^i_j: \mathsf{B}^3\to\mathsf{B}^3$ a bounded linear operator on $\mathsf{B}^3$, and by $\mathcal{A}^i_jf^j$ the image of a vector $f^i\in\mathsf{B}^3$. The operator norm $\|\mathcal{A}^i_j\|$ of the operator is defined as
\begin{equation}
\|\mathcal{A}^i_j\|=\sup\left\{\frac{\|\mathcal{A}^i_jf^j\|}{\|f^i\|}: f^i\in\mathsf{B}^3\ {\rm with}\ f^i\neq 0_{\mathsf{B}^3}\right\},
\label{norm_Op}
\end{equation}
where on the RHS, the norm of the Banach space is used, which is given by Eq.~(\ref{norm_B}) in our case (for details, see definition 5.8.1 and lemma 5.8.2 in \citet{Naylor}).

Now we can proceed with Eq.~(\ref{EoMdx_int}). We assume that for any values of $i,j$ the functions $C^i_j$ and $B^i_j-\D C^i_j/\D\hat{l}$ that appear in Eq.~(\ref{defKij}) are continuous on $[0,\hat{L}]$, and we define the linear operators $\mathcal{D}^i_j: \mathsf{B}^3\to\mathsf{B}^3_0$ and $\mathcal{K}^i_j: \mathsf{B}^3\to\mathsf{B}^3_0$ as 
\begin{eqnarray}
(\mathcal{D}^i_j f^j)(\hat{l}) &\coloneqq&\int^{\hat{L}}_0 \!\D\hat{l}^\prime \ D(\hat{l},\hat{l}^\prime)\delta^i_j f^j(\hat{l}^\prime),\label{Op_D}\\
(\mathcal{K}^i_j f^j)(\hat{l}) &\coloneqq&\int^{\hat{L}}_0 \!\D\hat{l}^\prime \ K^i_j(\hat{l},\hat{l}^\prime)f^j(\hat{l}^\prime)\label{Op_K}
\end{eqnarray}
where the vanishing of the image value in $\hat{l}=0$ and $\hat{l}=\hat{L}$ is ensured by the properties (\ref{Dbc}) and (\ref{Kbc}). 
Next, we denote by $b^i$ the linear function satisfying the boundary conditions (\ref{bcdx}). This function is given as
\begin{equation}
b^i(\hat{l})\coloneqq \Delta x^i_I+(\Delta x^i_F-\Delta x^i_I)\frac{\hat{l}}{\hat{L}}.\label{defb}
\end{equation}
Using these definitions and denoting by $\mathcal{I}^i_j$ the identity operator on $\mathsf{B}^3$, Eq.~(\ref{EoMdx_int}) can be written as
\begin{equation}
(\mathcal{I}^i_j-\mathcal{K}^i_j)\Delta x^j = b^i+\mathcal{D}^i_j S^j.
\label{EoMdx_Op}
\end{equation}

According to theorem 6.7.1 in \citet{Naylor}, for a continuous linear operator $\mathcal{K}^i_j$ satisfying $\|\mathcal{K}^i_j\|<1$, an inverse operator to $\mathcal{I}^i_j-\mathcal{K}^i_j$ exists, and it is given by geometric series as 
\begin{equation}
(\mathcal{I}^i_j-\mathcal{K}^i_j)^{-1}=\sum_{n=0}^{\infty}(\mathcal{K}^i_j)^n,
\label{invOp}
\end{equation}
where $(\mathcal{K}^i_j)^n$ is the operator given by $n$ consecutive applications of $\mathcal{K}^i_j$, so that $(\mathcal{K}^i_j)^0=\mathcal{I}^i_j$, $(\mathcal{K}^i_j)^1=\mathcal{K}^i_j$, and for $n>1$, we have $(\mathcal{K}^i_j)^n=\mathcal{K}^{i}_{k_{n-1}}\dots\mathcal{K}^{k_3}_{k_2}\mathcal{K}^{k_2}_{k_1}\mathcal{K}^{k_1}_{j}$. The continuity of the linear operator is equivalent to its boundedness (see definition 5.6.3 and theorem 5.6.4 in \citet{Naylor}), and the boundedness of $\mathcal{K}^i_j$ can be proven. The assumption $\|\mathcal{K}^i_j\|<1$ should be verified for the physical situation of interest. For the operator (\ref{Op_K}), the operator norm (\ref{norm_Op}) can be expressed in terms of the function $K^i_j(\hat{l},\hat{l}^\prime)$ as
\begin{equation}
\|\mathcal{K}^i_j\|\!=\!{\rm max}\left\{\sum_{j=1}^{3}\!\int_0^{\hat{L}}\!\!\D\hat{l}^\prime |K^i_j(\hat{l}, \hat{l}^\prime)|\!\!:\hat{l}\in[0,\hat{L}],i\!\in\!\{1,2,3\}\!\right\}\!.
\end{equation}

Using Eq.~(\ref{invOp}), the solution of Eq.~(\ref{EoMdx_Op}) can be written as
\begin{eqnarray}
\Delta x^i &=& (\mathcal{I}^i_j-\mathcal{K}^i_j)^{-1} \left(b^j+\mathcal{D}^j_k S^k\right)\nonumber\\
&=&\sum_{n=0}^{\infty}(\mathcal{K}^i_j)^n \left(b^j+\mathcal{D}^j_k S^k\right) ,\label{dx_sol}
\end{eqnarray}
where the boundary conditions (\ref{bcdx}) are ensured by the definition of $b^i$ and by the properties of the operators $\mathcal{D}^i_j, \mathcal{K}^i_j$ given by Eqs.~(\ref{Dbc}) and (\ref{Kbc}).
For the derivative $\D \Delta x^i/\D\hat{l,}$ which is also needed below, we obtain
\begin{eqnarray}
\frac{\D \Delta x^i}{\D\hat{l}}\!\!\!\!&=&\!\!\!\!\frac{\D}{\D\hat{l}}\left(b^i+\mathcal{D}^i_j S^j\right)
+\frac{\D}{\D\hat{l}}\left(\mathcal{K}^i_j\sum_{n=0}^{\infty}(\mathcal{K}^j_k)^n \left(b^k\!+\mathcal{D}^k_l S^l\right)\right)\\
&=&\!\!\!\!\frac{1}{\hat{L}}(\Delta x^i_F\!-\!\Delta x^i_I)+\dot{\mathcal{D}}^i_jS^j
\!+\dot{\mathcal{K}}^i_j\!\sum_{n=0}^{\infty}(\mathcal{K}^j_k)^n \left(b^k\!+\mathcal{D}^k_l S^l\right), \label{ddx/dl}
\end{eqnarray}
where
\begin{eqnarray}
(\dot{\mathcal{D}}^i_jS^j)(\hat{l})&\coloneqq&\frac{\D}{\D\hat{l}}(\mathcal{D}^i_j S^j)(\hat{l})\nonumber\\
&=&\int^{\hat{L}}_0\D \hat{l}^\prime\left(\hat{l}^\prime/{\hat{L}}- H(\hat{l}^\prime-\hat{l}) \right)S^i(\hat{l}^\prime) ,
\end{eqnarray}
and the operator $\dot{\mathcal{K}}^i_j$ is defined as 
\begin{eqnarray}
(\dot{\mathcal{K}}^i_jf^j)(\hat{l}) &\coloneqq& \frac{\D}{\D\hat{l}} ({\mathcal{K}}^i_jf^j)(\hat{l}) \nonumber\\
&=&C^i_j(\hat{l})f^j(\hat{l})+\int^{\hat{L}}_0 \!\D\hat{l}^\prime L^i_j(\hat{l},\hat{l}^\prime)f^j(\hat{l}^\prime) ,
\end{eqnarray}
with
\begin{equation}
L^i_j(\hat{l},\hat{l}^\prime)=\left(\hat{l}^\prime/{\hat{L}}- H(\hat{l}^\prime-\hat{l}) \right)\left[B^i_j-\frac{\D}{\D\hat{l}}C^i_j\right](\hat{l}^\prime)-\frac{1}{\hat{L}}C^i_j(\hat{l}^\prime).
\end{equation}

\subsubsection{Equation for $\Delta t$ and its solution}

For further analysis, it is convenient to split the solution (\ref{dx_sol}) for $\Delta x^i$ into a part that does not depend on $\Delta t$ and a part that  does depend on it, using Eq.~(\ref{defS}). Defining 
\begin{eqnarray}
\Delta x^i_0&=&\sum_{n=0}^{\infty}(\mathcal{K}^i_j)^n \left(b^j+\mathcal{D}^j_k S^k_0\right),\label{dx0}\\
\Delta x^i_t&=&\sum_{n=0}^{\infty}(\mathcal{K}^i_j)^n \mathcal{D}^j_k (S^k_t \Delta t),\label{dxt}
\end{eqnarray}
we can write
\begin{equation}
\Delta x^i=\Delta x^i_0+\Delta x^i_t.
\label{split_dx0t}
\end{equation}

Inserting Eqs.~(\ref{defDx}) and (\ref{defDt}) into Eq.~(\ref{eqmT}), using Eq.~(\ref{LtoLhat}) for reparameterization to the parameter $\hat{l}$, expanding the equation as Taylor series in $\Delta x^\alpha$ and $\D \Delta x^i/\D\hat{l}$ and using Eqs.~(\ref{dx0}), (\ref{dxt}) and (\ref{split_dx0t}), we obtain the following equation for $\Delta t(\hat{l})$:
\begin{equation}
c\frac{\D \Delta t}{\D \hat{l}}=\hat{\sigma}(\hat{l})+\Delta\sigma(\hat{l})+\partial_t n(\hat{x}^\alpha(\hat{l}))\Delta t ,
\label{EoMdt_sig}
\end{equation}
with $\hat{\sigma}(\hat{l})$ and $\Delta\sigma(\hat{l})$ given by
\begin{eqnarray}
\hat{\sigma}(\hat{l})\!\!\!\!&=&\!\!\!\!n+\frac{1}{c}\!\left(v_{Ri}\!+\!{{A}}_i\right)\!\frac{\D\hat{x}^i}{\D\hat{l}}
+\frac{2W}{c^2}+\frac{v_R^2}{2c^2}+\frac{1}{2c^2}\left(v_{Ri}\frac{\D\hat{x}^i}{\D\hat{l}}\right)^2\!\!,\label{hatsigma}\\
\Delta\sigma(\hat{l})\!\!\!\!&=&\!\!\!\!\partial_i n\Delta x^i_0+n\delta_{ij}\frac{\D\hat x^i}{\D\hat{l}}\frac{\D\Delta x^j_0}{\D\hat{l}}\nonumber\\
&+&\!\!\!\!\frac{1}{2}\frac{\D\Delta x^i_0}{\D\hat{l}}\frac{\D\Delta x^j_0}{\D\hat{l}}\left(\delta_{ij}-\delta_{ik}\delta_{jl}\frac{\D\hat{x}^k}{\D\hat{l}}\frac{\D\hat{x}^l}{\D\hat{l}}\right)\nonumber\\
&+&\!\!\!\!\frac{1}{c}v_{Ri}\frac{\D\Delta x^i_0}{\D\hat{l}}+\frac{1}{c}(\partial_j v_{Ri})\Delta x^j_0\frac{\D\hat x^i}{\D\hat{l}} ,\label{Dsigma}
\end{eqnarray}
where the fields $n, W, v_{Ri}, \text{and } {{A}}_i$ and their derivatives $\partial_i$ are evaluated in $\hat{x}^\alpha(\hat{l})$. On the RHS of Eq.~(\ref{EoMdt_sig}), we kept all terms on the order of $\mathcal{P}(2)$, taking into account that the largest order of $\Delta t$ is $c^{-1}$ $(p_{c^{-1}}=1, p_{\hat{N}}=p_\alpha=0)$. This is sufficient to reach the required approximation level for the signal propagation time, as described in Sect.~\ref{sec_accuracy}. All terms containing $\Delta x^i_t$ are on a negligible order.

Now we denote $\hat{l}_0$ the value of $\hat{l}$ for which $\Delta t(\hat{l}_0)=0$, for example, if we select $\hat{t}=t_I$, we have $\hat{l}_0=0$, or if we select $\hat{t}=t_F$, we have $\hat{l}_0=\hat{L}$. Integrating Eq.~(\ref{EoMdt_sig}) from $\hat{l}_0$ to $\hat{l}$, we obtain
\begin{equation}
c\Delta t(\hat{l})=\int_{\hat{l}_0}^{\hat{l}}\!\!\D\hat{l}^{\prime}(\hat{\sigma}(\hat{l}^\prime)+\Delta\sigma(\hat{l}^\prime))+\int_{\hat{l}_0}^{\hat{l}}\!\!\D\hat{l}^{\prime}\ \partial_t n(\hat{x}^\alpha\!(\hat{l}^\prime))\Delta t(\hat{l}^\prime).\label{dt_int}
\end{equation}

In principle, we could express the exact solution of Eq.~(\ref{dt_int}) in terms of operator series following a similar procedure as for the spatial correction $\Delta x^i$. However, an approximate solution corresponding to the approximation level of Eq.~(\ref{EoMdt_sig}) itself is sufficient for our application. To obtain this solution, we express the $\Delta t$ on the RHS of Eq.~(\ref{dt_int}) to its largest order $c^{-1}$ only: $\Delta t(\hat{l}) \approx (\hat{l}-\hat{l}_0)/c$, which we obtain by setting $\hat{\sigma}+\Delta\sigma \approx 1$ and $\partial_t n\approx 0$ in Eq.~(\ref{dt_int}). Equation (\ref{dt_int}) then gives
\begin{equation}
\Delta t(\hat{l})=\frac{1}{c}\!\int_{\hat{l}_0}^{\hat{l}}\!\!\D\hat{l}^{\prime}(\hat{\sigma}(\hat{l}^\prime)+\Delta\sigma(\hat{l}^\prime))
+\frac{1}{c^2}\!\int_{\hat{l}_0}^{\hat{l}}\!\!\D\hat{l}^{\prime}(\hat{l}^\prime\!-\hat{l}_0)\partial_t n(\hat{x}^\alpha(\hat{l}^\prime)).\label{dt_intI}
\end{equation}

\subsection{Coordinate time transfer}

\subsubsection{One-way time transfer}

In one-way time transfer, we wish to express the coordinate time $t_F-t_I$ of the propagation of a signal that is emitted from a position $x^i_I$ at a time $t_I$ and is received in a position $x^i_F$ at a time $t_F$. Based on Eqs.~(\ref{bcT}) and (\ref{defDt}), we obtain $t_F-t_I=\Delta t(\hat{L})-\Delta t(0)$. Using Eq.~(\ref{dt_intI}), this leads to
\begin{equation}
t_F-t_I=\frac{1}{c}\!\int_0^{\hat{L}}\!\!\D\hat{l}\ (\hat{\sigma}(\hat{l})+\Delta\sigma(\hat{l}))
+\frac{1}{c^2}\!\int_0^{\hat{L}}\!\!\D\hat{l}\ (\hat{l}-\hat{l}_0) \partial_t n(\hat{x}^\alpha\!(\hat{l})).\label{dt1}
\end{equation}

Now we proceed with the integration of $\Delta\sigma$. We set $\Delta x^i_I=0$, and we assume $\Delta x^i_F\neq 0$ in the boundary conditions (\ref{bcdx}). We keep nonvanishing $\Delta x^i_F$ as a preparation for later use in the two-way time transfer, where we set for the back-propagating signal $\Delta x^i_F$ equal to a shift in the position of the observer $A$ on the ground in the corotating frame during the back and forth propagation of the signal: $\Delta x^i_F=x^i_{A2}-x^i_{A1}\approx v^i_{A1} (2\hat{L}/c)$, with $v_{A1}^i$ being the velocity of observer $A$ in the corotating frame in emission event $A1$. Thus, we consider $\Delta x^i_F$ to be on the order of $c^{-1}$. In the time-transfer formulas, however, we keep only the highest-order term containing $\Delta x^i_F$, which is sufficient for an accuracy of 1 ps in case of a stationary ground clock for which the velocity $v_{A1}^i$ is given by the deformations of Earth, such as solid Earth tides, and by the nonuniformity of the rotation of Earth. In one-way transfer, we can set $\Delta x^i_F=0$ at the end.

Using the properties (\ref{Dbc}) and (\ref{Kbc}) in Eq.~(\ref{dxt}), we obtain $\Delta x^i_t(0)=\Delta x^i_t(\hat{L})=0$, and therefore, the boundary conditions for $\Delta x^i_0$ are the same as for $\Delta x^i$, namely
\begin{equation}
\Delta x_0^i(0)=0, \ \Delta x_0^i(\hat{L})=\Delta x^i_F.\label{bcDx0_owtt}
\end{equation}
Integrating terms containing $\D \Delta x_0^i/\D\hat{l}$ in $\Delta\sigma$ by parts with use of Eq.~(\ref{bcDx0_owtt}), using Eqs.~(\ref{eqmX0}) and (\ref{EoMdx_SBC}) to express the second derivatives and using $v_{Ri}=\epsilon_{ijk}\omega^j x^k$, we obtain the following expression including all terms on the order of $\mathcal{P}(2)$ except the $p_T=2$ order terms containing~$\Delta x^i_F$,
\begin{eqnarray}
\int_0^{\hat{L}}\D \hat{l}\Delta\sigma(\hat{l})&=&\frac{1}{2}\int_0^{\hat{L}}\!\D \hat{l}\ \delta_{ij}S^i_{\!(1)}(\hat{l})\Delta x_0^j(\hat{l})\nonumber\\
&+&\delta_{ij}\frac{\D\hat{x}^i}{\D\hat{l}}|_{\hat{L}}\Delta x_F^j , \label{intDsigma1}
\end{eqnarray}
where $S^i_{\!(1)}$ is the first-order $(p_T=1)$ approximation of $S^i$, which is given as
\begin{equation}
S^i_{\!(1)}(\hat{l})=\partial_j \alpha(\hat{x}^\beta(\hat{l}))\left(\delta^{ij}\!-\!\frac{\D\hat{x}^i(\hat{l})}{\D\hat{l}}\frac{\D\hat{x}^j(\hat{l})}{\D\hat{l}}\right)
+\frac{2}{c}\epsilon^i_{\ jk}\omega^k\frac{\D \hat{x}^j(\hat{l})}{\D \hat{l}}.\label{S1}
\end{equation}
The $\Delta x_0^j$ in Eq.~(\ref{intDsigma1}) can be expressed using Eq.~(\ref{dx0}). In this expansion, only the first term with $S_0^k$ approximated by $S^k_{\!(1)}$ is relevant for the required accuracy. Thus, we obtain 
\begin{equation}
\Delta x^j_0\approx b^j+\mathcal{D}^j_k S^k_{\!(1)},\label{Dx0approx}
\end{equation}
where $b^j=\Delta x^j_F{\hat{l}}/{\hat{L}}$ does not contribute to the considered order in Eq.~(\ref{intDsigma1}). Using the operator definition (\ref{Op_D}), we obtain the following result including all terms on the order of $\mathcal{P}(2)$ except the $p_T=2$ order terms containing $\Delta x^i_F$,
\begin{eqnarray}
\int_0^{\hat{L}}\D \hat{l}\Delta\sigma(\hat{l})&=&
\frac{1}{2}\int_0^{\hat{L}}\!\!\D\hat{l}\int_0^{\hat{L}}\!\!\D\hat{l}^{\prime}D(\hat{l},\hat{l}^{\prime})\delta_{ij}S^i_{\!(1)}(\hat{l})S^j_{\!(1)}(\hat{l}^{\prime})\nonumber\\
&+&\delta_{ij}\frac{\D\hat{x}^i}{\D\hat{l}}|_{\hat{L}}\Delta x_F^j.\label{intDsigma2}
\end{eqnarray}
In this formula, with $S^i_{\!(1)}$ given by Eq.~(\ref{S1}), the approximation (\ref{tan_ap}) is sufficient to obtain the required accuracy for the resulting time-transfer formulas. In this case, using Eq.~(\ref{defDij}), some of the integrations can be performed explicitly. In the last term of Eq.~(\ref{hatsigma}), the approximation (\ref{tan_ap}) is also sufficient. Therefore, using Eq.~(\ref{dt1}) with Eqs.~(\ref{hatsigma}) and (\ref{intDsigma2}), we obtain the following formula for the propagation time, including all terms on the order of $\mathcal{P}(3)$ except the $p_T=3$ order terms containing $\Delta x^i_F$:
\begin{eqnarray}
t_F-t_I\!\!\!&=&\!\!\!\frac{1}{c}\int_0^{\hat{L}}\!\!\!\D\hat{l}\ n +\frac{1}{c^3}\int_0^{\hat{L}}\!\!\!\D\hat{l}\ 2W \nonumber\\
&+&\!\!\!\frac{1}{c^2}\int_0^{\hat{L}}\!\!\D\hat{l}\left(v_{Ri}+{{A}}_i\right)\frac{\D\hat{x}^i}{\D\hat{l}}\nonumber\\
&+&\!\!\!\frac{1}{2c^3}\int_0^{\hat{L}}\!\!\D\hat{l} \left(v_R^2+(v_{Ri}\chi^i)^2\right)\nonumber\\
&-&\!\!\!\frac{1}{6}\frac{\hat{L}^3}{c^3}(\omega^2-(\omega_i\chi^i)^2)\nonumber\\
&+&\!\!\!\frac{1}{c^2}\epsilon^i_{\ jk}\chi^j\omega^k \int_0^{\hat{L}}\!\!\D\hat{l}\ \hat{l}(\hat{l}-\hat{L})\partial_i\alpha\nonumber\\
&+&\!\!\!\frac{1}{c^2}\int_0^{\hat{L}}\!\D\hat{l}\ (\hat{l}-\hat{l}_0) \partial_t \alpha\nonumber\\
&+&\!\!\!\frac{1}{2c}(\delta^{ij}\!-\!\chi^i\chi^j)\!
\int_0^{\hat{L}}\!\!\!\!\D\hat{l}\!\int_0^{\hat{L}}\!\!\!\!\D\hat{l}^{\prime}D(\hat{l},\hat{l}^{\prime})\ \partial_i\alpha(\hat{x}^\beta(\hat{l}))\ \partial_j\alpha(\hat{x}^\beta(\hat{l}^\prime))\nonumber\\
&+&\!\!\!\frac{1}{c}\chi_i\Delta x_F^i , \label{dt2}
\end{eqnarray}
where the fields $n, \alpha, W, v_{Ri},\text{ and } {{A}}_i$ and their partial derivatives with respect to the spacetime coordinates are evaluated in $\hat{x}^\alpha(\hat{l})$ if not stated explicitly otherwise.
Selecting $\Delta x^i_F=0$ and $\hat{l}_0=0$, which corresponds to $\hat{t}=t_I$, approximating the terms in the third line by integration over a straight line connecting $\hat{x}^i_I$ and $\hat{x}^i_F$ instead of $\hat{x}^i(\hat{l})$, which leads to a negligible difference in terrestrial conditions, and approximating $\hat{L}=D$ in the fourth line (see Eq.~(\ref{L_ap})), we obtain Eq.~(\ref{dt2_su}) in the summary.

\subsubsection{Definitions in two-way transfer } \label{sec_TWT_def}

In two-way time transfer, and later also in two-way frequency transfer, we consider a signal that is emitted by a stationary observer on the ground $A$ from the position $x^i_{A1}$ at the coordinate time $t_{A1}$ to an observer $B$, who receives the signal in the position $x^i_B$ at a coordinate time $t_B$ and immediately sends it back to $A$, who receives it in the position $x^i_{A2}$ at the coordinate time $t_{A2}$. We denote the quantities related to the signal from $A1$ to $B$ by the index +, whereas the quantities related to the signal from $B$ to $A2$ are denoted by the index -. Using this notation, we can write $t_{I+}=t_{A1}, x^i_{I+}=x^i_{A1}, t_{F+}=t_{I-}=t_B, x^i_{F+}=x^i_{I-}=x^i_B, t_{F-}=t_{A2},\text{ and } x^i_{F-}=x^i_{A2}$. Next, we define $\hat{x}^i_{+}(\hat{l})$ the solution of Eq.~(\ref{eqmX0}) with the boundary conditions $\hat{x}^i_{+}(0)=x^i_{A1}, \hat{x}^i_{+}(\hat{L})=x^i_B$, and $\hat{x}^i_{-}(\hat{l})$ the solution of Eq.~(\ref{eqmX0}) with the boundary conditions $\hat{x}^i_{-}(0)=x^i_{B}, \hat{x}^i_{-}(\hat{L})=x^i_{A1}$. It can be shown that
\begin{equation}
\hat{x}^i_{+}(\hat{l})=\hat{x}^i_{-}(\hat{L}-\hat{l}).\label{pm_trans}
\end{equation}
From the boundary conditions for $\hat{x}^i_{+}(\hat{l})$ and $\hat{x}^i_{-}(\hat{l})$ described above, it follows that $\Delta x^i_{I+}=\Delta x^i_{F+}=\Delta x^i_{I-}=0$ and $\Delta x^i_{F-}=x^i_{A2}-x^i_{A1}$. The shift in the position of $A$ during the back and forth propagation of the signal we denote $x^i_{A2}-x^i_{A1}\coloneqq \Delta x^i_A$.
For $\hat{t}$ we choose 
\begin{equation}
\hat{t}_{+}=\hat{t}_{-}=t_B,\label{hatt_val}
\end{equation}
which leads to $\hat{l}_{0+}=\hat{L}$ and $\hat{l}_{0-}=0$.

\subsubsection{Two-way time transfer} \label{sec_TWTT}

Now we can express the time $\Delta t_+\coloneqq t_{F+}-t_{I+}$ by setting $\hat{x}^\alpha(\hat{l})=\hat{x}_{+}^\alpha(\hat{l}), \hat{l}_0=\hat{L}$ and $\Delta x^i_F=0$ in Eq.~(\ref{dt2}) and the time $\Delta t_-\coloneqq t_{F-}-t_{I-}$  by setting $\hat{x}^\alpha(\hat{l})=\hat{x}_{-}^\alpha(\hat{l}), \hat{l}_0=0$ and $\Delta x^i_F=\Delta x^i_A$ in Eq.~(\ref{dt2}), respectively.

Denoting by  $x^i_A(t)$ the trajectory of the observer $A$ in the corotating coordinates parameterized by the coordinate time, expanding the trajectory as a Taylor series at $t_{A1}$, denoting by $v^i_{A1}=\D x^i_A/\D t|_{t_{A1}}$ and expressing $t_{A2}-t_{A1}=\Delta t_+ + \Delta t_-$ using Eq.~(\ref{dt2}), we obtain $\Delta x^i_A$ up to the highest order,
\begin{eqnarray}
\Delta x^i_A&=&x^i_{A}(t_{A2})-x^i_{A}(t_{A1})\nonumber\\
&=&v^i_{A1}\frac{2\hat{L}}{c}+O(2).\label{dxA}
\end{eqnarray}
Transforming the integration variables in $\Delta t_-$ as $\hat{l}=\hat{L}-s$, $\hat{l}^\prime =\hat{L}-s^\prime$, using
\begin{eqnarray}
\hat{x}^\alpha_-(\hat{L}-s)&=&\hat{x}^\alpha_+(s), \label{trans_xhat}\\
\frac{\D\hat{x}^i_{-}}{\D\hat{l}}(\hat{L}-s)&=&-\frac{\D\hat{x}^i_{+}(s)}{\D s}, \label{trans_dxhat}
\end{eqnarray}
which follows from Eqs.~(\ref{pm_trans}) and (\ref{hatt_val}) and using the properties (\ref{Dsymm}) and (\ref{Dback}) of the function $D(\hat{l},\hat{l}^\prime)$ we arrive at the following result for the two-way time transfer correction including all terms on the order of $\mathcal{P}(3)$ except the $p_T=3$ order terms originating from $\Delta x^i_A$:
\begin{eqnarray}
\Delta t_{-}-\Delta t_{+}&=&-\frac{2}{c^2}\int_0^{\hat{L}}\!\!\D\hat{l}\ (v_{Ri+}+{{A}}_{i+})\frac{\D \hat{x}^i_+}{\D\hat{l}}\nonumber\\
&+&\frac{2}{c^2}\epsilon^i_{\ jk}\chi^j_+\omega^k\int_0^{\hat{L}}\!\!\D\hat{l}\ \hat{l}(\hat{L}-\hat{l})\partial_i\alpha_{+}\nonumber\\
&+&\frac{2}{c^2}\int_0^{\hat{L}}\!\!\D\hat{l}(\hat{L}-\hat{l})\partial_t \alpha_{+}\nonumber\\
&-&\frac{2}{c^2}\hat{L}\chi_{i+}v^i_{A1},\label{TWTT_th}
\end{eqnarray}
where the fields with the index + are evaluated in $\hat{x}^\alpha_+(\hat{l})$, for example, $v_{Ri+}=v_{Ri}(\hat{x}^\alpha_+(\hat{l}))$, and $\chi^i_+$ is a unit vector in the direction from $x^i_{A1}$ to $x^i_B$. This is Eq.~(\ref{TWTT}) in the summary.

\subsection{Frequency transfer}

\subsubsection{One-way frequency transfer}

In this section, we proceed with the calculation of the frequency shift of the signal that propagates in the atmosphere of Earth according to Eq.~(\ref{fratioX}).

An observer moving with the velocity ${v}^i=\D {x}^i/\D t$ in the corotating coordinates has the four-velocity components (in the same coordinates) given by Eq.~(\ref{udef}) with 
\begin{equation}
u^0=1+\frac{W}{c^2}+\frac{1}{2c^2}|{v}^i_R+{v^i}|^2+O(c^{-4}).\label{u0}
\end{equation}
Denoting by $W_I\!=\!W(x^\alpha_I)$, $W_F\!=\!W(x^\alpha_F)$, $v_{RI}^i\!=\!v_R^i(x^\alpha_I)$, and $v_{RF}^i\!=\!v_R^i(x^\alpha_F)$, with $x^\alpha_I, x^\alpha_F$ being the spacetime points of the signal emission and reception, respectively, and denoting by $v_I^i, v_F^i$ the observers' velocities at the emission and reception events, we can express the first part of Eq.~(\ref{fratioX}) as
\begin{equation}
\frac{u^0_I}{u^0_F}=\frac{1+\frac{W_I}{c^2}+\frac{1}{2c^2}|{v}^i_{RI}+{v^i_I}|^2}{1+\frac{W_F}{c^2}+\frac{1}{2c^2}|{v}^i_{RF}+{v^i_F}|^2}+O(c^{-4}).
\label{fratio_rs}
\end{equation}

Next, we proceed with the Doppler part of Eq.~(\ref{fratioX}). The components $\bar{l}_i$ in the corotating coordinates can be obtained using Eqs.~(\ref{Hamx_inv}), (\ref{dledla}), and  (\ref{mag_l}). Thus, we obtain the following expression including all terms on the order of $\mathcal{P}(2)$:
\begin{eqnarray}
\bar{l}_i&=&\frac{\D x^i}{\D l_E}\left(n+\frac{2W}{c^2}+\frac{v_R^2}{2c^2}\right)+\frac{1}{c}(v^i_R+{{A}}^i) \nonumber\\
&+&\frac{1}{c^2}v_{Rj}v_{Rk}\frac{\D x^k}{\D l_E}\left(\delta^{ij}-\frac{1}{2}\frac{\D x^i}{\D l_E}\frac{\D x^j}{\D l_E}\right).\label{barl}
\end{eqnarray}

The normalized tangent to the light ray $\D x^i/\D l_E$ can be expressed using Eqs.~(\ref{defDx}) and (\ref{LtoLhat}) up to the second order in $\D\Delta x^i/\D\hat{l}$ as 
\begin{eqnarray}
\frac{\D x^i}{\D l_E}&=&\frac{\D \hat{x}^i}{\D\hat{l}}+\frac{\D \Delta x^j}{\D\hat{l}}\!\left(\delta^i_j\!-\!\delta_{jk}\frac{\D\hat{x}^i}{\D\hat{l}} \frac{\D\hat{x}^k}{\D\hat{l}}\right)\nonumber\\
&-&\frac{1}{2}\frac{\D\hat{x}^i}{\D\hat{l}}\delta_{jk}\!\frac{\D\Delta x^j}{\D\hat{l}}\!\frac{\D\Delta x^k}{\D\hat{l}}\nonumber\\
&+&\frac{3}{2}\frac{\D\hat{x}^i}{\D\hat{l}}\!\left(\!\delta_{jk}\frac{\D\hat{x}^j}{\D\hat{l}}\frac{\D\Delta x^k}{\D\hat{l}}\right)^{\!\!2}\!-\!\frac{\D\Delta{x}^i}{\D\hat{l}}\delta_{jk}\!\frac{\D\hat{x}^j}{\D\hat{l}}\!\frac{\D\Delta x^k}{\D\hat{l}} ,
\end{eqnarray}
where, at the end, the terms in the third line do not contribute to the required order of $\bar{l}_i$. 
In order to obtain the boundary values $(\bar{l}_i)_I$ and  $(\bar{l}_i)_F$, we need to express the boundary values of $\D\Delta x^i/\D\hat{l}$. We introduce the parameter $\eta =0$ or $1$, and we evaluate Eq.~(\ref{ddx/dl}) in $\eta\hat{L}$. We assume $\Delta x^i_I=0$, and we keep nonvanishing $\Delta x^i_F$ as a preparation for the two-way frequency transfer, similarly as in the case of the two-way time transfer. Denoting
\begin{equation}
Q^i_j=B^i_j-\frac{\D}{\D\hat{l}}C^i_j ,\label{def_Q}
\end{equation}
we obtain 
\begin{eqnarray}
\frac{\D \Delta x^i}{\D\hat{l}}|_{\eta\hat{L}}&=&\frac{1}{\hat{L}}\Delta x^i_F+\frac{1}{\hat{L}}\int^{\hat{L}}_0\!\!\!\D \hat{l}\ (\hat{l}-(1\!-\!\eta)\hat{L}) S^i(\hat{l})\nonumber\\
&+&\frac{1}{\hat{L}}\int^{\hat{L}}_0\!\!\!\D \hat{l}\left\{\left((\hat{l}-(1\!-\!\eta)\hat{L})Q^i_j(\hat{l})-C^i_j(\hat{l})\right)\right.\nonumber\\
&&\left.\times\left[\sum_{n=0}^{\infty}(\mathcal{K}^j_k)^n \left(b^k+\mathcal{D}^k_l S^l\right)\right](\hat{l})\right\}\nonumber\\
&+&\eta C^i_j(\hat{L})\Delta x^j_F.\label{ddxdlL}
\end{eqnarray}

We denote by $\bar{l}_{i|\eta}$ the boundary values of $\bar{l}_i$, namely $\bar{l}_{i|0}=(\bar{l}_i)_I$ and $\bar{l}_{i|1}=(\bar{l}_i)_F$. We write the resulting expression for $\bar{l}_{i|\eta}$ as the sum 
\begin{equation}
\bar{l}_{i|\eta}=\bar{l}_{i(c^0)|\eta}+\bar{l}_{i(c^{-1})|\eta}+\bar{l}_{i(c^{-2})|\eta}+\bar{l}_{i(\Delta x_F)|\eta} ,\label{li_split}
\end{equation}
including all terms on the order of $\mathcal{P}(2)$ except terms containing $\Delta x^i_F$, which are given up to the largest order only.
The terms in Eq.~(\ref{li_split}) are sorted according to their power of $c^{-1}$, as indicated by their index, and the term containing $\Delta x^i_F$ is given separately. These terms are given by Eqs.~(\ref{li_0})-(\ref{li_dx}) below,
%
\begin{subequations}
\begin{eqnarray}
\bar{l}_{i(c^0)|\eta}\!\!\!&=&\!\!\!n|_{\eta \hat{L}}\frac{\D \hat{x}^i}{\D\hat{l}}|_{\eta \hat{L}}\nonumber\\
&+&\!\!\!n|_{\eta \hat{L}}\left(\delta^i_j-\delta_{jq}\frac{\D\hat{x}^i}{\D\hat{l}}|_{\eta \hat{L}}\frac{\D\hat{x}^q}{\D\hat{l}}|_{\eta \hat{L}}\right)\nonumber\\
&\times&\!\!\!\frac{1}{\hat{L}}\int_0^{\hat{L}}\!\!\!\D\hat{l}\ (\hat{l}-(1\!-\!\eta)\hat{L})\ \partial_k\alpha\left(\delta^{jk}-\frac{\D\hat{x}^j}{\D\hat{l}}\frac{\D\hat{x}^k}{\D\hat{l}}\right)\nonumber\\
&+&\!\!\!(\delta^{ij}(\delta^{kq}\!-\!\chi^k\chi^q)+\chi^j(\chi^k\delta^{iq}\!-\!\chi^i\delta^{kq}))\nonumber\\
&\times&\!\!\!\frac{1}{\hat{L}}\int_0^{\hat{L}}\!\!\!\!\D\hat{l}\!\int_0^{\hat{L}}\!\!\!\!\D\hat{l}^\prime \left\{ D(\hat{l},\hat{l}^\prime)(\hat{l}-(1\!-\!\eta)\hat{L})\right.\nonumber\\
&&\!\!\!\times\ \left.\partial_j\partial_k(\hat{n}+\alpha)(\hat{x}^\beta(\hat{l}))\ \partial_q\alpha(\hat{x}^\beta(\hat{l}^\prime))\right\}\nonumber\\
&+&\!\!\!(\delta^{ij}\!-\!\chi^i\chi^j)\nonumber\\
&\times&\!\!\!\frac{1}{\hat{L}}\int_0^{\hat{L}}\!\!\!\!\D\hat{l}\!\int_0^{\hat{L}}\!\!\!\!\D\hat{l}^\prime D(\hat{l},\hat{l}^\prime)\ \chi^k \partial_k(\hat{n}+\alpha)(\hat{x}^\beta(\hat{l}))\ \partial_j\alpha(\hat{x}^\beta(\hat{l}^\prime))\nonumber\\
&-&\!\!\!(\delta^{ij}\!-\!\chi^i\chi^j)\frac{1}{\hat{L}}\int_0^{\hat{L}}\!\!\!\D\hat{l}\ (\hat{l}-(1\!-\!\eta)\hat{L})\ \alpha\partial_j\alpha \nonumber\\
&-&\!\!\!\frac{1}{2}\chi^i\left|(\delta^{jk}\!-\!\chi^j\chi^k)\frac{1}{\hat{L}}\int_0^{\hat{L}}\!\!\!\D\hat{l}\ (\hat{l}-(1\!-\!\eta)\hat{L})\ \partial_k\alpha\right|^2 \label{li_0}
\end{eqnarray}
\begin{eqnarray}
\bar{l}_{i(c^{-1})|\eta}\!\!\!&=&\!\!\!\frac{1}{c}(v^i_{R}+{{A}}^i)|_{\eta \hat{L}}\nonumber\\
&+&\!\!\!\frac{1}{c} n|_{\eta \hat{L}}\left(\delta^{ij}-\frac{\D\hat{x}^i}{\D\hat{l}}|_{\eta \hat{L}}\frac{\D\hat{x}^j}{\D\hat{l}}|_{\eta \hat{L}}\right)\nonumber\\
&\times&\!\!\!\frac{1}{\hat{L}}\int_0^{\hat{L}}\!\!\!\D\hat{l}\ (\hat{l}-(1\!-\!\eta)\hat{L})\frac{1}{n}\epsilon_{jkl}\left(2\omega^l+{\rm curl}{{A}}^l\right)\frac{\D\hat{x}^k}{\D\hat{l}}\nonumber\\
&+&\!\!\!\hat{n}|_{\eta \hat{L}}\left(\delta^i_j-\delta_{jq}\frac{\D\hat{x}^i}{\D\hat{l}}|_{\eta \hat{L}}\frac{\D\hat{x}^q}{\D\hat{l}}|_{\eta \hat{L}}\right)\nonumber\\
&\times&\!\!\!\frac{1}{\hat{L}}\int_0^{\hat{L}}\!\!\!\D\hat{l} \left\{\left((\hat{l}-(1\!-\!\eta)\hat{L})Q^j_{k({\scriptstyle\hat{N}}^p)}(\hat{l})-C^j_{k({\scriptstyle\hat{N}}^p)}(\hat{l})\right)\right.\nonumber\\
&&\!\!\!\times\left.\left[\sum_{n=0}^\infty(\mathcal{K}^k_{l({\scriptstyle\hat{N}}^p)})^n\mathcal{D}^l_m S^m_{({\scriptstyle\hat{N}}^p\!c^{-1})}\right](\hat{l})\right\}\nonumber\\
&+&\!\!\!(\delta^{ij}\!-\!\chi^i\chi^j)\frac{1}{c}\frac{1}{\hat{L}}\int_0^{\hat{L}}\!\!\!\D\hat{l}\ (\hat{l}-(1\!-\!\eta)\hat{L})(\hat{l}-\hat{l}_0)\partial_t\partial_j\alpha\nonumber\\
&+&\!\!\!\frac{1}{c}\epsilon_{klm}\omega^m\chi^l(\delta^{ia}\delta^{kb}+\chi^a\chi^b\delta^{ik})\nonumber\\
&\times&\!\!\!\frac{1}{\hat{L}}\int_0^{\hat{L}}\!\!\!\D\hat{l}\ \hat{l}(\hat{l}-\hat{L})(\hat{l}-(1\!-\!\eta)\hat{L})\partial_a\partial_b\alpha\nonumber\\
&+&\!\!\!\frac{1}{c}\epsilon_{mjk}\omega^j(\delta^{im}\delta^{kl}\!\!-2\chi^k(\delta^{im}\chi^l\!\!+\!\delta^{lm}\chi^i))\frac{1}{\hat{L}}\!\int_0^{\hat{L}}\!\!\!\D\hat{l}\ \hat{l}(\hat{l}-\hat{L})\partial_l\alpha \nonumber\\
&+&\!\!\!\frac{1}{c}\chi^i(\eta\hat{L}-\hat{l}_0)\partial_t\alpha|_{\eta \hat{L}} \label{li_1}
\end{eqnarray}
\begin{eqnarray}
\bar{l}_{i(c^{-2})|\eta}\!\!\!&=&\!\!\!\frac{1}{c^2}\chi^i\!\left(2W+\tfrac{1}{2}v_{Rj}v_{Rk}(\delta^{jk}\!-\!\chi^j\chi^k)\right)\!|_{\eta\hat{L}}
+\frac{1}{c^2} (v^i_{R} v_{Rj})|_{\eta \hat{L}}\chi^j\nonumber\\
&+&\!\!\!\frac{\hat{L}^2}{c^2}\left(\tfrac{1}{3}\omega^i\chi^k\omega_k+\tfrac{1}{6}\chi^i(\omega_j\chi^j)^2-\tfrac{1}{2}\chi^i\omega^2\right) \nonumber\\
&+&\!\!\!\frac{1}{c^2}\frac{1}{\hat{L}}\int_0^{\hat{L}}\!\!\!\D\hat{l}\ \left\{(\hat{l}-(1\!-\!\eta)\hat{L})\right. \nonumber\\
&&\!\!\!\times\left.\left((\delta^{ij}\!-\!\chi^i\chi^j)\left(2\partial_j\Delta W+\tfrac{1}{2}\partial_j v_R^2\right)+\partial_i(v_{Rk}v_{Rl})\chi^k\chi^l\right)\right\} \nonumber\\
&&\label{li_c2}\\
\bar{l}_{i(\Delta x_F)|\eta}\!\!\!&=&\!\!\!(\delta^i_j-\chi^i\chi_j)\frac{1}{\hat{L}}\Delta x^j_F\label{li_dx}
\end{eqnarray}
\label{li_components}
\end{subequations}
%
In Eqs.~(\ref{li_components}), all the fields $n,\hat{n},\alpha, v_{Ri}, {{A}}_i, W,\text{ and } \Delta W$ and their partial derivatives with respect to the spacetime coordinates are evaluated in $\hat{x}^\alpha(\hat{l})$, if not stated explicitly otherwise. For example, we denote $\hat{n}=\hat{n}(\hat{x}^\alpha(\hat{l}))$, $\partial_k\hat{n}=(\partial_k\hat{n})(\hat{x}^\alpha(\hat{l}))$. The functions evaluated in $\hat{l}=\eta \hat{L}$ are denoted by $|_{\eta\hat{L}}$, for example, $\hat{n}|_{\eta\hat{L}}=\hat{n}(\hat{x}^\alpha(\eta\hat{L}))$. The functions $Q^i_{j({\scriptstyle\hat{N}}^p)}$ and $C^i_{j({\scriptstyle\hat{N}}^p)}$ in $\bar{l}_{i(c^{-1})|\eta}$ are the functions $Q^i_j$ and $C^i_j$ as given by Eqs.~(\ref{def_Q}), (\ref{def_B}), and (\ref{def_C}), with the level of approximation including all terms with $p_{c^{-1}}=p_\alpha=0$ and arbitrary $p_{\hat{N}}$, namely
\begin{eqnarray}
Q^i_{j({\scriptstyle\hat{N}}^p)}(\hat{l})&=&\left(\frac{1}{\hat{n}}\partial_k\partial_l\hat{n}-\frac{2}{\hat{n}^2}\partial_k\hat{n}\partial_l\hat{n}\right)\nonumber\\
&\times&\left(\delta^{ik}\delta^l_j-2\delta^{ik}\frac{\D\hat{x}^l}{\D\hat{l}}\frac{\D\hat{x}^j}{\D\hat{l}}+\delta^i_j\frac{\D\hat{x}^k}{\D\hat{l}}\frac{\D\hat{x}^l}{\D\hat{l}}\right)\nonumber\\
&+&\frac{1}{\hat{n}^2}\partial_k\hat{n}\partial_l\hat{n}\left(\delta^i_j\delta^{kl}-\delta^l_j\frac{\D\hat{x}^i}{\D\hat{l}}\frac{\D\hat{x}^k}{\D\hat{l}}\right),\label{Qij_N}\\
C^i_{j({\scriptstyle\hat{N}}^p)}(\hat{l})&=&\frac{1}{\hat{n}}\partial_k\hat{n}\left(2\delta^{ik}\frac{\D\hat{x}^j}{\D\hat{l}}-\delta^k_j\frac{\D\hat{x}^i}{\D\hat{l}}-\delta^i_j\frac{\D\hat{x}^k}{\D\hat{l}}\right),\label{Cij_N}
\end{eqnarray}
with $\hat{n}$ and its partial derivatives being evaluated in $\hat{x}^\alpha(\hat{l})$. The operator $\mathcal{K}^i_{j({\scriptstyle\hat{N}}^p)}$ in $\bar{l}_{i(c^{-1})|\eta}$ is then generated by the corresponding function
\begin{equation}
K^i_{j({\scriptstyle\hat{N}}^p)}(\hat{l},\hat{l}^\prime)=D(\hat{l},\hat{l}^\prime)Q^i_{j({\scriptstyle\hat{N}}^p)}\!(\hat{l}^\prime)
+(-\hat{l}/\hat{L}+H(\hat{l}-\hat{l}^\prime))C^i_{j({\scriptstyle\hat{N}}^p)}\!(\hat{l}^\prime),\label{Kij_N}
\end{equation}
and $S^i_{({\scriptstyle\hat{N}}^p\! c^{-1})}$ is $S^i$ as given by Eq.~(\ref{defS}) with the approximation level including all terms with $p_{c^{-1}}=1, p_\alpha=0$ and arbitrary $p_{\hat{N}}$, namely 
\begin{equation}
S^i_{({\scriptstyle\hat{N}}^p\! c^{-1})}(\hat{l})=\frac{1}{\hat{n}c}\epsilon^i_{\ jk}\left(2\omega^k+{\rm curl}\hat{{{A}}}^k\right)\frac{\D\hat{x}^j}{\D\hat{l}},\label{Si_Nc}
\end{equation}
with $\hat{{{A}}}^k=(1-\hat{n}^2)V^k$, and the fields $\hat{n}$ and ${\rm curl}\hat{{{A}}}^k$ evaluated along $\hat{x}^\alpha(\hat{l})$.

Now we proceed with the integral in the exponent in Eq.~(\ref{fratioX}), which we denote $I$. We parameterize the signal trajectory by the Euclidean length $l_E\in[0,L]$, and using Eqs.~(\ref{goptrot}), (\ref{l^0}), (\ref{mag_l}) and (\ref{eqm_ray_t2_lE}), we obtain the following expression including all terms on the order of $\mathcal{P}(3)$:
\begin{eqnarray}
I&\equiv&\int_0^L\frac{1}{2}l^\alpha\partial_0\bar{g}_{\alpha\beta}\frac{\D x^\beta}{\D l_E}\D l_E \nonumber\\
&=&\int_0^L\!\!\D l_E \left(\frac{1}{c}\partial_t n+\frac{1}{c^2}(\partial_t {{A}}_i)\frac{\D x^i}{\D l_E}+\frac{2}{c^3}\partial_t W\right), \label{I}
\end{eqnarray}
where the fields $\partial_t n, \partial_t {{A}}_i,\text{and } \partial_t W$ are evaluated along the signal trajectory $x^\alpha(l_E)$. 

Next, we transform the integration variable to $\hat{l}$ using the first-order expansion of Eq.~(\ref{LtoLhat}) in $\D\Delta x^i/\D\hat{l}$ and using Eqs.~(\ref{defDx}) and (\ref{defDt}). Then, taking the Taylor expansion of the integrand in $\Delta x^\alpha$, assuming $\Delta x_I^i=0$, and using the approximations
\begin{eqnarray}
\Delta x^i &\approx& \Delta x^i_F\frac{\hat{l}}{\hat{L}}+\mathcal{D}^i_j S^j_{(1)},\\
\Delta t &\approx& \frac{1}{c}(\hat{l}-\hat{l}_0),
\end{eqnarray}
we obtain the following formula including all terms on the order of $\mathcal{P}(3)$ except terms containing $\Delta x^i_F$, which are given up to the largest order only:
\begin{subequations}
\begin{eqnarray}
I\!\!\!&=&\!\!\!I_{(c^{-1})}+I_{(c^{-2})}+I_{(c^{-3})}+I_{(\Delta x_F)},\\
I_{(c^{-1})}\!\!\!&=&\!\!\!\frac{1}{c}\int_0^{\hat{L}}\!\!\!\D\hat{l}\ \partial_t n \nonumber\\
&+&\!\!\!\frac{1}{c}(\delta^{ij}\!\!-\!\!\chi^i\chi^j)\!\int_0^{\hat{L}}\!\!\!\!\D\hat{l}\!\int_0^{\hat{L}}\!\!\!\!\D\hat{l}^\prime D(\hat{l},\hat{l}^\prime)\ \partial_i\partial_t\alpha(\hat{x}^\beta(\hat{l}))\ \partial_j\alpha(\hat{x}^\beta(\hat{l}^\prime)),\nonumber\\
&&\\
I_{(c^{-2})}\!\!\!&=&\!\!\!\frac{1}{c^2}\int_0^{\hat{L}}\!\!\!\D\hat{l}\ (\partial_t {{A}}_i)\frac{\D \hat{x}^i}{\D \hat{l}}\nonumber\\
&+&\!\!\!\frac{1}{c^2}\epsilon^i_{\ jk}\chi^j\omega^k\!\int_0^{\hat{L}}\!\!\!\D\hat{l}\ \hat{l}(\hat{l}-\hat{L})\partial_i\partial_t \alpha \nonumber\\
&+&\!\!\!\frac{1}{c^2}\!\int_0^{\hat{L}}\!\!\!\D\hat{l}\ (\hat{l}-\hat{l}_0)\partial_t\partial_t\alpha ,\\
I_{(c^{-3})}\!\!\!&=&\!\!\!\frac{2}{c^3}\int_0^{\hat{L}}\!\!\!\D\hat{l}\ \partial_t W,\\
I_{(\Delta x_F)}\!\!\!&=&\!\!\!\frac{1}{c\hat{L}}\Delta x^i_F\int_0^{\hat{L}}\!\!\!\D\hat{l}\ (\hat{l}\partial_i\partial_t\alpha +\chi_i\partial_t\alpha),
\end{eqnarray}
\label{I2}
\end{subequations}
where the fields $n, \alpha, {{A}}_i,\text{and } W$ and their partial derivatives with respect to the spacetime coordinates are evaluated in $\hat{x}^\alpha(\hat{l})$ if not stated explicitly otherwise. 
Selecting $\Delta x^i_F=0$ and $\hat{l}_0=0$, which corresponds to $\hat{t}=t_I$, we obtain Eq.~(\ref{I2_su}) in the summary.

\subsubsection{Two-way frequency transfer}

In two-way frequency transfer, we consider the experimental setup as already described in Sect.~\ref{sec_TWT_def}, and we admit the definitions introduced there.
Furthermore, we define $\bar{l}_{i+|\eta}$ and $I_+$ by setting $\hat{x}^\alpha(\hat{l})=\hat{x}_{+}^\alpha(\hat{l})$, $\hat{l}_0=\hat{L}$, $\Delta x^i_F=0$ in Eqs.~(\ref{li_split})-(\ref{Si_Nc}) and (\ref{I2}), and we define $\bar{l}_{i-|\eta}$ and $I_-$  by setting $\hat{x}^\alpha(\hat{l})=\hat{x}_{-}^\alpha(\hat{l})$, $\hat{l}_0=0$, $\Delta x^i_F=\Delta x^i_A$ in Eqs.~(\ref{li_split})-(\ref{Si_Nc}) and (\ref{I2}). The proper frequencies $\nu$ and the observer velocity quantities $v^i$, $u^0$ for the particular emission and reception events are denoted by the corresponding index $A1$, $A2$, or $B$. Similarly, fields evaluated in the spacetime points $x^\alpha_{A1}$, $x^\alpha_{B}$, or $x^\alpha_{A2}$ of the emission and reception events are denoted by the corresponding index, for example, $W_{A1}=W(x^\alpha_{A1})$, $v^i_{R|A1}=v^i_R(x^\alpha_{A1})$. 

The goal of this section is to express the correction $\Delta$ defined by Eq.~(\ref{defDelta}). From this definition, it follows that
\begin{equation}
\Delta=\frac{\nu_{A2}}{\nu_B}\left(1-\frac{1}{2}\frac{\nu_B}{\nu_{A1}}\right)-\frac{1}{2}. \label{Delta1}
\end{equation}
Using Eq.~(\ref{fratioX}) for the frequency shifts in the particular directions, we obtain
\begin{eqnarray}
\frac{\nu_{A1}}{\nu_B}&=&\frac{u^0_{A1}}{u^0_B}\ \frac{1-\frac{1}{c}v_{A1}^i \bar{l}_{i+|0}}{1-\frac{1}{c}v_B^i \bar{l}_{i+|1}}\ \exp I_+ ,\label{fratioXAB}\\
\frac{\nu_B}{\nu_{A2}}&=&\frac{u^0_B}{u^0_{A2}}\ \frac{1-\frac{1}{c}v_B^i \bar{l}_{i-|0}}{1-\frac{1}{c}v_{A2}^i \bar{l}_{i-|1}}\ \exp I_- .\label{fratioXBA}
\end{eqnarray}

In order to proceed with the correction (\ref{Delta1}), it is useful to express the relations for the contributions to $\bar{l}_{i|\eta}$ given by Eqs.~(\ref{li_components}) and for the contributions to $I$ given by Eqs.~(\ref{I2}) between the two directions of the light-ray propagation. Transforming the integration variables in the contributions to $\bar{l}_{i-|(1-\eta)}$ and in the contributions to $I_-$ as $\hat{l}=\hat{L}-s$, $\hat{l}^\prime=\hat{L}-s^\prime$, and so on, using Eqs.~(\ref{trans_xhat}) and (\ref{trans_dxhat}), taking into account that $\chi^i_+=-\chi^i_-$, using the properties (\ref{Dsymm}), (\ref{Dback}) and taking into account that $\Delta x^i_{F+}=0$, we obtain the following relations:
\begin{subequations}
\begin{eqnarray}
\bar{l}_{i+(c^0)|\eta}&=&-\bar{l}_{i-(c^0)|(1-\eta)}\\
\bar{l}_{i+(c^{-1})|\eta}&=&\bar{l}_{i-(c^{-1})|(1-\eta)}\\
\bar{l}_{i+(c^{-2})|\eta}&=&-\bar{l}_{i-(c^{-2})|(1-\eta)}\\
\bar{l}_{i+(\Delta x_F)|\eta}&=&0
\end{eqnarray}
\label{li_odd_even}
\end{subequations}
and
\begin{subequations}
\begin{eqnarray}
I_{+(c^{-1})}&=&I_{-(c^{-1})}\\
I_{+(c^{-2})}&=&-I_{-(c^{-2})}\\
I_{+(c^{-3})}&=&I_{-(c^{-3})}\\
I_{+(\Delta x_F)}&=&0.
\end{eqnarray}
\label{I_odd_even}
\end{subequations}

The $u^0$ terms in Eqs.~(\ref{fratioXAB}) and (\ref{fratioXBA}) are expressed using Eq.~(\ref{u0}). Quantities located at event $A2$ (except the potential $W_{A2}$) are expressed in terms of the quantities located at $A1$ using Eq.~(\ref{dxA}) and expressing the velocity difference $\Delta v^i_A=v^i_{A2}-v^i_{A1}$ in terms of the acceleration $a^i_{A1}=\D^2 x^i_{A}/\D t^2|_{t_{A1}}$ of the observer $A$ as
\begin{equation}
\Delta v_A^i=a_{A1}^i\frac{2\hat{L}}{c}+O(2).\label{dvA}
\end{equation}

Then, inserting Eqs.~(\ref{fratioXAB}) and (\ref{fratioXBA}) into Eq.~(\ref{Delta1}), we obtain the following result including all terms on the order of $\mathcal{P}(3)$ except the $p_T\geq~\!\!3$ order terms originating from the motion of the observer on the ground $A$ in the corotating frame and except the terms quadratic in $v^i_{A1}/c$:
\begin{eqnarray}
\Delta\!\!\!&=&\!\!\!\frac{1}{2c^2}\!\left[W_{A1}\!+\!W_{A2}\!-\!2W_B\!+\!|v_{R|A1}^i|^2\!-\!|v_{R|B}^i|^2\!-\!|v^i_{B}|^2\right]
\left(1-\frac{1}{c}v_B^j\chi_{j+}\right)\nonumber\\
&+&\!\!\!\frac{1}{c}v_B^i\beta_i-\frac{1}{c^2}v_B^i\beta_{i(1)} v^j_B\chi_{j+}\nonumber\\
&+&\!\!\!\frac{1}{c^2}D\chi_{i+}(a_{A1}^i+\epsilon^i_{\ jk}\omega^j v_{A1}^k)+\frac{1}{c^2}v_{Bi}v_{A1}^i\nonumber\\
&+&\!\!\!I_{+(c^{-2})}-\frac{1}{c^2}v_B^i\chi_{i+}\int_0^{\hat{L}}\!\!\!\D\hat{l}\ \partial_t n(\hat{x}^\alpha_+(\hat{l})),\label{Delta2}
\end{eqnarray}
where $\beta_i=\bar{l}_{i+(c^{-1})|1}-\tfrac{1}{c}v^i_{R|B}$ is given by Eq.~(\ref{beta}) below, and $\beta_{i(1)}$ is its first-order $(p_T=1)$ part given as 
\begin{equation}
\beta_{i(1)}=-\frac{1}{c}\epsilon_{ijk}\omega^j(x_B^k-x_{A1}^k)=-\frac{1}{c}(v^i_{R|B}-v^i_{R|A1}), \label{beta(1)}
\end{equation}
and $I_{+(c^{-2})}$ is given by Eq.~(\ref{Ic2}).

\begin{eqnarray}
\beta_i\!\!\!&=&\!\!\!\frac{1}{c}{{A}}^i_B \nonumber\\
&+&\!\!\!\frac{1}{c}{n}_B \left(\delta^{ij}\!-\!\frac{\D\hat{x}^i_+}{\D\hat{l}}|_{\hat{L}}\frac{\D\hat{x}^j_+}{\D\hat{l}}|_{\hat{L}}\right)\frac{1}{\hat{L}}\!\int_0^{\hat{L}}\!\!\!\D\hat{l}\ \hat{l}\frac{1}{{n}_+}\epsilon_{jkl}\left(2\omega^l\!+{\rm curl}{{A}}^l_+\right)\frac{\D\hat{x}^k_+}{\D\hat{l}}\nonumber\\
&+&\!\!\! \hat{n}_B \left(\delta^i_j\!-\delta_{jq}\frac{\D\hat{x}^i_+}{\D\hat{l}}|_{\hat{L}}\frac{\D\hat{x}^q_+}{\D\hat{l}}|_{\hat{L}}\right)\frac{1}{\hat{L}}\!\int_0^{\hat{L}}\!\!\!\D\hat{l} \left\{\left(\hat{l}Q^j_{k+({\scriptstyle\hat{N}}^p)}(\hat{l})-C^j_{k+({\scriptstyle\hat{N}}^p)}(\hat{l})\right)\right. \nonumber\\
&&\!\!\!\times\left.\left[\sum_{n=0}^\infty(\mathcal{K}^k_{l+({\scriptstyle\hat{N}}^p)})^n\mathcal{D}^l_m S^m_{+({\scriptstyle\hat{N}}^p\!c^{-1})}\right](\hat{l})\right\}\nonumber\\
&+&\!\!\!\frac{1}{c} (\delta^{ij}\!-\!\chi^i_+\chi^j_+)\frac{1}{\hat{L}}\int_0^{\hat{L}}\!\!\!\D\hat{l}\ \hat{l}(\hat{l}-\hat{L})\partial_t\partial_j\alpha_+\nonumber\\
&+&\!\!\!\frac{1}{c}\ \epsilon_{klm}\omega^m\chi^l_+(\delta^{ia}\delta^{kb}+\chi^a_+\chi^b_+\delta^{ik})\frac{1}{\hat{L}}\int_0^{\hat{L}}\!\!\!\D\hat{l}\ \hat{l}^2(\hat{l}-\hat{L})\partial_a\partial_b\alpha_+\nonumber\\
&+&\!\!\!\frac{1}{c}\ \epsilon_{mjk}\omega^j(\delta^{im}\delta^{kl}\!-2\chi^k_+(\delta^{im}\chi^l_+ +\delta^{lm}\chi^i_+))\frac{1}{\hat{L}}\!\int_0^{\hat{L}}\!\!\!\D\hat{l}\ \hat{l}(\hat{l}-\hat{L})\partial_l\alpha_+ \nonumber\\
&&\label{beta}
\end{eqnarray}
\begin{eqnarray}
I_{+(c^{-2})}\!\!\!&=&\!\!\!\frac{1}{c^2}\int_0^{\hat{L}}\!\!\!\D\hat{l}\ (\partial_t {{A}}_{i+})\frac{\D \hat{x}^i_+}{\D \hat{l}} \nonumber\\
&+&\!\!\!\frac{1}{c^2}\epsilon^i_{\ jk}\chi^j_+\omega^k\int_0^{\hat{L}}\!\!\D\hat{l}\ \hat{l}(\hat{l}-\hat{L})\partial_i\partial_t \alpha_+ \nonumber\\
&+&\!\!\!\frac{1}{c^2}\int_0^{\hat{L}}\!\!\D\hat{l}\ (\hat{l}-\hat{L})\partial_t\partial_t\alpha_+  \label{Ic2}
\end{eqnarray}
%
In Eqs.~(\ref{beta}) and (\ref{Ic2}), the index $+$ for fields denotes that the field is evaluated along the $\hat{x}^\alpha_{+}(\hat{l})$ path, for example, $\alpha_+=\alpha(\hat{x}^\alpha_+(\hat{l}))$, $\partial_k\alpha_+=(\partial_k\alpha)(\hat{x}^\alpha_+(\hat{l}))$. The functions $Q^j_{k+({\scriptstyle\hat{N}}^p)}$, $C^j_{k+({\scriptstyle\hat{N}}^p)}$, $S^m_{+({\scriptstyle\hat{N}}^p\!c^{-1})}$ and the generating function ${K}^k_{l+({\scriptstyle\hat{N}}^p)}$ for the operator $\mathcal{K}^k_{l+({\scriptstyle\hat{N}}^p)}$ in the third and fourth line of Eq.~(\ref{beta}) are given by Eqs.~(\ref{Qij_N})-(\ref{Si_Nc}) with $\hat{x}^\alpha(\hat{l})=\hat{x}^\alpha_+(\hat{l})$.

To quantify the effects in the examples in this paper and for most of the satellite applications, it is practical to express Eq.~(\ref{beta}) explicitly up to the second order $(p_T\leq2)$, leading to the contribution up to the third order $(p_T\leq 3)$ in $\Delta$. Moreover, the terms in Eq.~(\ref{beta}) that are proportional to $c^{-1}N_B$ are negligible for satellite applications because the atmospheric refractivity $N_B$ in the satellite position is very low for satellite altitudes of several hundred kilometers. Thus, up to the second order and neglecting the terms proportional to $c^{-1}N_B$, we can write
\begin{equation} 
\beta_i=\beta_{i(1)}+\beta_{i(2)}, \label{beta(12)}
\end{equation}
with $\beta_{i(1)}$ given by Eq.~(\ref{beta(1)}) and $\beta_{i(2)}$ given by Eq.~(\ref{beta(2)}). Applying Eq.~(\ref{beta(12)}) in Eq.~(\ref{Delta2}) leads to Eq.~(\ref{Delta2_su}) in the summary section.

Some steps in the calculation of $\beta_i$ up to the second order are described below. For the end-point tangent, we use Eq.~(\ref{tanL_eF}) expanded up to the first order, 
\begin{equation}
\frac{\D\hat{x}^i_+}{\D\hat{l}}|_{\hat{L}}=\chi^i_+ + \varsigma^i_+ \varepsilon_B +O(2),
\end{equation}
where the index $B$ replaces the general index for a final point $F$. Where possible, the approximation (\ref{tan_ap}) is used. 
After expanding $1/n_+=1-\hat{N}_+ -\alpha_+ + O(2)$ in the integral term in the second line of Eq.~(\ref{beta}), we use the following way of integrating the term containing $\omega^l$:
\begin{eqnarray}
\int^{\hat{L}}_0 \!\!\! \D\hat{l}\ \hat{l}\ \frac{\D\hat{x}^i_+}{\D\hat{l}}\!\!\!&=&\!\!\!
\frac{1}{2}\int^{\hat{L}}_0 \!\!\! \D\hat{l}\left[\frac{\D}{\D\hat{l}}\!\left(\hat{l}(\hat{l}-\hat{L})\frac{\D\hat{x}^i_+}{\D\hat{l}}\right)\!+\!\hat{L}\frac{\D\hat{x}^i_+}{\D\hat{l}}\!-\!\hat{l}(\hat{l}-\hat{L})\frac{\D^2\hat{x}^i_+}{\D\hat{l}^2}\right]\nonumber\\
&=&\!\!\!\frac{\hat{L}}{2}(x_B^i\!-\!x_{A1}^i)\nonumber\\
&-&\!\!\!\frac{1}{2}\!\int^{\hat{L}}_0 \!\!\!\! \D\hat{l}\ \hat{l}(\hat{l}-\hat{L})(\delta^{ij}\!-\!\chi^i_+\chi^j_+)\partial_j\hat{n}_+  +O(2),
\end{eqnarray}
where the second derivative of the path $\hat{x}^i_+(\hat{l})$ was expressed using Eq.~(\ref{eqmX0}) expanded up to the $p_T=1$ order.

Expanding the term in the third and fourth line of Eq.~(\ref{beta}) with the use of Eqs.~(\ref{Qij_N})-(\ref{Si_Nc}) and (\ref{Op_D}), one of the terms coming from $Q^j_{k+({\scriptstyle\hat{N}}^p)}$ can be processed as follows:
\begin{eqnarray}
\hspace{-7mm}&&\int^{\hat{L}}_0 \!\!\! \D\hat{l}\ \hat{l}^2(\hat{l}-\hat{L})\chi^i_+\chi^j_+\partial_i\partial_j\hat{n}_+ =\nonumber\\
\hspace{-7mm}&&=\int^{\hat{L}}_0 \!\!\! \D\hat{l}\ \hat{l}^2(\hat{l}-\hat{L})\frac{\D}{\D\hat{l}}(\chi^j_+\partial_j\hat{n}_+) + O(2) \nonumber\\
\hspace{-7mm}&&=-\int^{\hat{L}}_0 \!\!\! \D\hat{l}\ (2\hat{l}(\hat{l}-\hat{L})+\hat{l}^2)\chi^j_+\partial_j\hat{n}_+ + O(2) \nonumber\\
\hspace{-7mm}&&=-2\!\int^{\hat{L}}_0 \!\!\! \D\hat{l}\ \hat{l}(\hat{l}-\hat{L})\chi^j_+\partial_j\hat{n}_+ -\int_0^{\hat{L}}\!\!\!\D\hat{l}\ \hat{l}^2\frac{\D}{\D\hat{l}}\hat{N}_+ + O(2)\nonumber\\
\hspace{-7mm}&&=-2\!\int^{\hat{L}}_0 \!\!\! \D\hat{l}\ \hat{l}(\hat{l}-\hat{L})\chi^j_+\partial_j\hat{n}_+ +2\!\int_0^{\hat{L}}\!\!\!\D\hat{l}\ \hat{l}\hat{N}_+ \!-\! \hat{L}^2\hat{N}_B + O(2),
\end{eqnarray}
where in the second and fourth line, the approximation (\ref{tan_ap}) was used and the terms were integrated by parts. The same procedure can be applied to the analogous term in Eq.~(\ref{beta}) with $\alpha_+$ instead of $\hat{n}_+$. The use of these relations leads to the cancellation of several terms in the expression for $\beta_{i(2)}$.

\section{Conclusion}

A relativistic model of one-way and two-way time and frequency transfer through flowing media was developed in this paper. The main focus was on applications to the atmosphere of Earth. The model includes gravitational and atmospheric effects given by the fields of the scalar gravitational potential $W(x^\alpha)$, the atmospheric refractive index $n(x^\alpha)$, and the wind speed $V^i(x^\alpha)$. The nonstationarity of the fields and deviations from spherical symmetry are also included in the model.

The method used in this paper is based on solving the equation of motion for a light ray in coordinates $x^\alpha=(ct, x^i)$ corotating with the Earth. This equation is given as the null geodesic equation of the Gordon optical metric \citep{Gordon}. First, the exact solution $\hat{x}^i(\hat{l})$ in a spherically symmetric, static part of $n(x^\alpha)$ and $W(x^\alpha)$ is found, which is parameterized by its Euclidean length in the corotating frame $\hat{l}$. This solution is defined by the spatial coordinates of the emission and the reception events as boundary values. Then, a complete solution $x^i(\hat{l}), t(\hat{l})$ is found using a perturbation method that takes the inertial forces, wind speed, and deviations of $n(x^\alpha)$ and $W(x^\alpha)$ from sphericity and staticity into account. The time- and frequency-transfer corrections are then expressed in terms of integrals of the fields along the path $\hat{x}^i(\hat{l})$ at a certain constant coordinate time.

In the numerical examples evaluated in this paper, we focused on the two-way ground-to-satellite transfer with a satellite altitude similar to that of the ISS. In the two-way time transfer, the main contribution of the atmosphere appears in the Sagnac effect and is given by a change in the Sagnac area that is spanned by the light path $\hat{x}^i(\hat{l})$ when the path changes from the vacuum case to the case influenced by the atmospheric refraction. In the example studied in this paper, the contribution of the atmosphere to the Sagnac effect is below 0.1~ps for the satellite zenith angle $\theta$ (the angle between the vertical line from a ground station and the line connecting the ground station with the satellite position at the reception/re-emission time) below $78^\circ$, it reaches 1~ps for $\theta\approx 87^\circ$, and it increases to approximately 5~ps as $\theta$ approaches $90^\circ$. Therefore, for time transfer at an accuracy level of 1~ps, this effect is only significant for large zenith angles.

The effect of the wind in the two-way time transfer is given by the difference in the propagation times caused by the Fresnel-Fizeau effect of light dragging. It is much smaller than 1~ps for normal atmospheric conditions.

In the two-way frequency transfer, the correction $\Delta$ of \citet{Ashby} and \citet{Blanchet} is generalized to include the atmospheric effects. In the example presented in this paper, the effect of the spherically symmetric, static part of the refractive index gives a contribution that starts at $10^{-17}$ for $\theta=0$, reaches $10^{-16}$ for $\theta\approx 56^{\circ}$, and ends at $10^{-13}$ for $\theta$ approaching $90^\circ$.

Remarkably, in the equation of motion for a light ray, the quantity ${{A}}^i=(1-n^2)V^i$ related to the wind speed acts similarly as the magnetic potential in the equation of motion for a charged particle. In particular, the opposite direction of the propagation leads to opposite acceleration. In the two-way frequency transfer, it may lead to a significant contribution to the correction $\Delta$. For the example of a constant horizontal wind field, the effect reaches $10^{-17}$ already for a wind speed of about 11~m/s (the wind speed in the corotating frame is meant).

The effects of the deviation of $n(x^\alpha)$ from sphericity and staticity are also included in the resulting corrections. They are not evaluated numerically in this paper, however.

The evaluation of the atmospheric effects in the time- and frequency-transfer corrections including the effect of the wind shows that they need to be considered in the data analysis of the forthcoming clock-on-satellite experiments such as ACES or I-SOC.
The results obtained in this paper can be used not only to calculate the time- and frequency-transfer corrections from given atmospheric data, but also inversely to determine atmospheric properties such as the refractive index profile or the flow rate from the time or frequency measurements. This will be a subject of a future research.

\begin{acknowledgements}

This research was funded by grants of Institutional Financing IF2105021501, IF2205021501, and IF2309021501 provided by Ministry of Industry and Trade of the Czech Republic to the Czech Metrology Institute. The author is also thankful to Adrien Bourgoin and Pac\^{o}me Delva for their valuable comments about the paper manuscript.

\end{acknowledgements}

\begin{appendix}
\section{Refractive index profiles}

\subsection{Refractive index as a function of the density of air} \label{appendix1}

According to Eq. (4) of \citet{Owens} the refractive index of air~$n$ depends on the densities of its components as
\begin{equation}
\frac{n^2-1}{n^2+2}=R_1\rho_1+R_2\rho_2+R_3\rho_3 , \label{n_Owens}
\end{equation}
where $\rho_1$, $\rho_2$, and $\rho_3$ are the partial densities of dry, ${\rm CO}_2$-free air, of water vapor, and of carbon dioxide in the mixture, and $R_1$, $R_2$, and $R_3$ are the corresponding specific refractions of the components of the mixture that only depend on wavelength. Therefore, to obtain the field $n(x^\alpha)$ in spacetime, the fields $\rho_1(x^\alpha)$, $\rho_2(x^\alpha)$, and $\rho_3(x^\alpha)$ need to be known.

An approximate formula for the refractivity $N\equiv n~-~1$ can be derived from Eq.~(\ref{n_Owens}). This formula is given, for example, by Eq. (5) of \citet{Ciddor} which states 
\begin{equation}
N=\frac{\rho_a}{\rho_{axs}}N_{axs}+\frac{\rho_w}{\rho_{ws}}N_{ws} , \label{n_Ciddor}
\end{equation}
where $\rho_{axs}$ is the density of pure dry air at 15~$^\circ$C and 101~325~Pa with certain molar fraction $x_c$ of ${\rm CO}_2$, $\rho_{ws}$~is the density of pure water vapor at 20~$^\circ$C and 1333~Pa and $N_{axs}$, and $N_{ws}$ are the refractivities of the dry air and water vapor at the conditions corresponding to $\rho_{axs}$, $\rho_{ws}$. The refractivities as functions of the vacuum wavelength are known for these specific conditions, and they are given by \citet{Ciddor}. Similarly, $\rho_a$ and $\rho_w$ are the densities of the dry air component and of the water vapor component of moist air for the actual conditions. The densities $\rho_{a}$ and $\rho_{w}$ can be derived from the equation of state (Eq. (4) of \citet{Ciddor}) as functions of the temperature $T$, the pressure $p$, and the air composition, represented, for example, by the molar fractions $x_c$ of ${\rm CO}_2$ and $x_w$ of the water vapor. Therefore, the refractivity field $N(x^\alpha)$ for a given fixed vacuum wavelength can be determined when the fields $T(x^\alpha)$, $p(x^\alpha)$, $x_c(x^\alpha)$, and $x_w(x^\alpha)$ are known.

For the special case when the air composition does not change with spacetime position (i.e., $x_c$ and $\rho_w/\rho_a$ are constants in spacetime), we can divide Eq.~(\ref{n_Ciddor}) by the total air density $\rho=\rho_a+\rho_w$, and we obtain that $N/\rho$ is constant in spacetime. In this case, we can write
\begin{equation}
N(x^\alpha)=N_0\frac{\rho(x^\alpha)}{\rho_0} , \label{n_simple}
\end{equation}
where $N_0$ and $\rho_0$ are the air refractivity and the density in a certain reference spacetime point.

\subsection{Air density field of the atmosphere in hydrostatic equilibrium} \label{appendix2}

The temperature of the atmosphere of Earth as a function of altitude and the related pressure and density fields derived under the assumption of hydrostatic equilibrium up to altitudes of about 80~ km are discussed, for example, in \citet{ICAO} and \citet{USatm} and we summarize them in this section.

Assuming that the atmosphere rigidly rotates together with the gravitating body and is in hydrostatic equilibrium with its gravitational and centrifugal forces, the atmospheric pressure~$p$ satisfies
\begin{equation}
\partial_i p =\rho \partial_i \phi , \label{HEE}
\end{equation}
with 
\begin{equation}
\phi=W+v_R^2/2 \label{grav_and_centr}
\end{equation}
being the gravitational plus centrifugal potential. Next, we assume the perfect gas equation of state for the air, 
\begin{equation}
p= \frac{RT}{M_a}\rho , \label{EoS}
\end{equation}
with $R$ being the universal gas constant, $T$ thermodynamic temperature, and $M_a$ the molar mass of the air, which is assumed to be constant. 
Following \citet{ICAO} we define the geopotential altitude 
\begin{equation}
H=-(\phi - \phi_s)/g , \label{geoalt}
\end{equation} 
where $\phi_s$ is the sea-level value of $\phi$, and $g$ is the standard acceleration due to gravity (constant). We note that $H$ has the dimension of length. We assume that the atmosphere can be divided into intervals in $H$ in which the temperature depends on $H$ approximately linearly as
\begin{equation}
T=T_b+\beta (H-H_b), \label{THdep}
\end{equation}
where $H_b$ is the boundary value of $H$ between two intervals, $T_b$ is the temperature at this boundary, and $\beta$ is the constant that is different for each of the intervals (see Table D of \citet{ICAO}).
The solution of the hydrostatic equilibrium equation (\ref{HEE}) for $\rho$ using Eqs.~(\ref{EoS}) and (\ref{THdep}) with $\beta\neq 0$ reads 
\begin{equation}
\rho=\rho_b\left(1+\frac{\beta}{T_b}(H-H_b)\right)^{-\left(1+\frac{M_a g}{\beta R}\right)}, \label{rho_lin}
\end{equation}
and for $\beta =0$, it reads
\begin{equation}
\rho=\rho_b \exp\left(-\frac{M_a g}{RT_b}(H-H_b)\right). \label{rho_isotherm}
\end{equation}

\subsection{Spherically symmetric, static component of the atmospheric refractivity} \label{appendix3}

The time- and frequency-transfer model developed in this paper uses a spherically symmetric, static component of the refractive index $\hat{n}(r)$ (see Eq.~(\ref{def_alpha})) or the related refractivity $\hat{N}(r)=\hat{n}(r)-1$. One option how to define this field is to use Eqs.~(\ref{rho_lin}) and (\ref{rho_isotherm}) together with Eq.~(\ref{n_simple}) and to replace the potential $\phi$ given by Eq.~(\ref{grav_and_centr}) by its monopole part $\hat{W}=GM/r$.

In Sects.~\ref{TT_examples} and \ref{FT_examples}, where we evaluated the magnitudes of the effects for specific examples, we assumed for sake of simplicity an isothermal atmosphere with a constant temperature $T_0$.  From Eqs.~(\ref{rho_isotherm}) and (\ref{n_simple}), we then obtain
\begin{eqnarray}
\rho&=&\rho_0 \exp\left(\frac{M_a}{RT_0}(\phi-\phi_0)\right), \label{barfo}\\
N&=&N_0\exp\left(\frac{M_a}{RT_0}(\phi-\phi_0)\right), \label{n-W}
\end{eqnarray}
where $\rho_0$, $N_0$, and $\phi_0$ are the air density, air refractivity, and the potential in a certain reference spacetime point.
The refractivity $\hat{N}(r)$ is then given as
\begin{equation}
\hat{N}=N_0\exp\left(\frac{M_a}{RT_0}(\hat{W}-\hat{W}_0)\right), \label{hatN_ex}
\end{equation}
where $\hat{W}_0$ is the value of $\hat{W}$ in the reference point.

\end{appendix}

\end{document}